\documentclass[sort&compress]{elsarticle}
\usepackage[utf8]{inputenc}
\usepackage{graphics}

\newcommand{\ai}{\emph{ab initio}}
\newcommand{\EinA}{Einstein \emph{A} coefficients}
\usepackage{xcolor}
\usepackage{longtable}
\usepackage{geometry}
\usepackage{comment}
\newcommand{\red}[1]{{\color{red} #1}}

\title{Analysis of the calculated \emph{X-X} ro-vibrational transition intensities in molecular hydrogen}

\author{V.~G.~Ushakov$^a$}
\author{S.~A.~Balashev$^b$}
\author{E.~S.~Medvedev$^a$}

\date{April 2023}
\address{$^a$Federal Research Center of Problems of Chemical Physics and Medicinal Chemistry (former Institute of Problems of Chemical Physics), Russian Academy of Sciences, 142432 Chernogolovka, Russian Federation}

\address{$^b$A.~F.~Ioffe Physical-Technical Institute, Russian Academy of Sciences, St. Petersburg, Russian Federation}

\begin{document}

\begin{abstract}
    The potential-energy and quadrupole-moment functions of the H$_2$ ground electronic state are well known in literature (Komasa \emph{et al.}, 2019; Wolniewicz \emph{et al.}, 1998), and the line list of the vibrational-rotational transitions was calculated (Roueff \emph{et al.}, 2019). In this paper, we analyze the calculated intensities in order to learn how the intensities will change when analytic quadrupole-moment functions fitted to the \ai\ and experimental data are used instead of spline-interpolated functions. We found that the use of splines does not deteriorate the intensities and does not lead to nonphysical saturation, as in heavier molecules, owing to the high precision of the \ai\ data and the high density of the grid. The accuracy of the calculated intensities is estimated up to high overtones. Extraction of new spectroscopic information from the observational data that supplements the laboratory measurements is performed. The laboratory and observational data do not help increase the quality of the analytic functions. Numerous anomalies resulting from the destructive interference are identified in the calculated line lists, some of them being situated within the recently observed spectral regions, 1.5-2.5 $\mu$m. The intensities of these anomalies can be sensitive to the form of the molecular functions as well as to the proton-to-electron mass ratio. In this connection, the similar Le Roy anomalies (Brown and LeRoy, 1973; Le Roy and Vrscay, 1975) also arising due to the destructive interference in the Lyman and Werner systems are discussed.
\end{abstract}

\maketitle

\section{Introduction}

The significance of H$_2$ for astrophysics has been described in Refs. \cite{HITRAN2020,Ubachs16,Pike16}. The high-quality analytic potential-energy function (PEF) \cite{Piszczatowski09,Pachucki10,Komasa19} and \ai\ quadrupole-moment function (QMF) were created and used to calculate the line lists \cite{Turner77,Wolniewicz98,Campargue12,Roueff19}.

In order to calculate the \EinA, the \ai\ QMF needs, as a rule, to be interpolated in a functional form. Based on our experience with CO \cite{Medvedev15,Medvedev22}, PN \cite{Ushakov23}, and other diatomics \cite{Medvedev22c}, we suspected that using popular splines might introduce inaccuracies to the calculated transition probabilities. In the case of H$_2$, an alternative exists owing to the available high-precision QMFs that permit direct evaluation of the transition-quadrupole-moment (TQM) integral by the sinc-DVR method applied over the dense enough grid of the \ai\ points.
We  were also concerned with the fact that no experimental or observational data on the H$_2$ transition intensities have been used for refining the calculated \EinA.

In this paper, we show that in H$_2$, in contrast to the molecules with heavier atoms, the interpolation of the \ai\ QMF with splines does not have any significant effect on the calculated intensities owing to a very high precision of the \ai\ calculations and a very dense grid of the \ai\ points. It is important that the \ai\ data were calculated with eight significant digits because, as we will see in Sec. \ref{splines}, rounding-off to four digits results in saturation of transition intensities at $\Delta v>6$. 

A drawback of the spline interpolation is that it does not permit evaluation of the accuracy of the calculated intensities. In contrast, we use analytic functions to interpolate the \ai\ QMF, 
analyze the calculated intensities, and provide for estimates of the errors involved. To this end, we constructed a few analytic QMFs that were used, along with the \textbf{Roueff19} \cite{Roueff19} data, for comparison of the calculated intensities in order to obtain the relevant estimates. 

Our analysis of the calculated overtone-transition intensities is based on the NIDL\footnote{Normal Intensity Distribution Law.} theory \cite{Medvedev12}, according to which the intensities must decay exponentially with the overtone number, the rate of the exponential fall-off being dependent only on the steepness of the repulsive branch of the potential. This result was confirmed by both experimental data and calculations for a number of diatomic molecules. A new feature discovered here in H$_2$ is that the rate of the exponential fall-off depends not only on the potential but also on the form of the QMF. We show that this feature is intrinsic for the molecules with small reduced mass like H$_2$.

We constructed two analytic QMFs fitted to all available \ai\ data from Refs. \cite{Wolniewicz98,Campargue12,Roueff19} and one QMF fitted also to the laboratory data \cite{Campargue12,Fink65,Margolis73,Chackerian75a,Chackerian75b,Bergstralh78,Reid78,Trauger78,Brault80,Bragg82,Jennings82,Ferguson93,Reuter94,Gupta06,Robie06,Hu12,Kassi14},  and then we compared the intensities calculated in the present study using these analytic QMFs and the H$_2$\_spectre PEF of \cite{Pachucki10,Komasa19} with both the experimental intensities from the laboratory measurements  and astrophysical observations \cite{Pike16,Oh16,Geballe17,Kaplan17,Le17}.

The paper is organized as follows. Section \ref{analytic} presents our analytic forms to be fitted to the available \ai\ and experimental data described in Sec. \ref{data}. The various fitting procedures detailed in Sec. \ref{fitting} were used to obtain more scatter for the calculated intensities needed to estimate the expected uncertainties of the calculated intensities. Section \ref{splines} discusses the question of how the sinc-DVR method and the spline interpolation of the \ai\ QMF work in H$_2$ as opposed to the molecules with heavy atoms. Our method to analyze the calculated intensities is briefly described in Sec. \ref{Method}. In Sec. \ref{pecul}, important features of the intensity distribution in H$_2$ are outlined. The calculation errors estimated with use of a number of comparison QMFs proposed in Sec. \ref{analytic} are presented in Sec. \ref{determ}. Sections \ref{CompLab} and \ref{CompAstr} are devoted to comparison of the present results with those of Roueff \emph{et al.} \cite{Roueff19} and with the experimental and observational data. Section \ref{anom} describes the anomalies in the calculated \emph{X-X} spectrum; similar Le Roy anomalies in the electronic spectra of H$_2$ are discussed with the emphasis on the possible application to the problem of variation of the proton-to-electron mass ratio. The results are summarised in Sec. \ref{concl}.

\section{The analytic QMFs}
\label{analytic}

We explored two analytic forms whose analytic properties in the complex plane differ significantly.

Function irregK with \emph{K} independent variable parameters is given by
\[
\textrm{irregK}=\frac{\left( 1-e^{-c_{2}r}\right)
^{6}}{r^{4}\sqrt{\left( r^{2}-c_{3}^{2}\right)
^{2}+c_{4}^{2}}}\sum_{i=0}^{N}b_{i}z^{i}\,,
\]%
\[
z=1-2e^{-c_{1}r}\,,
\]%
where $c_{1},...,c_{4,}$ $b_{0},...,b_{N},$ total of $K=N+5$, are the
parameters.

The second function is quadrK:%
\[
\textrm{quadrK}=\left( \frac{a_{1}}{r^{6}+a_{2}}+\frac{a_{3}}{r^{6}+a_{4}}+%
\frac{a_{5}}{r^{6}+a_{6}}\right) \sum_{i=1}^{N}q_{i}y^{i}\,,
\]%
\[
y=\frac{r^{2}}{r^{2}+r_{0}^{2}}\,,
\]%
\[
\sum_{i=1}^{N}q_{i}=1,
\]%
where $r_{0},$ $a_{1},...,a_{6},$ $q_{1},...,q_{N},$ are variable parameters, of which $K=N+6$ are independent. 

The number of parameters depends on the details of the fitting procedure, in particular, the data sets selected for fitting and the weights and uncertainties assigned to the data.

For our purposes, we constructed two functions with different analytic properties: irregK has branching points in the complex plane but no poles; on the contrary, quadrK has poles but no branching points. Both kinds of singularities can affect the intensities of higher overtones, and it is essential that functions with different properties gave the intensities with a minimized scatter.

\section{The input data sets} 
\label{data}

Wolniewicz \emph{et al.} \cite{Wolniewicz98} calculated the \ai\ QMF at $r=0.2$-20 au (data set I, grid of 257 points) with the estimated relative precision of $10^{-5}$, which constitutes the absolute value of $1\cdot10^{-5}$ au. Komasa with coworkers \cite{Campargue12,Roueff19} calculated another \ai\ QMF at $r=0.1$-10 au (data set II, grid of 31 points. We are grateful to J. Komasa for sending us the unpublished original data file \textbf{Q2ofR}, which was used to interpolate and average the QMF in several works \cite{Campargue12,Roueff19,Pachucki11}. These data were obtained using explicitly correlated Gaussian functions.).
Within the interval of $1<r<2$ au, the spline-interpolated set II QMF differs from the set I \ai\ QMF by $10^{-6}$-$10^{-7}$ au, but outside this interval the difference reaches $10^{-3}$ au (see further discussion in supplementary file 1 and in Sec. \ref{CompLab}). The spline-interpolated
set II \ai\ QMF was used in \cite{Wolniewicz98,Campargue12,Roueff19} to calculate the \emph{X-X} transition probabilities.

The experimental data used in the fits (set III) include the laboratory measurements prior 2012 summarised in Ref. \cite{Campargue12} plus data of Ref. \cite{Kassi14}. They are collected in supplementary file 2.

\section{The fitting procedure}\label{fitting}

Fitting was performed by minimization of the following functional:
\begin{eqnarray}
   \chi^2=\sum_{\alpha}\chi_\alpha^2,\label{chi2}\\ 
   \chi_\alpha^2=\sum_i\left(\frac{y_{\alpha i}^{\textrm{fit}}-y_{\alpha i}^{\textrm{dat}}}{\sigma_{\alpha i}}\right)^2,
    \label{chi2a}
\end{eqnarray}
where $y^\textrm{dat}_{\alpha i}$ are the \emph{ab initio} QMF points of sets I and II or measured intensities of set III and $\chi_\alpha^2$ are contributions of the individual data sets. We also will use the standard deviations,
\begin{equation}
 \textrm{std}=\sqrt{\frac{1}{N_\alpha}\sum_i\left({y_{\alpha i}^{\textrm{fit}}-y_{\alpha i}^{\textrm{dat}}}\right)^2},
\end{equation}
where $N_\alpha$ are the individual numbers of the data points. 

Both analytic forms from Sec. \ref{analytic} with various numbers of variable parameters \emph{K} were fitted to various combinations of the data sets and various assignments of the point uncertainties, $\sigma_{\alpha i}$. 
The following functions will be used further. 

The irreg15 QMF was fitted to sets I and II with $\sigma_{i,\alpha}=$ const (the least-squares method); the std of the fit is $2\cdot10^{-5}$ au, which corresponds to the claimed uncertainty of the  \ai\ QMFs. This function gives the intensities very close to those of \textbf{Roueff19} and will be used to obtain an estimate of the uncertainties of the calculated intensities, see Sec. \ref{determ}.

The irreg15exp QMF was fitted to sets I-III with $\sigma_{i,\alpha}=0.001$ au for the \ai\ data and the experimental uncertainties for set III. The increased $\sigma_{i,\alpha}$ with respect to the claimed uncertainty of the \ai\ data was used in order to minimize the deviation of the theory from experiment at the highest $v$, see Sec. \ref{CompLab}.

The quadr16 and quadr19 functions where fitted to sets I and II with $\sigma_{\alpha i}$ being inversely proportional to the square root of the \ai\ QMF in order to increase the accuracy of reproducing the \ai\ data in the range 1--4 au essential for the TQM calculation. 

The FORTRAN codes for the irreg15, irreg15exp, and quadr16 functions along with their outputs are given in supplementary files 3-6. The functions, bond length, and fitted parameters are in au. We use the definition of QMF as in \cite{Wolniewicz98}, \emph{i.e.} the QMF is two times larger than in \cite{Komasa19}. The output files contain the calculated QMFs and their comparison with the \ai\ data as well as the calculated Einstein \emph{A} coefficients in comparison with the laboratory data. Supplementary file 7 contains the TQMs calculated with three QMFs for the S(0), Q(1), and O(2) lines. The fitted parameters of three functions are given in supplementary file 8 in the ASCII format.

The intensity calculations with all the above QMFs except for quadr19 were performed with the H2\_spectre potential of \textbf{Komasa19} \cite{Komasa19}; for quadr19, the PEF of \textbf{Piszczatowski09} \cite{Piszczatowski09} was used. 

\section{Splines \emph{vs} sinc-DVR}\label{splines}

Our method of analysis of the computed data is based on the NIDL theory, which was verified by numerous measurements on diatomic molecules and quasi-diatomic local vibrations in polyatomic molecules, see detailed review \cite{Medvedev12} and references therein.

In order to check the validity of the dipole- (or quadrupole-) moment interpolation using splines, we plot the calculated intensities in the NIDL coordinates in order to see possible nonphysical saturation \cite{Medvedev22c}. In the case of H$_2$, an additional test is available due to a very dense uniform grid of the \ai\ points, which permits direct application of the sinc-DVR method\footnote{As always, we use our home-made codes to fit the analytic QMFs, to solve the Schr\"odinger equation, and to calculate the \EinA.} using this grid to evaluate the TQM integral directly, without interpolation of the QMF. 

\begin{figure}[htbp]
    \centering
    \includegraphics[scale=0.2]{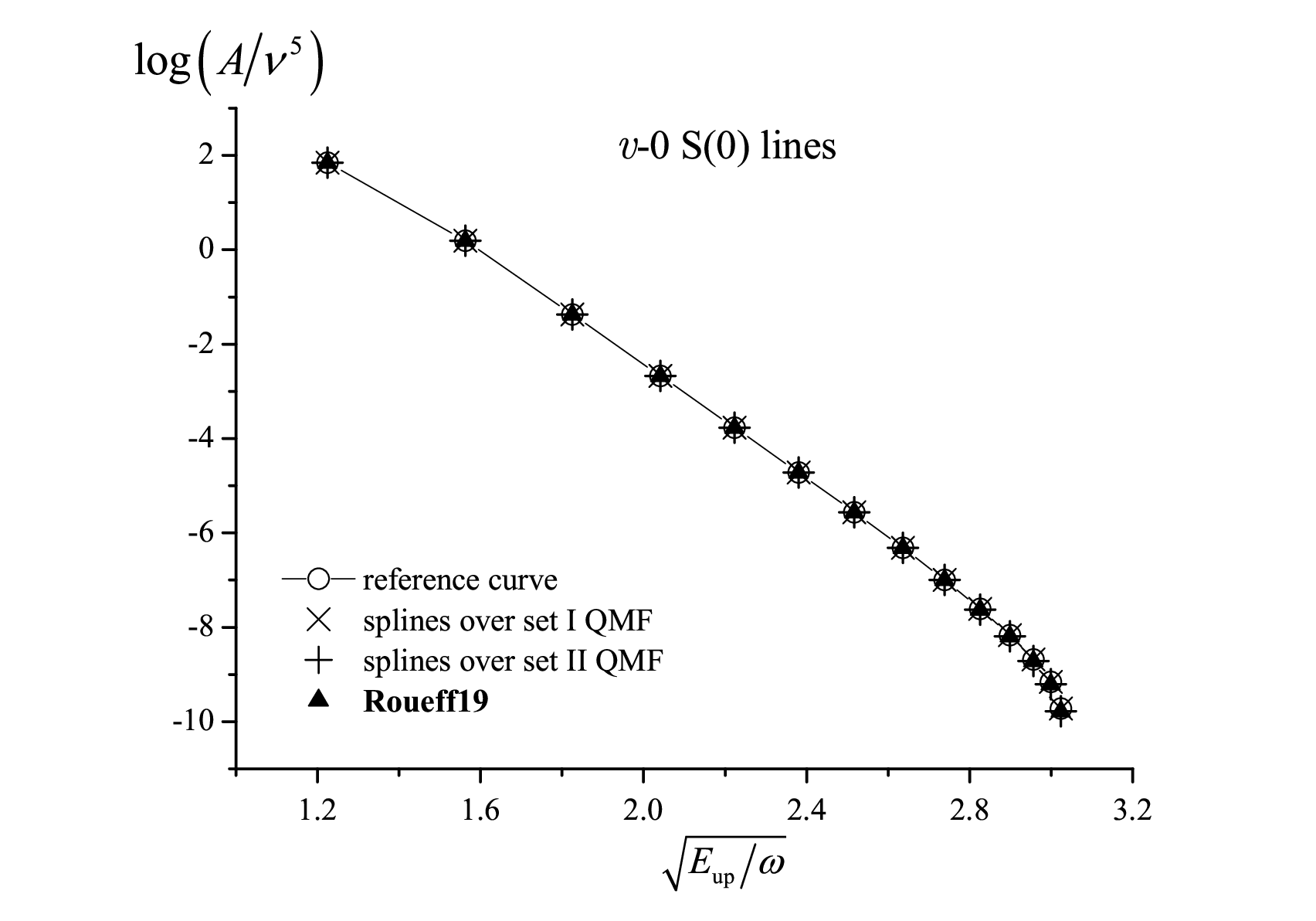}
    \includegraphics[scale=0.2]{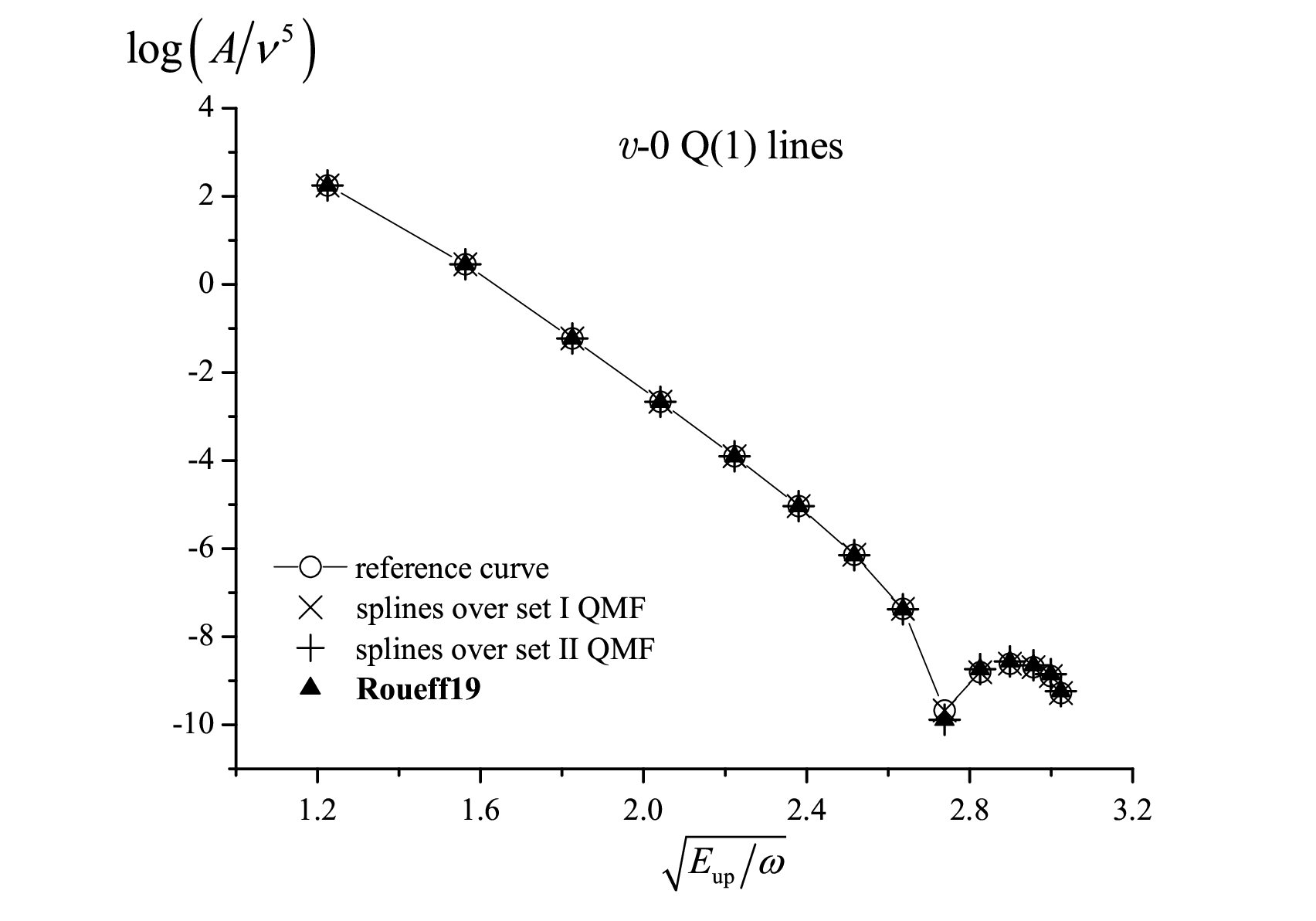}
    \includegraphics[scale=0.4]{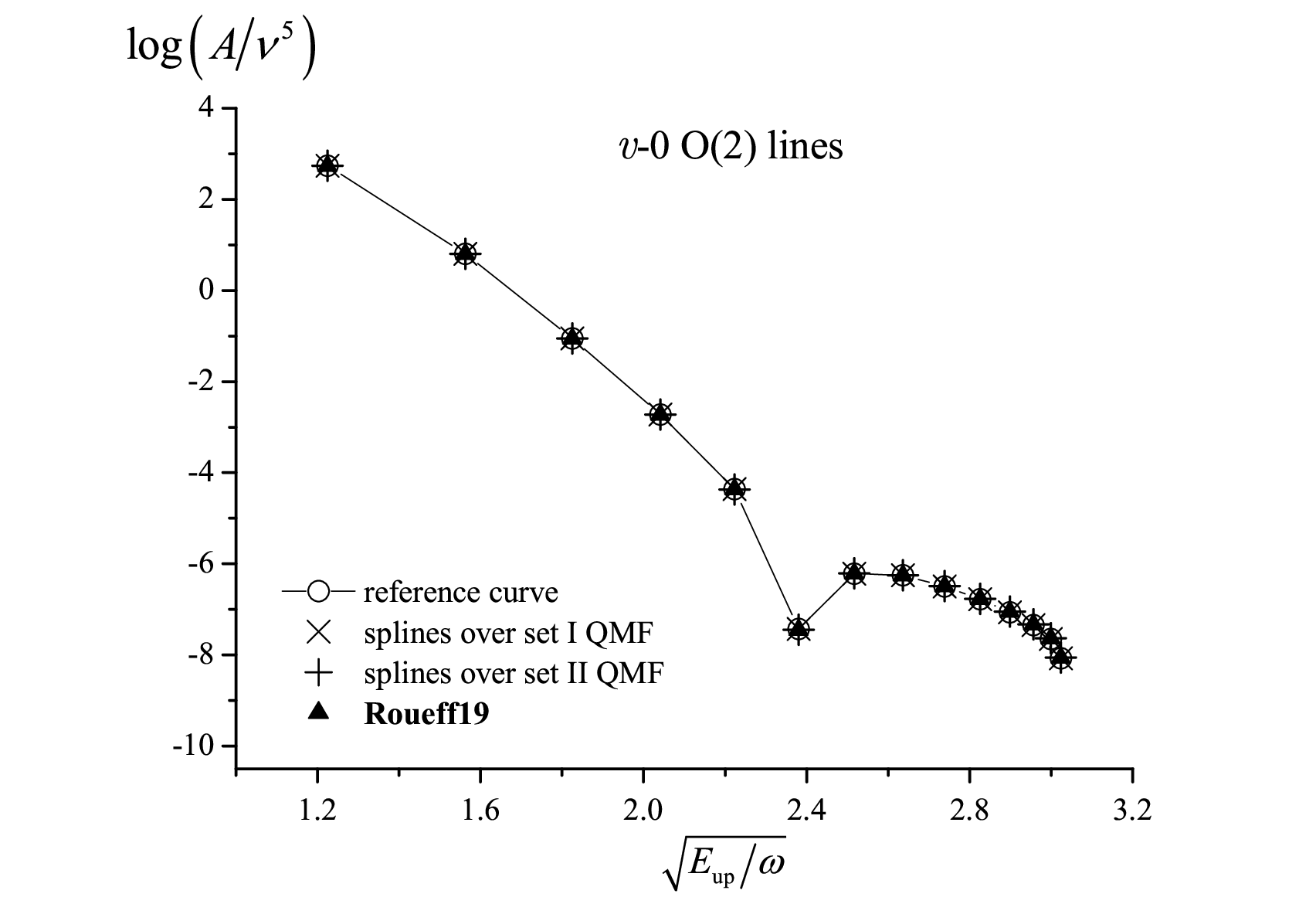}
    \caption{Comparison of the calculated $v$--0 transition intensities with the results of \textbf{Roueff19} \cite{Roueff19}. Calculations in the present study, if not specified otherwise, were performed with the H2\_spectre PEF of \textbf{Komasa19} \cite{Komasa19}. Circles with line, the sinc-DVR method using the \ai\ grid of the original (with 8 decimal places) set I QMF \cite{Wolniewicz98}, no interpolation; crosses, set I QMF interpolated with cubic splines; pluses, set II QMF \cite{Campargue12,Roueff19} interpolated with cubic splines; triangles, Ref. \cite{Roueff19}. $\nu$, transition frequency, au; $E_\textrm{up}$, the upper-state energy, cm$^{-1}$; $\omega\approx4160$ cm$^{-1}$, harmonic frequency; $v=1$--14.}
    \label{splDVR}
\end{figure}

The results for the $v$-0 transitions at $v=1$-14, \emph{i.e.} up to the highest bound vibrational level, are shown in Fig. \ref{splDVR} where ratio $A/\nu^5\propto$ TQM$^2$ is plotted in the NIDL coordinates because TQM$^2$ is expected to show the NIDL-like behavior (see Sec. \ref{Method}); this ratio will be called ``intensity" for brevity. 
The calculations were performed with the H2\_spectre potential of \textbf{Komasa19} \cite{Komasa19}. 

First, no signs of saturation is seen. 

Second, the sinc-DVR method using the grid of set I \ai\ QMF (circles with line), which is considered as ``exact" and will be further used as reference, is compared with two other ``exact" calculations with use of the set I QMF (crosses) or set II QMF (pluses, previously used in \textbf{Roueff19}) interpolated with splines. The close coincidence of the three calculations suggests that splines do work excellently. 
It should be noted that in the second and third curves in Fig. \ref{splDVR}, as well as in all subsequent calculations, as distinct from the calculation of the first (reference) curve, the Schr\"odinger equation was solved by the sinc-DVR method using the adaptive analytical mapping approach of Meshkov \emph{et al.} \cite{Meshkov06}.

Third, our exact sinc-DVR intensity calculations perfectly coincide with the results obtained with the set I spline QMF (crosses) at all transitions including the anomalies (see Sec. \ref{anom}) at $v=9$ for Q(1) lines and $v=6$ for O(2). Numerically, the differences between sinc-DVR and spline do not exceed 0.001 \% at $v\le13$ and 0.1--0.4\% at $v=14$. 

Thus, we come to the very important conclusions that in H$_2$, as distinct from the molecules with large reduced mass, splines do not deteriorate the high-overtone intensities and that the sinc-DVR method built on the extremely precise \ai\ data over the very dense grid is equivalent to the use of a certain analytic function, which is very close, yet not identical, to the true molecular QMF.

\begin{figure}[htbp]
    \centering
    \includegraphics[scale=0.2]{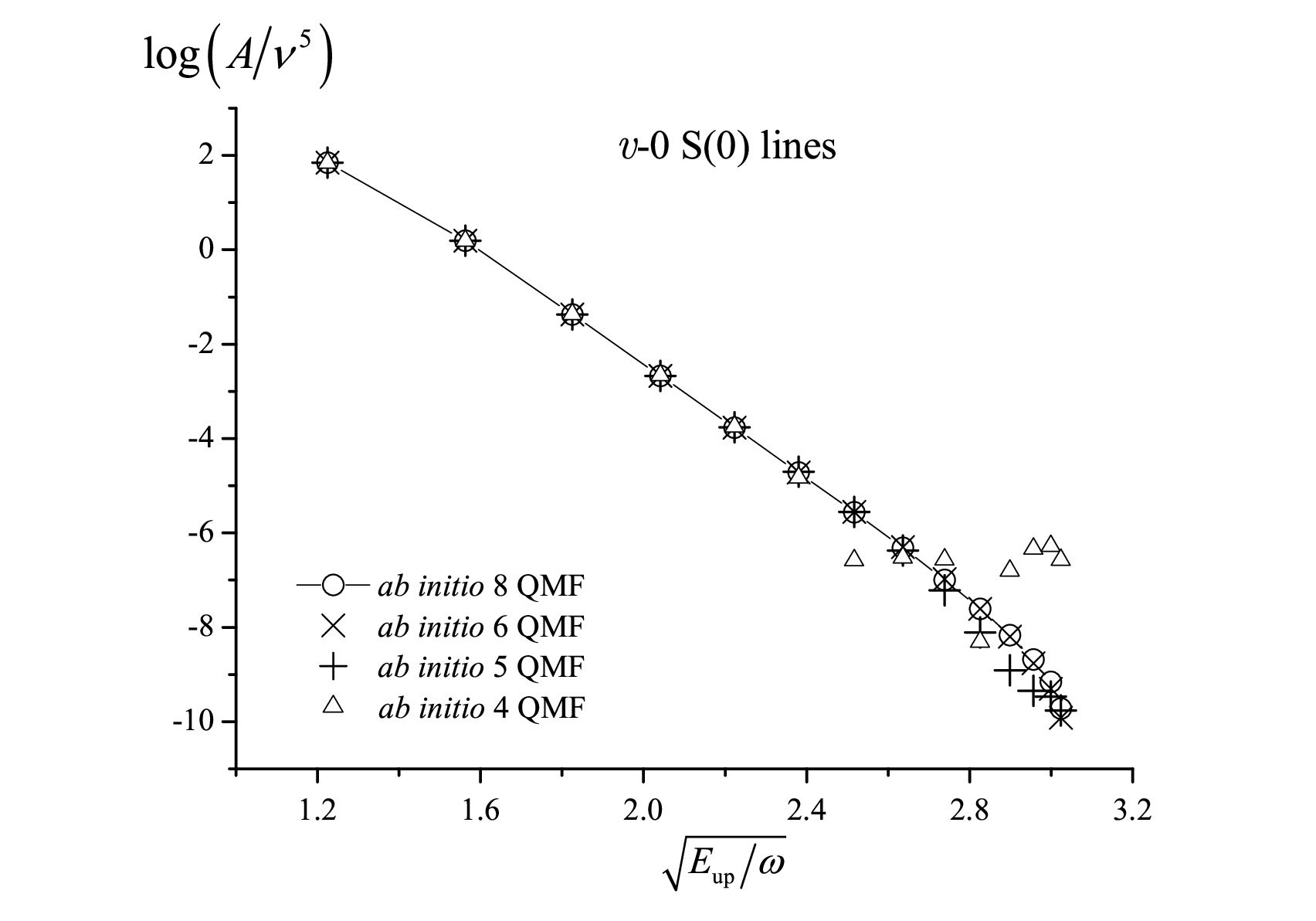}
    \includegraphics[scale=0.2]{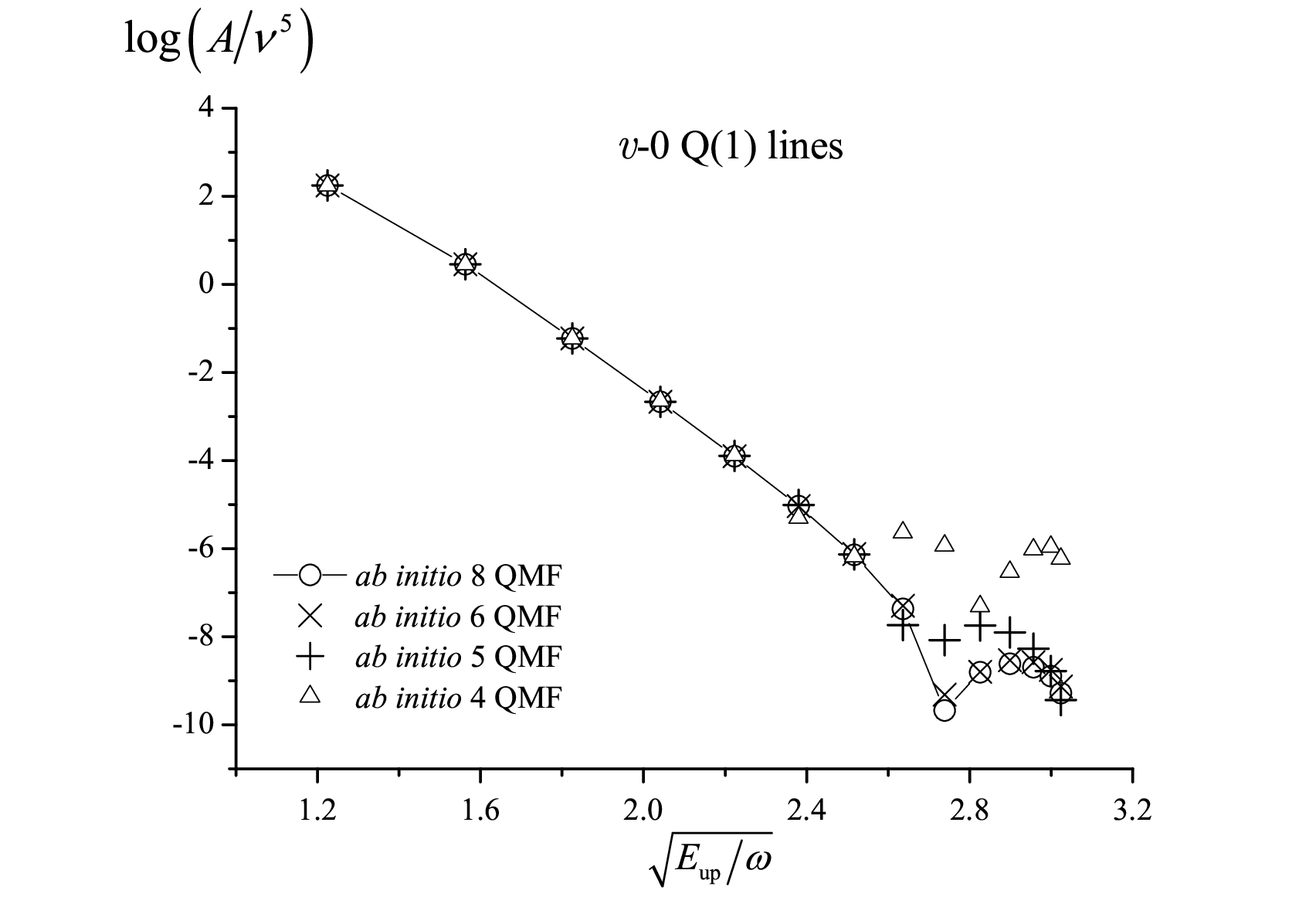}
    \includegraphics[scale=0.5]{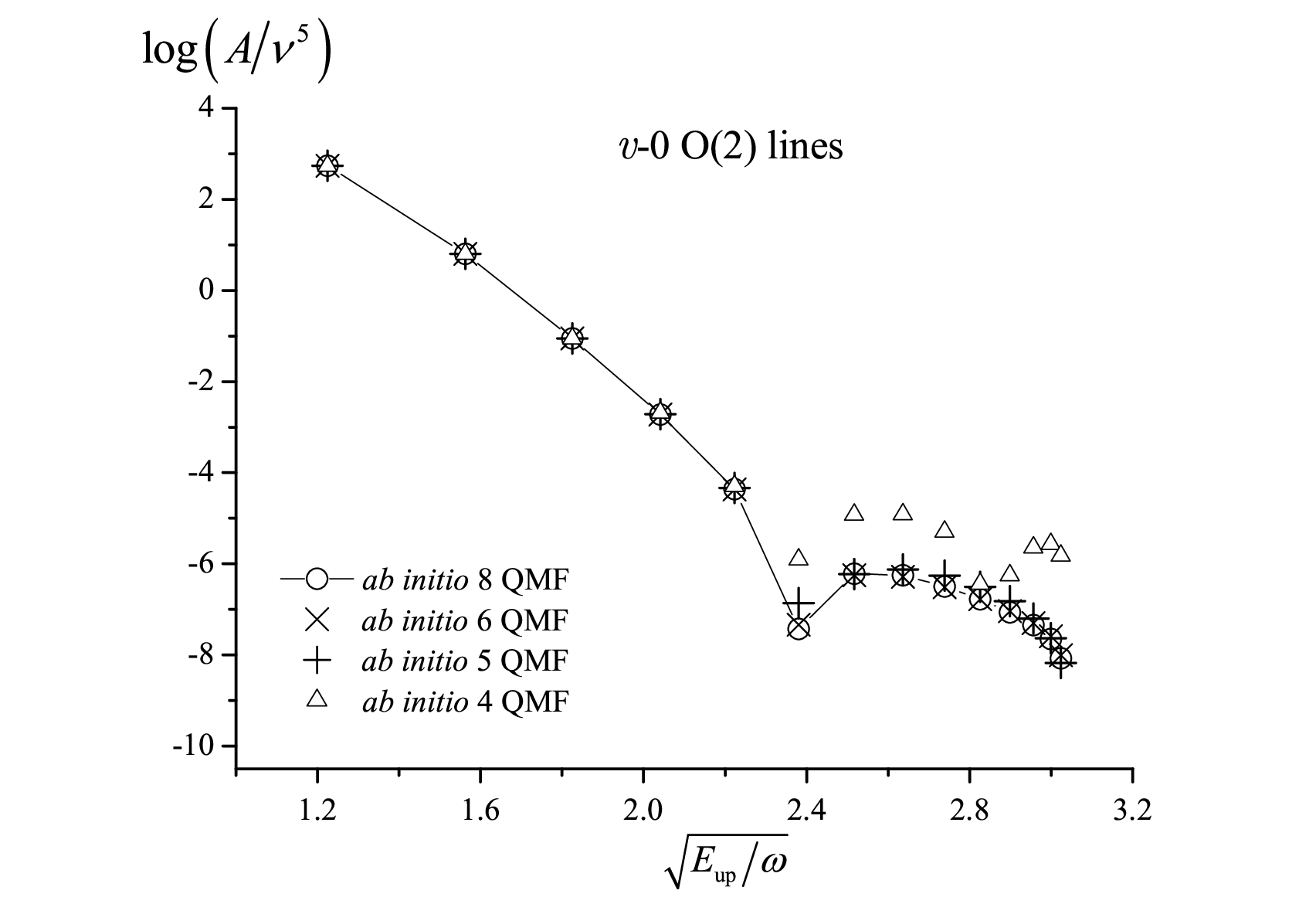}
    \caption{The $v$--0 transition intensities calculated by the sinc-DVR method over the \ai\ grid with use of the set I \ai\ QMF values rounded to 6, 5, or 4 decimal places (no interpolation). The reference curve (circles+line, the same as in Fig. \ref{splDVR}) is calculated with the 8-digit original QMF of Ref. \cite{Wolniewicz98}. $v=$1--14.}
    \label{round}
\end{figure}

The latter property deserves special consideration. The declared relative accuracy of the \ai\ calculated QMF is 10$^{-5}$ \cite{Wolniewicz98}, which is equivalent to five significant decimal places. Hence, decimal places 6-8 contain systematic errors associated with the \ai\ method used. If one would change the method, a different analytic function would be obtained and then could be used as a comparison function in order to estimate the uncertainties of the calculated intensities. 

Here, we explore a different method to construct the comparison functions. First, we introduce random errors into the \ai\ QMF by rounding-off the original data to 6, 5, or 4 decimal places in order to determine the actual number of the reliable digits. The results are shown in Fig. \ref{round}. It is seen that 6 decimal places is sufficient to recover the data obtained with the original 8-digit QMF; 5 places slightly modify the TQM, and 4 digits destroy it significantly. 

For determination of the expected uncertainties of the calculated overtone intensities in Sec. \ref{determ}, we will use, as comparison functions, the irreg15, quadr16, and (with caution) irreg15exp QMFs defined in Sec. \ref{fitting}.

\section{Method of analysis of the calculated intensities}\label{Method}

In our method 
based on the NIDL theory, the wave functions are written in the  quasi-classical form in terms of the classical momenta and actions \cite[Appendix]{Medvedev22}. Then the transition quadrupole moment (TQM) is obtained in the form \cite[Eq. (A.2)]{Ushakov23}
\begin{equation}\label{T0B0}
    \textrm{TQM}=T_0B_0,
\end{equation}
\begin{equation}\label{T0}
    T_0= \exp{\left(-\Delta\Phi_0\right)},
\end{equation}
where $\Delta\Phi_0$ is the difference between the imaginary parts of the action integrals (in units of $\hbar$),
\begin{equation}\label{Phi}
    \Phi_{vJ}=\hbar^{-1}\sqrt{2\mu}\int_{r_0}^{r^-_{vJ}}\sqrt{U_J(r)-E_{vJ}}dr,
\end{equation}
in the lower ($v^{\prime\prime}J^{\prime\prime}$) and upper ($v^\prime J^\prime$) states. The integral is taken from a point $r_0$ where the potential is very large, $U_J(r_0)\gg E_{vJ}$, till the left turning point at a given energy, $E_{vJ}$. Factor $T_0$ is responsible for the NIDL-like decay of the overtone intensities, factor $B_0$ can generate the anomalies. Factor $B_0$ is given by some integral in the complex plane along a contour circumventing the turning points in the upper complex half-plane. The integrand depends on the semiclassical wave functions and on the QMF calculated at the complex points along the integration contour. The explicit expression for $B_0$ is given in \cite[Eq. (A.9)]{Ushakov23}.

In fact, as we have demonstrated earlier \cite{Medvedev22,Ushakov23}, this quasi-classical method gives practically exact results for heavier molecules, CO and PN. In order to verify the validity of the method in application to a light molecule, H$_2$, we need to compare the calculations using Eqs. (\ref{T0B0})-(\ref{Phi}) with the exact results obtained by solving the Schr\"odinger equation. To this end, we used the analytic potential from \textbf{Piszczatowski09} \cite{Piszczatowski09} and the analytically interpolated \ai\ QMF of sets I and II, quadr19, to calculate the TQM integral both exactly and quasi-classically, by Eqs. (\ref{T0B0})-(\ref{Phi}). 

\begin{figure}[htbp]
    \centering
    \includegraphics[scale=0.2]{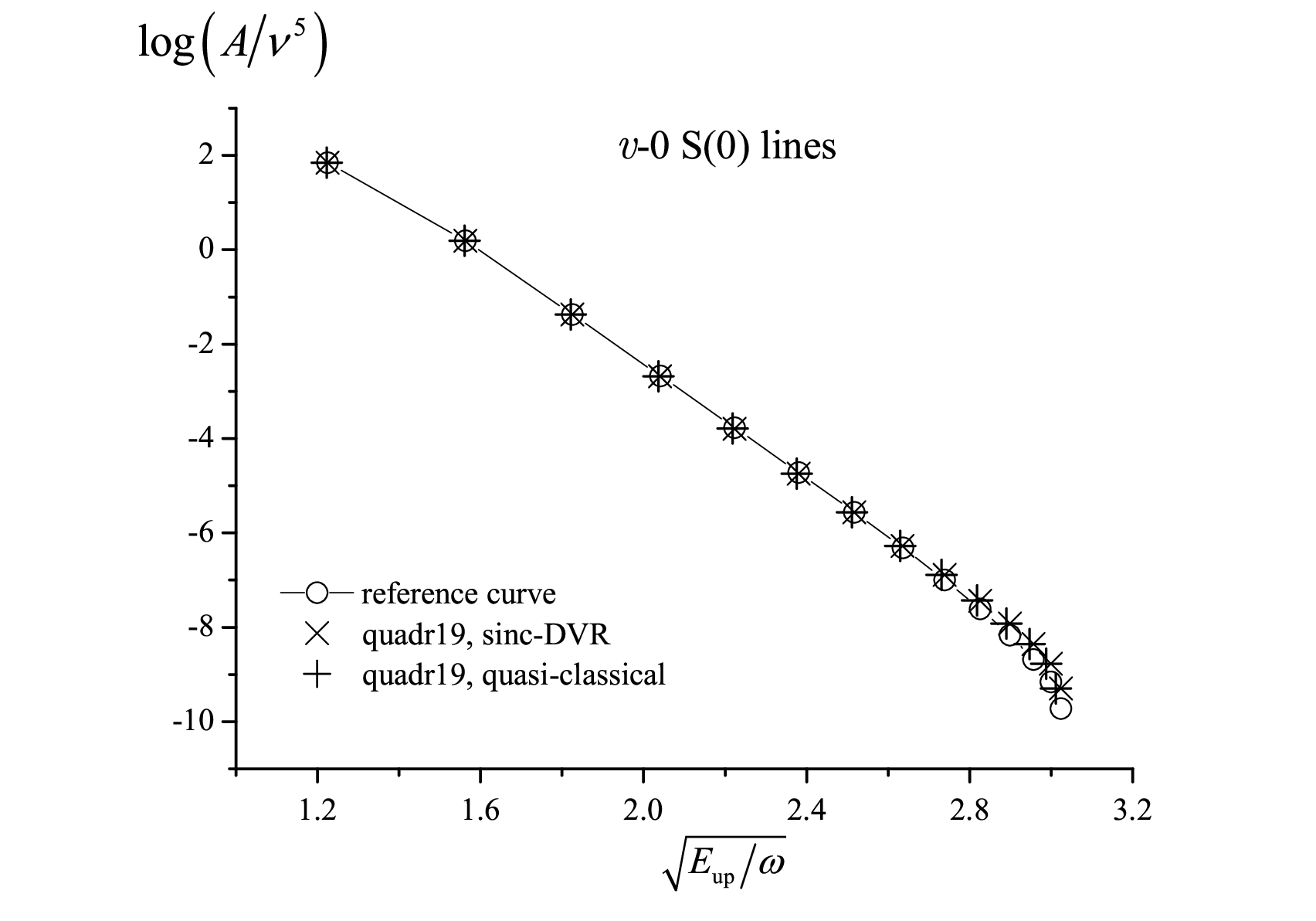}
    \includegraphics[scale=0.2]{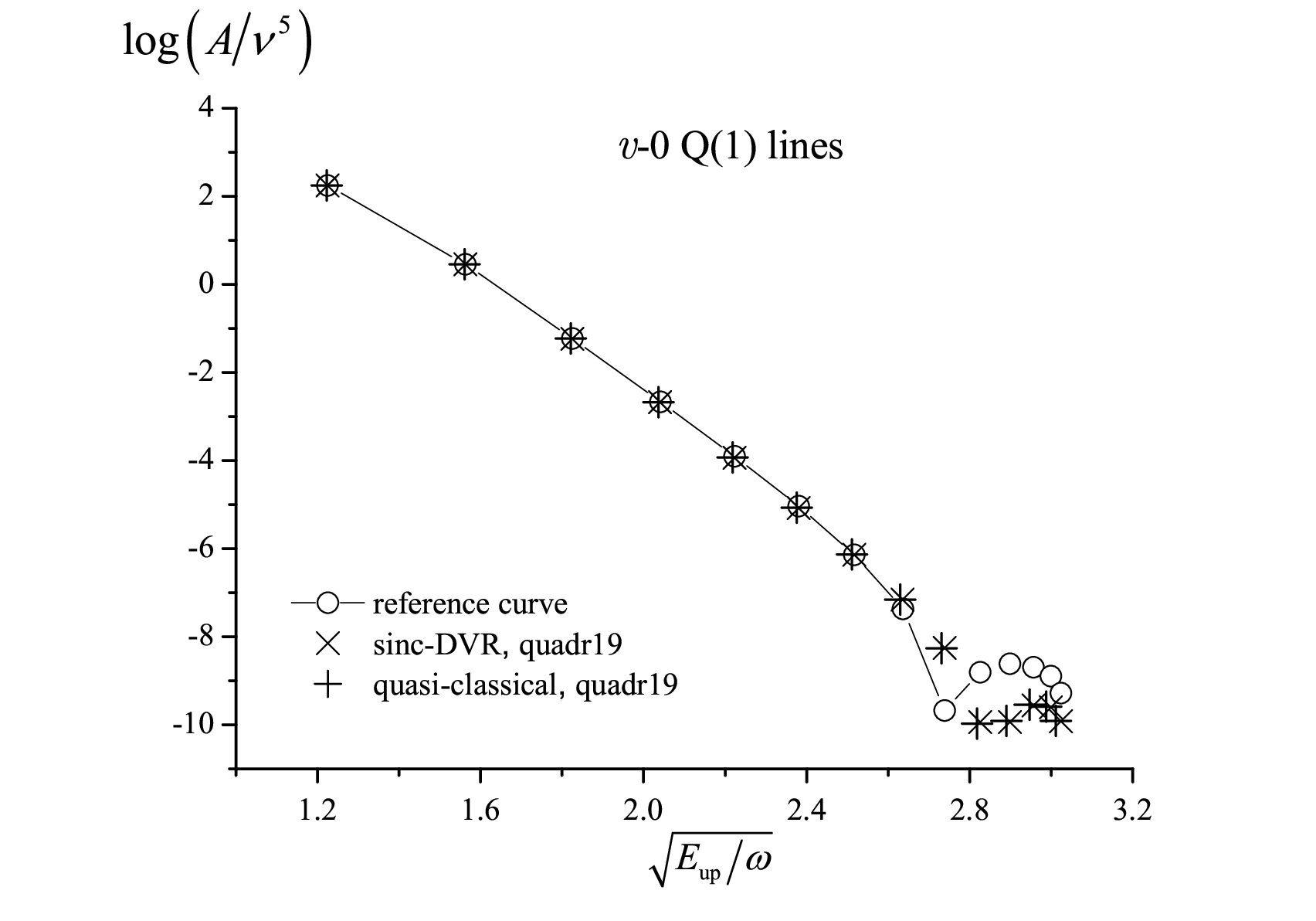}
    \includegraphics[scale=0.5]{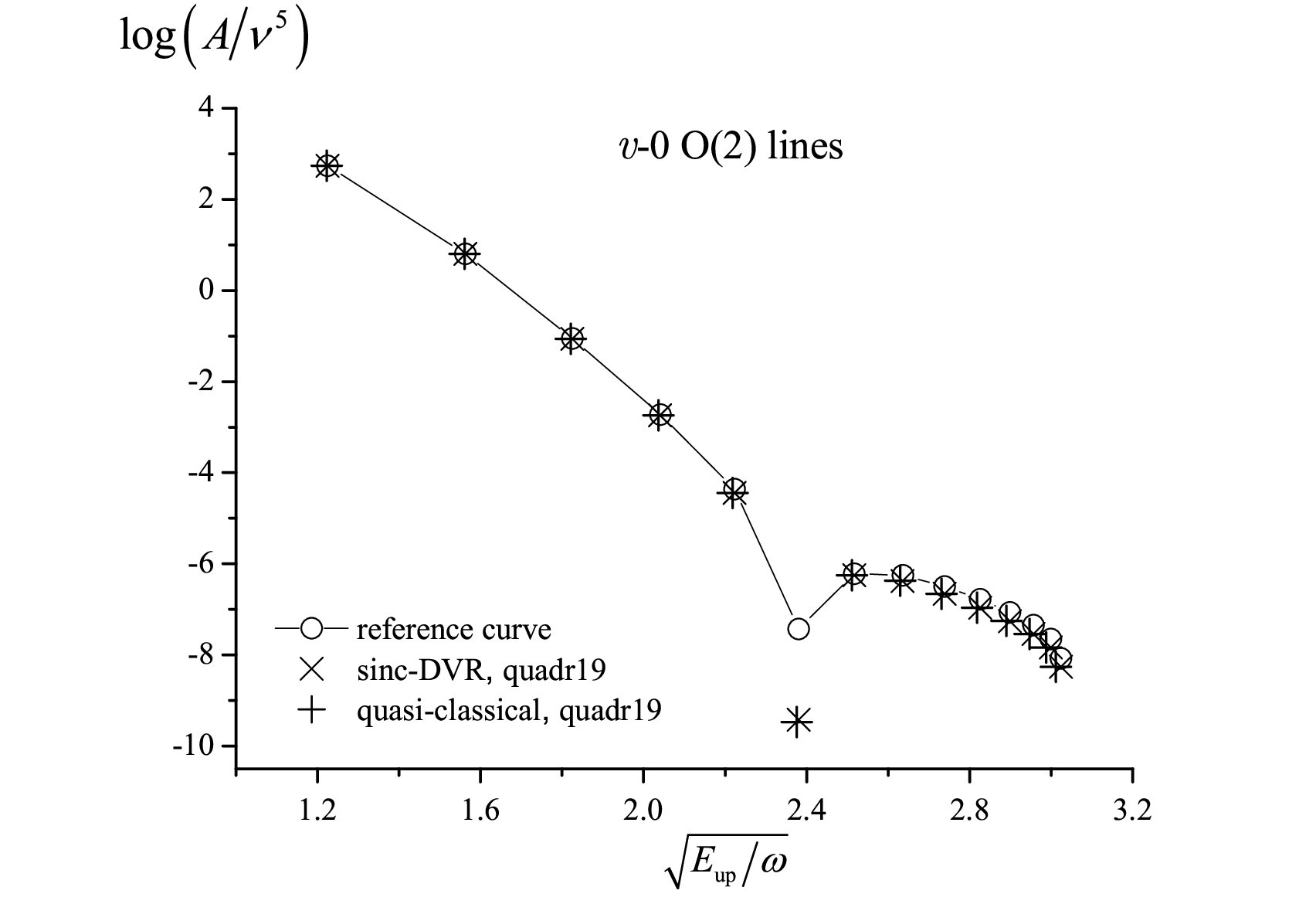}
    \caption{Comparison of the quasi-classical calculations of the intensities for the low-\emph{J} $v$--0 transitions with the exact results.
    Circles, the reference curve from Fig. \ref{splDVR}; crosses, the exact sinc-DVR calculation with the PEF of \textbf{Piszczatowski09} \cite{Piszczatowski09} and the analytic quadr19 QMF fitted to both sets I and II; pluses, the TQM integral is calculated by the quasi-classical Eqs. (\ref{T0B0})-(\ref{Phi}) with use of the same PEF and QMF as crosses. $v=$1--14.}
    \label{SQO}
\end{figure}

The results are shown in Fig. \ref{SQO}; as in Fig. \ref{splDVR}, we divided \emph{A} (in s$^{-1}$) by $\nu^5$ ($\nu$ is transition frequency in au) 
because  TQM$^2\propto A/\nu^5$ is expected to manifest the NIDL-like dependence in the form of a straight line. The exact calculations are presented by the circles+line data, our reference curve (H2\_spectre + sinc-DVR  \red{over the set I} uniform grid), and also by our model QMF, quadr19, fitted to both sets I and II (crosses). The quasi-classical calculation (pluses) was performed with the same quadr19 QMF. Crosses and pluses are calculated with the \textbf{Piszczatowski09} \cite{Piszczatowski09} potential, which slightly differs from H2\_spectre PEF of \textbf{Komasa19} \cite{Komasa19} used for the reference curve.

As is seen in the figure, the quasi-classical method (pluses) is highly accurate, there are no visible differences with the exact calculation (crosses) at the given scale; the differences with the reference curve at the anomalies are due to a different representation of the QMF.

The figure demonstrates once again that replacing splines with an analytic function changes the intensities very insignificantly except for the anomaly. The anomalies (see Sec. \ref{anom}) are a very special phenomenon because they are highly sensitive to the form of the moment function \cite{Medvedev18} that is never known for sure; therefore, only experiment can establish which one is closer to the true molecular function than the others.  The NIDL predicted by Eqs. (\ref{T0B0})-(\ref{Phi}) is indeed obeyed except for the anomalies.

As a side result, the effect of the small difference between the potentials of Refs. \cite{Piszczatowski09} and \cite{Komasa19} on the intensity of the 6-0 O(2) anomaly was estimated to be 10\% for the quadr19 QMF.  

The above-noted agreement between the quasi-classical and exact results means that the quasi-classical method can be used for the analysis of the calculated data. Yet, in what follows, all the TQM integrals, except where specially noted, are computed exactly, by solving the Schr\"odinger equation with use of the sinc-DVR method and the analytic functions of Sec. \ref{analytic} fitted to the data of Sec. \ref{data}.

\section{Peculiarities of the intensity distribution in H$_2$}
\label{pecul}

In the quasi-classical equations (\ref{T0}) and (\ref{Phi}), factor $T_0$ depends only on the repulsive branch of the potential-energy function (PEF), while factor $B_0$ depends on both the PEF and the QMF \cite{Medvedev22,Ushakov23}. In Eq. (\ref{T0B0}), $T_0$ is responsible for the fast intensity decay with increasing the overtone number whereas $B_0$ is expected to slowly vary with it. This feature was demonstrated for CO \cite{Medvedev22} and PN \cite{Ushakov23}. However, in the case of H$_2$, the situation is different. 

\begin{figure}[htbp]
    \centering
    \includegraphics[scale=0.2]{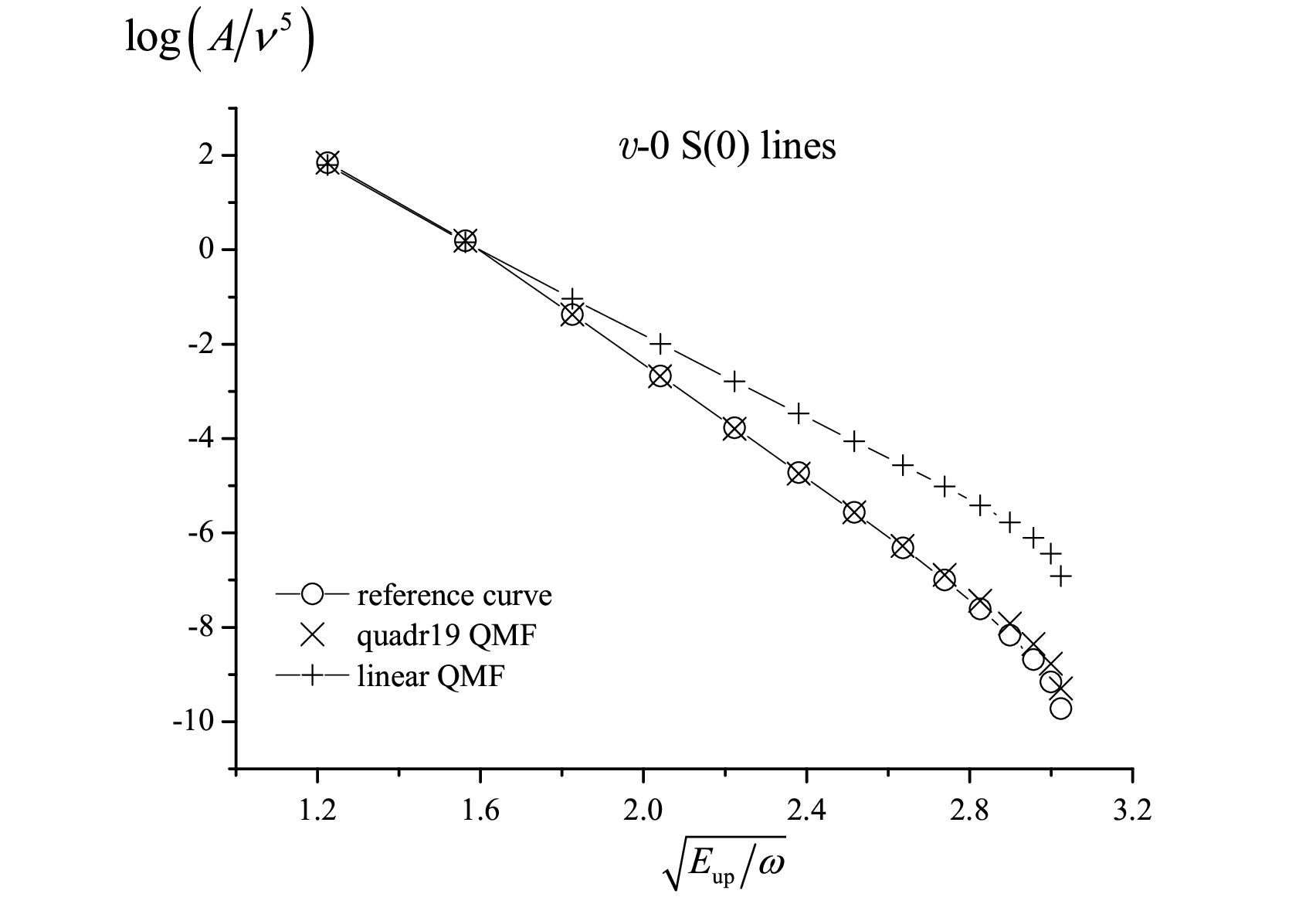}
    \includegraphics[scale=0.2]{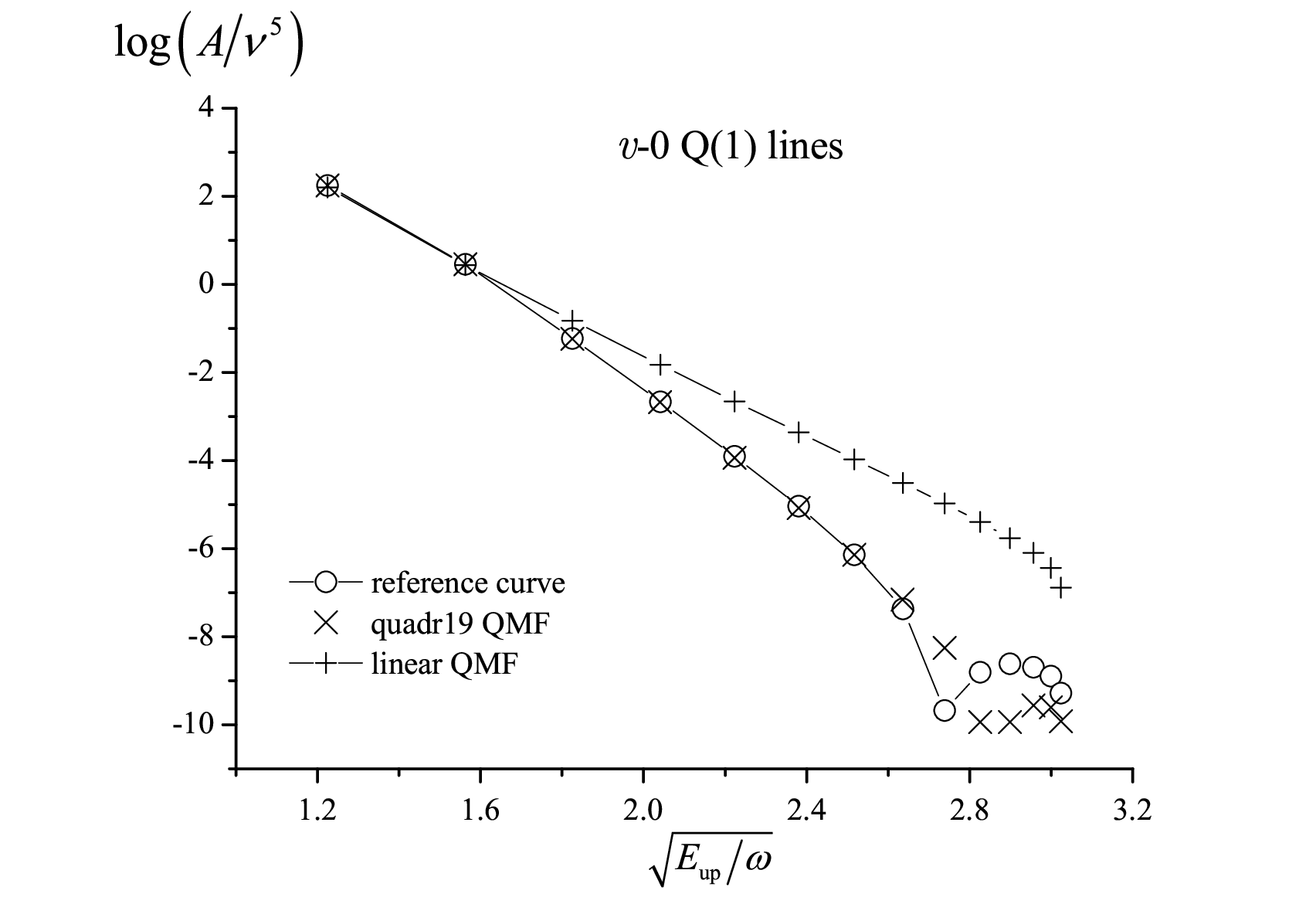}
    \includegraphics[scale=0.5]{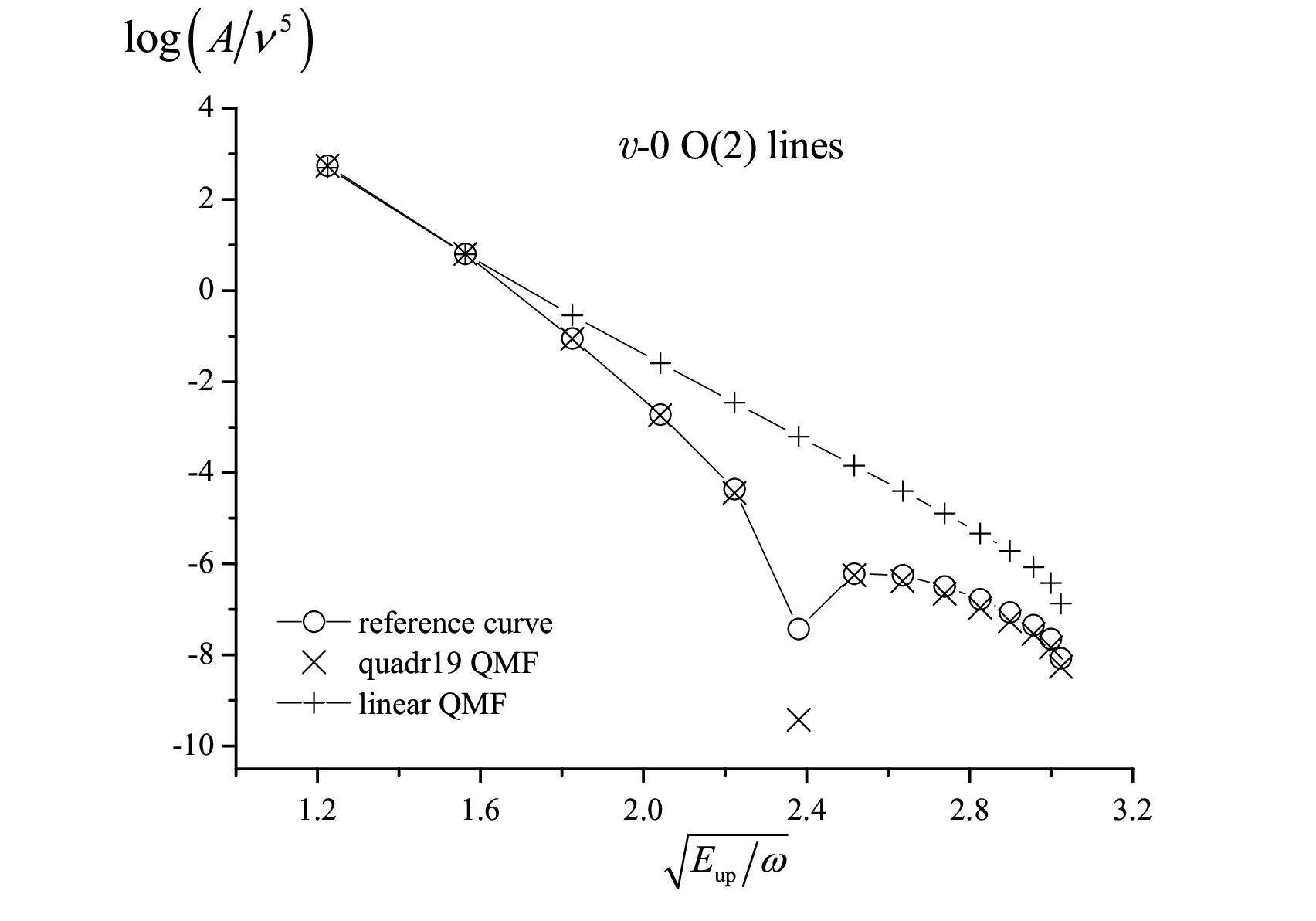}
    \caption{The $v$--0 intensities calculated with three QMFs, see text. $v=$1--14.}
    \label{quadr_linear}
\end{figure}

Figure \ref{quadr_linear} shows the intensities of three lines in the $v$--0 bands, each being calculated with three QMFs, namely, set I QMF as in Fig. \ref{splDVR} (our reference curve) and two analytic functions: the quadr19 QMF and the linear QMF. The latter is expected to show the pure NIDL with no anomalies that usually do not happen with the linear QMFs.

First, we note that all curves indeed manifest the NIDL-like behavior in the form of straight lines except for the anomalies at $v=6$ (the O(2) lines, the reference and quadr19 curves) or at $v=$ 9-10 (the Q(1) lines, the reference and quadr19 curves). Both the NIDL and the anomalies are in line with the general theory of the intensity distributions in the overtone vibrational spectra \cite{Medvedev12}.


Second, and most importantly, the NIDL slope for the linear QMF is significantly different from two other QMFs, which  contradicts to the statement of the general theory \cite{Medvedev12} that the NIDL slope depends mostly on the potential via factor $T_0^2$ whereas $B_0^2$ is a slow function of the overtone number, as took place for CO and PN. \textbf{In the case of H$_2$, $B_0^2$ is not a slow function}.

\begin{figure}[htbp]
    \centering
    \includegraphics[scale=0.3]{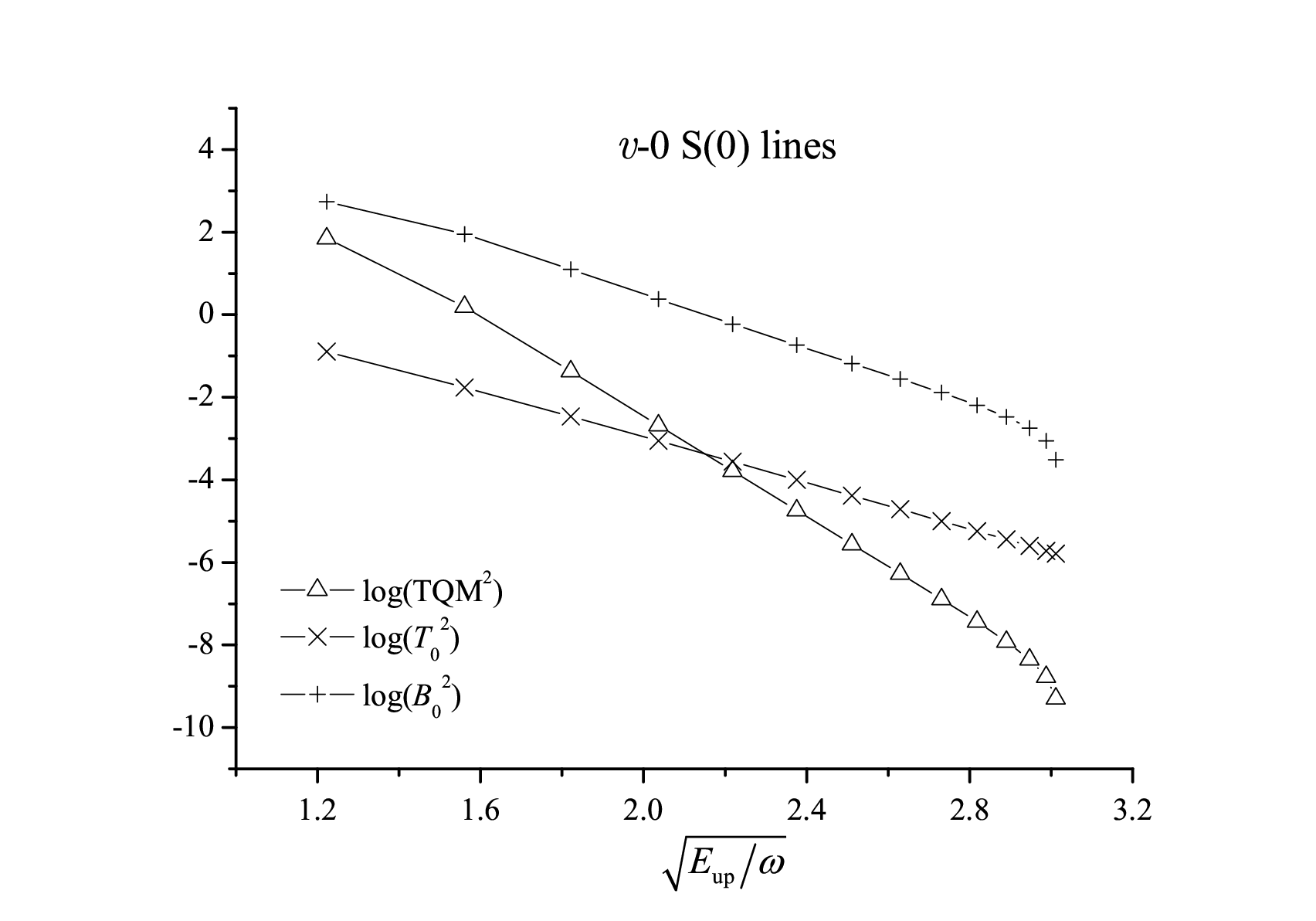}
    \includegraphics[scale=0.3]{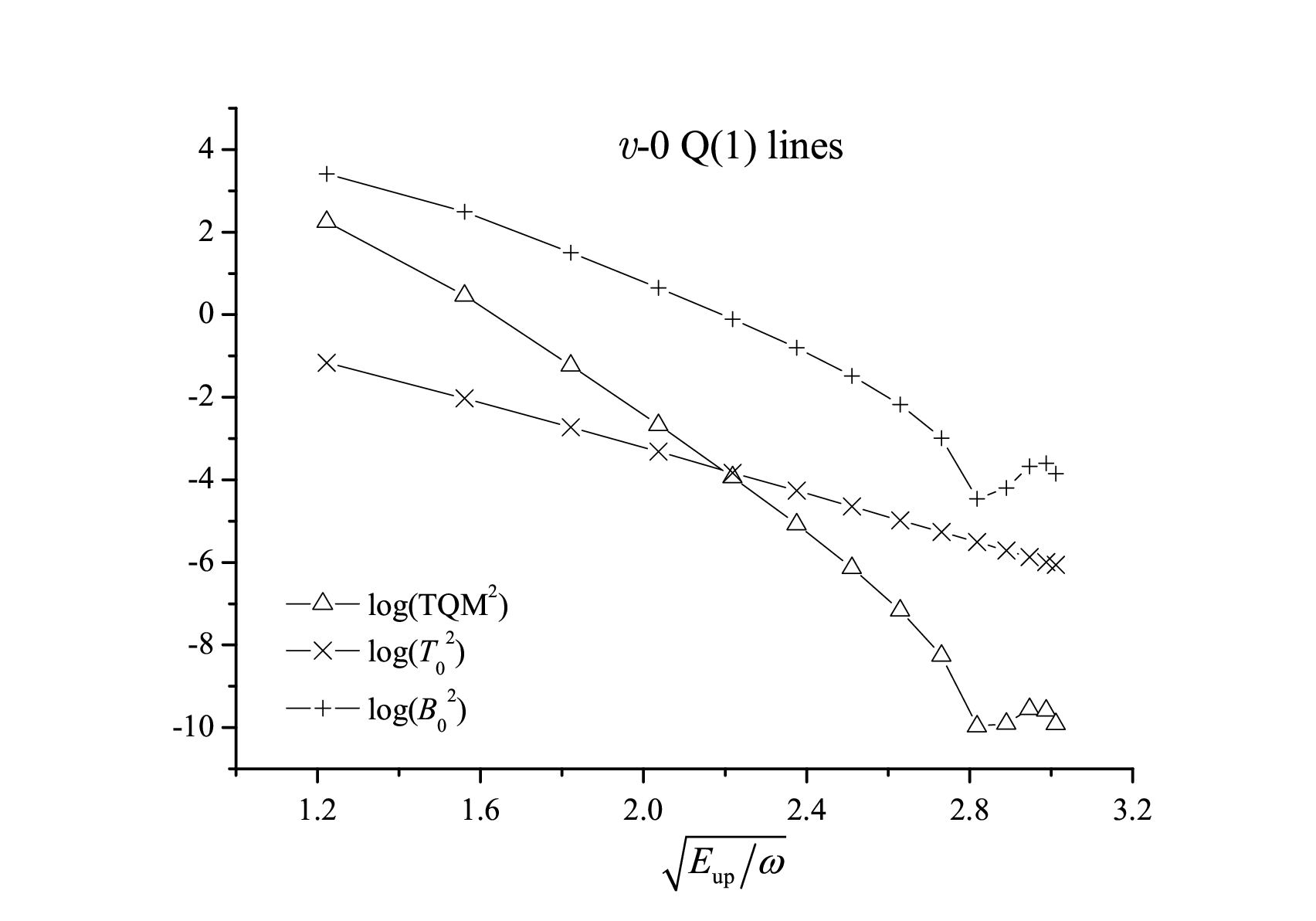}
    \includegraphics[scale=0.5]{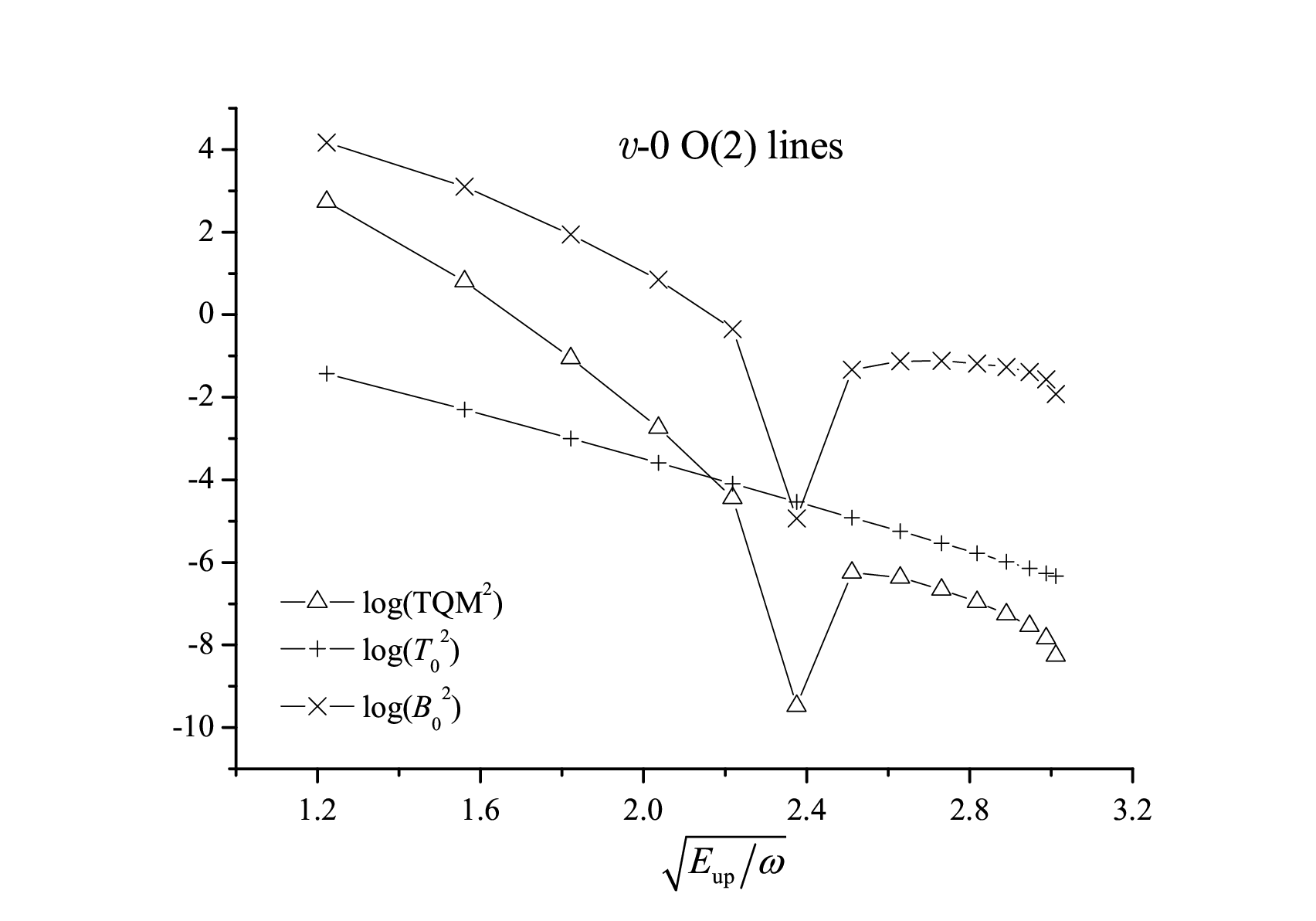}
    \caption{Plots of TQM$^2$, $T_0^2$, and $B_0^2$ for the quadr19 QMF, see text.}
    \label{_T0B0}
\end{figure}

In order to demonstrate this explicitly, we plot in Fig. \ref{_T0B0} both factors separately for the quadr19 QMF. It is seen that both plots have comparable slopes. This feature deserves special consideration. 

Close inspection of the calculated data reveals the reasons for such behavior. The imaginary parts of the classical action entering $T_0$ of Eqs. (\ref{T0}) and (\ref{Phi}) is very large, on the order of 1500 and 800 (in units of $\hbar$) for CO and PN, respectively, whereas it is only about 20 for H$_2$. Two features contribute to this effect: small reduced mass of hydrogen and its less steep potential in the repulsive region. Indeed, the dissociation energy of H$_2$ is about 36118 cm$^{-1}$ \cite{Roueff19}, whereas it is about 51000 in PN and 90000 in CO, which means (by analogy with the Morse potential) that the potential of H$_2$ increases more slowly at large and small $r$ than those of CO and PN. Also, it looks quite plausible that small number of electrons provides for weaker repulsion/attraction at small/large inter-atomic separations than in many-electron molecules. 

As a result of $\Phi$'s smallness, their difference, $\Delta\Phi_0$, is also small, hence, the steepness of the $T_0^2$ decay decreases and becomes comparable with that of $B_0^2$. Nevertheless, the product of these two factors still obeys the NIDL. Thus, despite the difference of the tunnelling actions is not large,
the quasi-classical approximation works perfect in H$_2$ as in other molecules, therefore our method to estimate the expected uncertainties of the calculations can be applied as well.

As a part of our strategy, comparison functions with correct behavior in the complex plane, see \cite{Medvedev22,Ushakov23}, have to be constructed that provide for similar NIDLs; then, the remaining differences will serve as estimates of the errors in the calculated intensities. In the case of molecules with heavy atoms, the reference NIDL is that of $T_0^2$. As follows from the above consideration, in the case of H$_2$, we should use the full TQM$^2$ (calculated by the sinc-DVR method using the grid of the set I QMF) as  reference.

\section{Determination of the calculation errors}\label{determ}

\begin{figure}[htbp]
    \centering
    \includegraphics[scale=0.2]{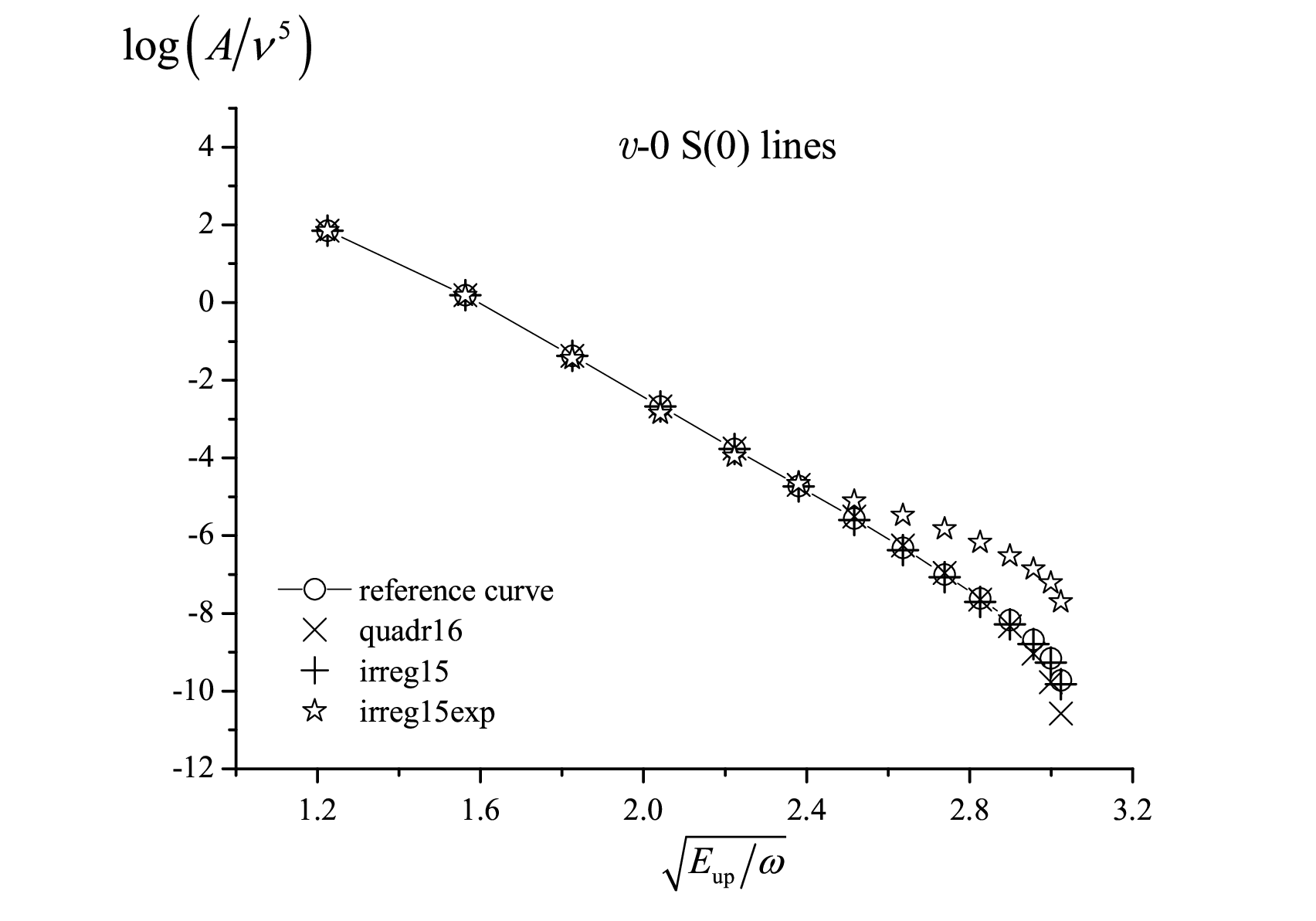}
    \includegraphics[scale=0.2]{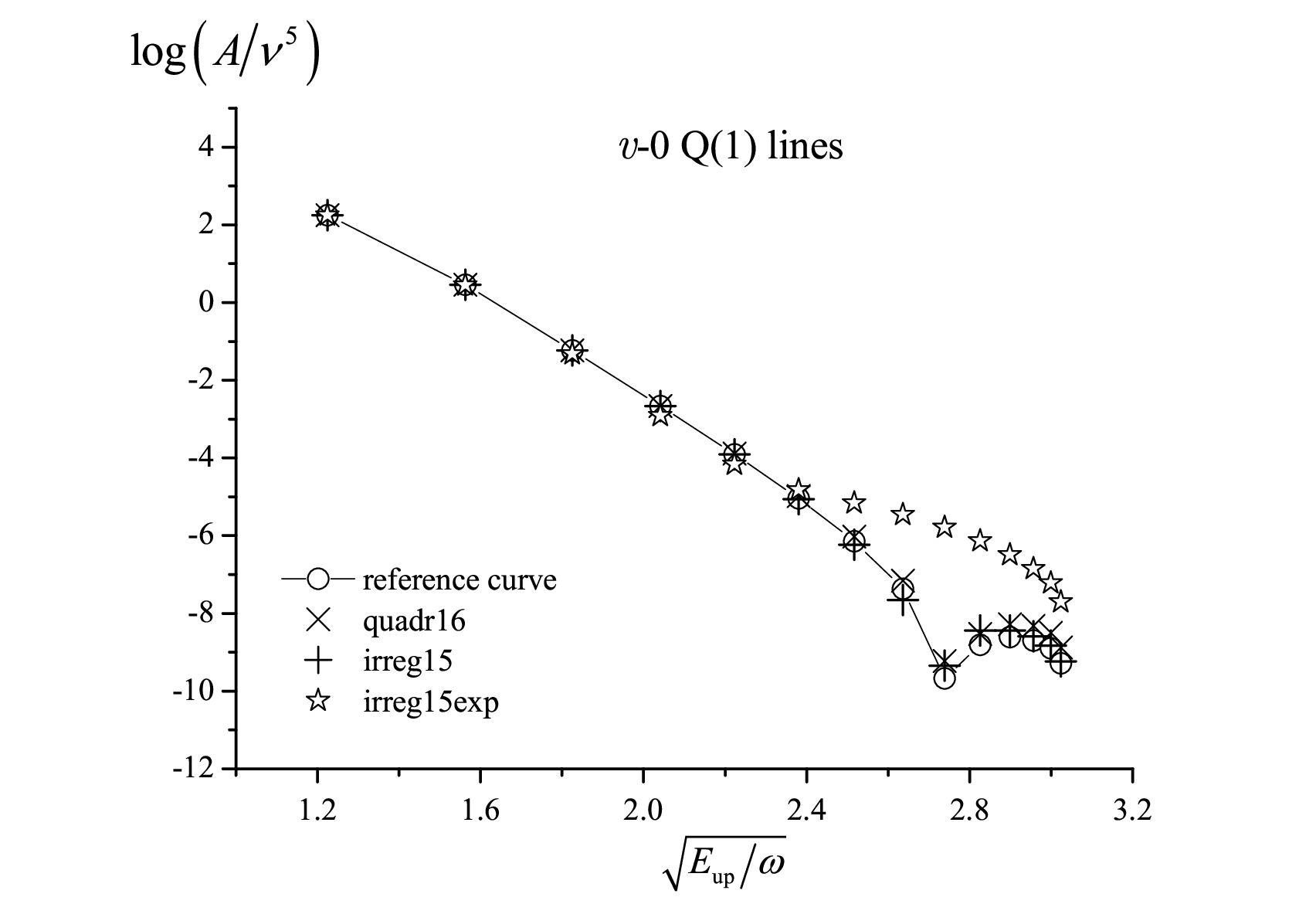}
    \includegraphics[scale=0.4]{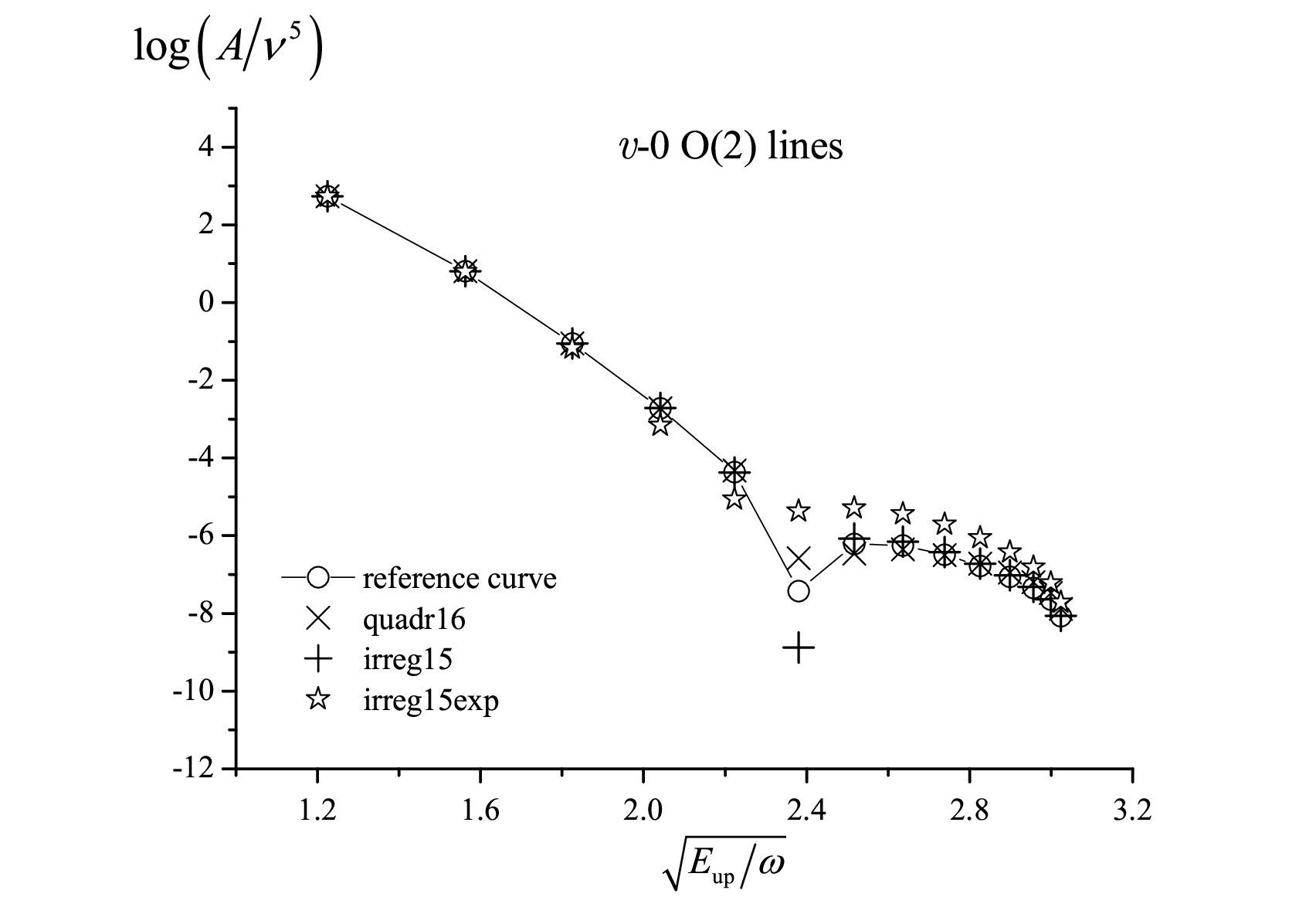}
    \caption{Intensities of the $v$--0 transitions calculated with three QMFs of different analytic forms fitted to different data sets are plotted along with the reference curve (circles with line) in order to determine the errors of the calculations. $v=1$--14.}
    \label{errors}
\end{figure}

We used our QMFs, Sec. \ref{analytic} and \ref{fitting},  and the H2\_spectre potential-energy function of \textbf{Komasa19} \cite{Komasa19} to calculate the \EinA\ for the \emph{X-X} ro-vibrational transitions between all bound states of H$_2$. The Schr\"odinger equation was solved by the sinc-DVR method using the adaptive analytical mapping approach of Meshkov \emph{et al.} \cite{Meshkov08}. 

In  Fig. \ref{errors}, we compare the reference intensities with those calculated with quadr16 and irreg15 fitted to the \ai\ data sets I and II,
as well as with the irreg15exp QMF fitted not only to sets I and II, but also to the laboratory experimental data, set III. The results for irreg15, quadr16, and irreg15exp are given in supplementary files 4-7.

Obviously, the intensities calculated with all three functions very weakly depend on the QMF form at not very high $v$, namely at $v<8$ for the S(0) lines, $v<7$ for the Q(1) lines, and $v<5$ for the O(2) lines. 
Even at the highest transitions, the variations of the data obtained with functions irreg15 and quadr16 are within 5--20\% over the interval where the intensity itself varies by twelve orders of magnitude. 
The difference between the results obtained with use of the irreg15exp QMF and the reference curve is larger, it can reach 1-2 orders of magnitude at the highest overtones. Even such values do not seem too large as compared to the variations of the intensity itself, and therefore they could be used to estimate the uncertainties of the calculations. However, it is not clear whether we can trust them because it poorly reproduces the \ai\ data.

At this junction, it is important to emphasize that the uncertainty of the set I \ai\ QMF declared by Wolniewicz \emph{et al.} \cite{Wolniewicz98} is ``likely" to be 10$^{-5}$ au, as written by the authors, \emph{i.e.} 100 times smaller than 
the uncertainty of 0.001 au assigned by us to these data in fitting the irreg15exp QMF. The main reason to consider the irreq15exp QMF is its better reproducing the experimantal data (see Sec. \ref{CompLab}).

The other reason can be seen in the figure included in supplementary file 1 where the spline-interpolated set II QMF built on a sparse grid is compared with the set I \ai\ QMF over the dense grid. The interpolated QMF closely coincides with the set I QMF at $r=$ 1-2 au but rapidly declines outside this interval in oscillating manner with the amplitude up to 0.001 au, which could affect the high-$v$ transitions; this indeed occurs in CO and PN. In H$_2$, however, the effect is negligible (not shown) because the TQMs themselves are relatively large as compared to the heavier molecules.

The important result is that three functions of very different analytic forms fitted to different data sets give relatively small  differences of the intensities of the normal lines in a wide range of transitions, which testifies that all of them, including irreg15exp, approach one and the same true molecular QMF, hence the remaining differences, about one order of magnitude over the 12-orders variations in TQM$^2$, can be interpreted as intrinsic uncertainties of the calculation. If the deviation of 0.001 au from the \ai\ QMF is considered impossible, the irreg15exp should be ignored.

\begin{figure}[htbp]
    \centering
    \includegraphics[scale=0.3]{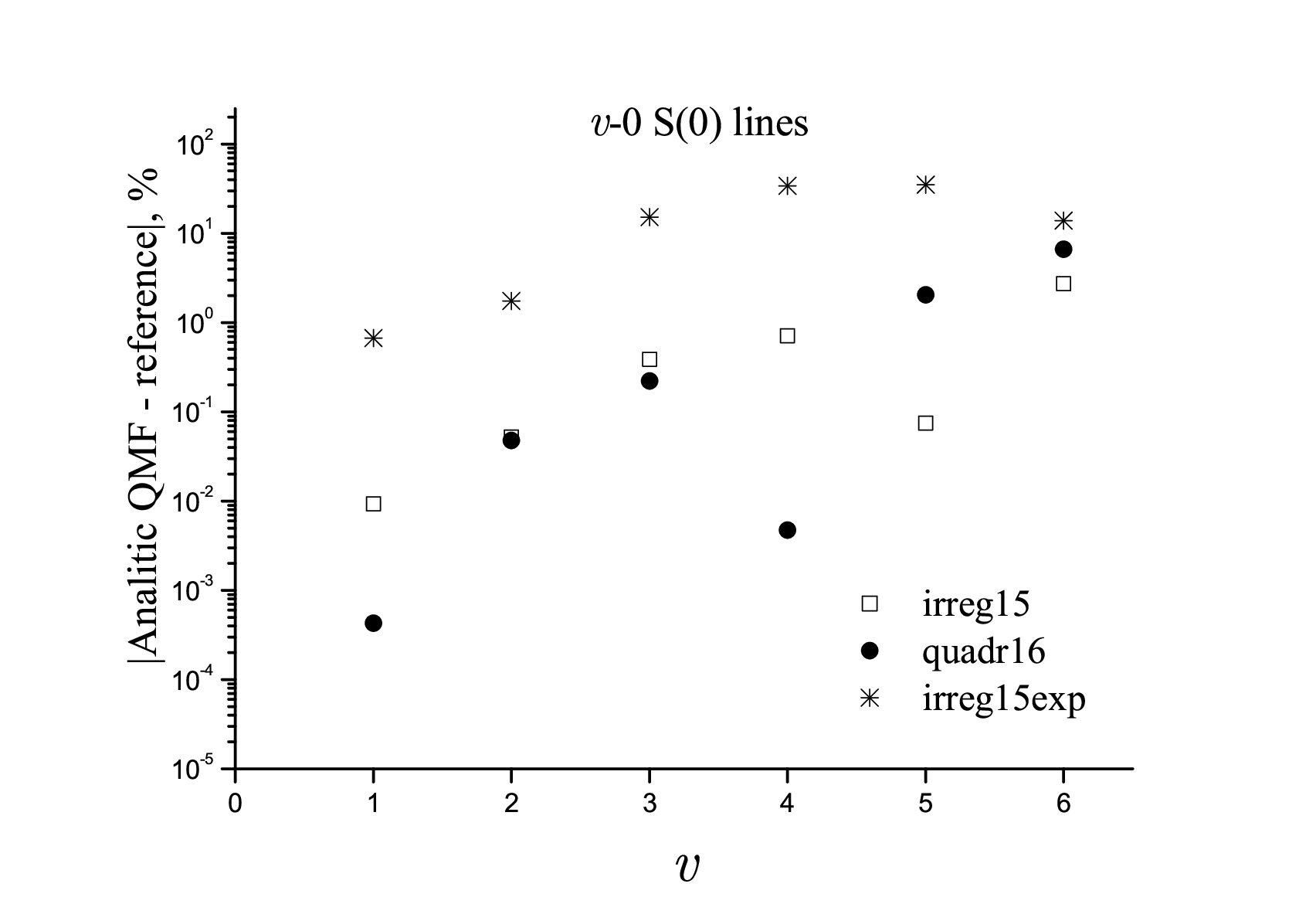}
    \includegraphics[scale=0.3]{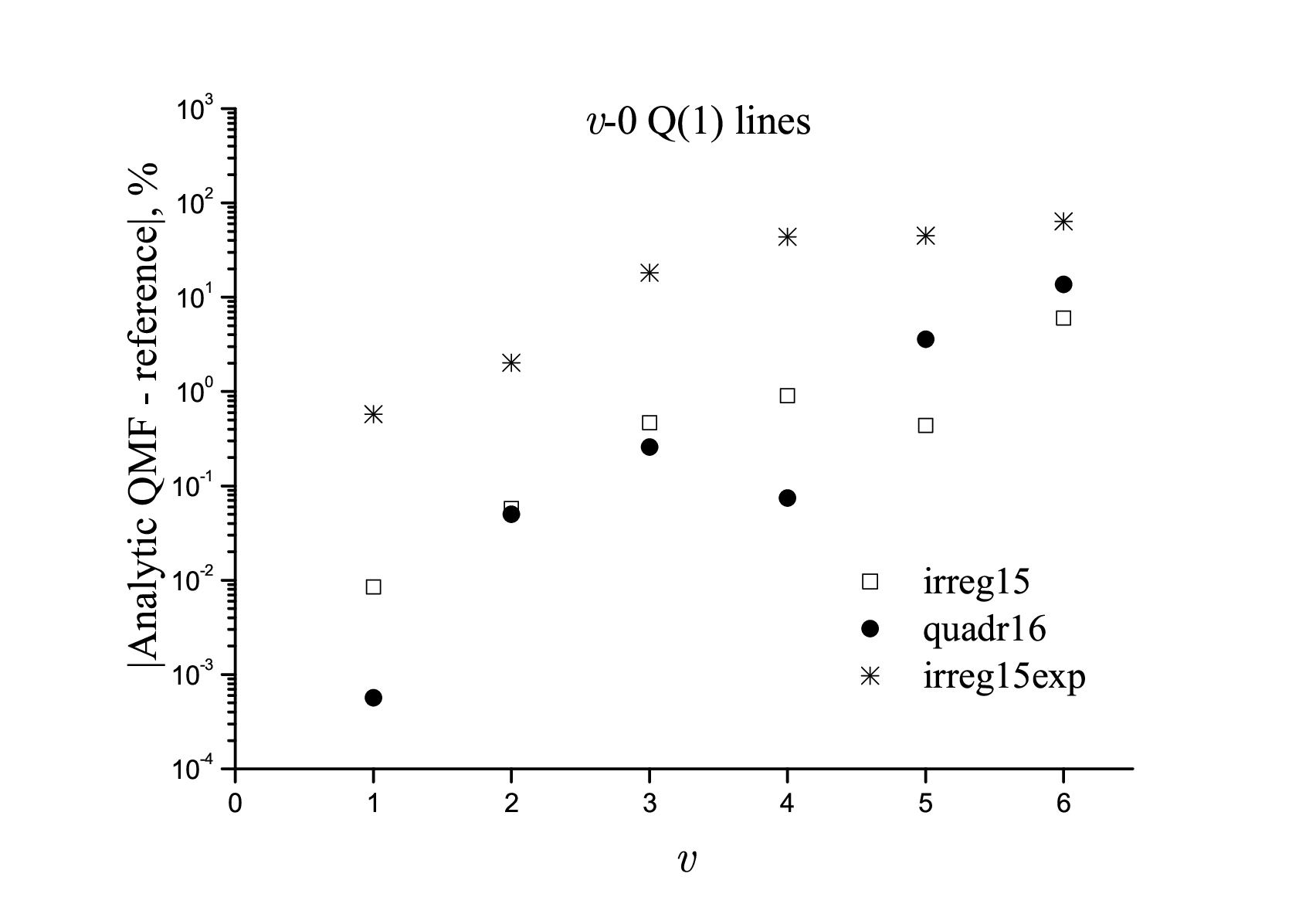}
    \includegraphics[scale=0.5]{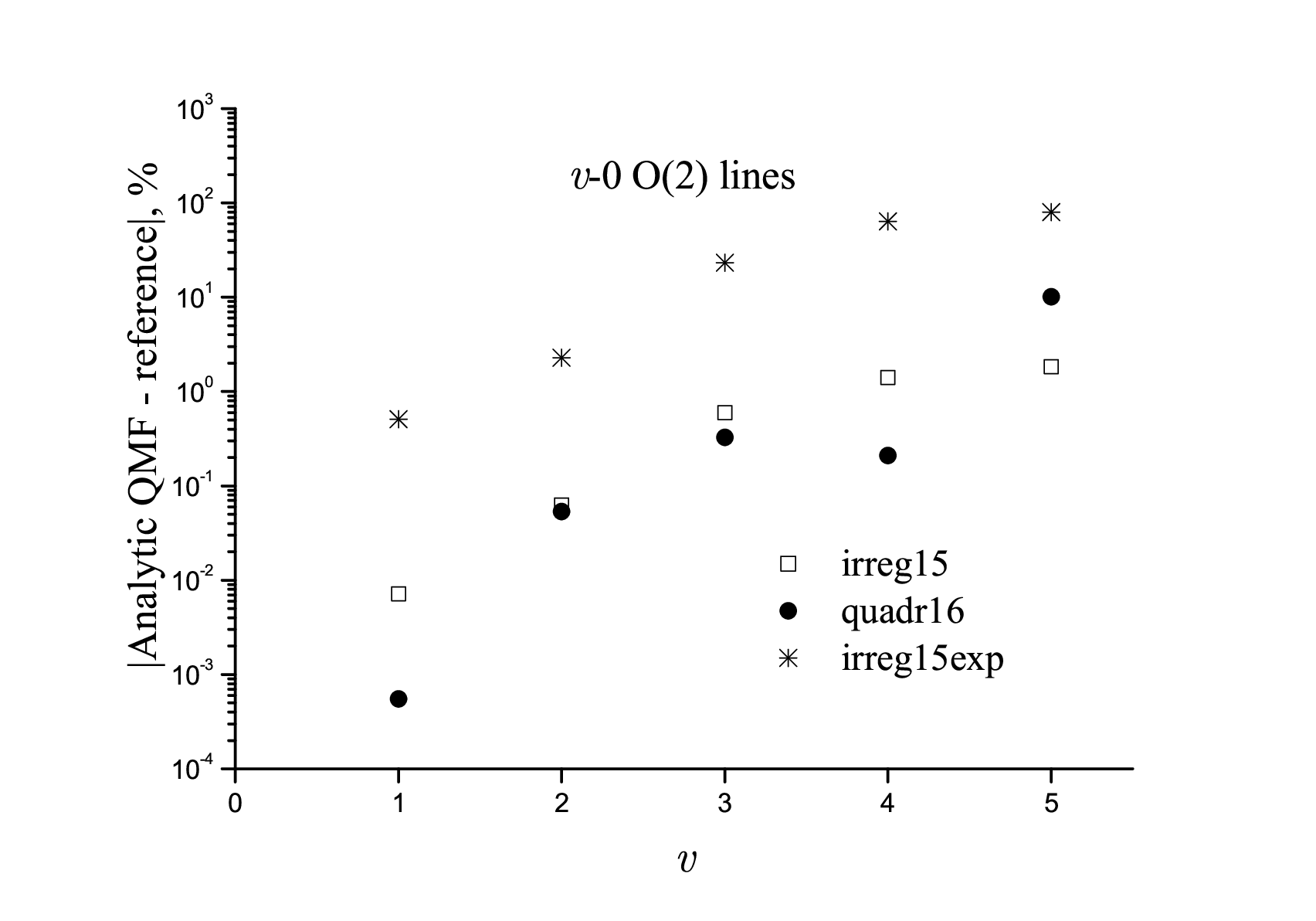}
    \caption{The relative differences between the reference intensities and those calculated with various analytic QMFs.}
    \label{diff}
\end{figure}

The differences obtained for the low-$v$ transitions are explicitly  shown in Fig. \ref{diff}. 
For irreg15 and quadr16, the differences with the reference values are at the sub-percent level at low $v$; they increase to a few percent at $v=5$, as seen in the figure, and to $20$\% at $v=14$ (not shown). At the anomalies, the differences reach 1-2 orders of magnitude depending on the anomaly depth. If the irreg15exp is included in comparison, the error at $v=5$ increases up to 60\%. 
These data can be considered as our estimates of the calculation errors.

The calculation errors obtained testify that the estimated current precision of the \ai\ calculation exceeds the precision of the experimental measurements of the intensities. The differences of the intensities obtained with use of quadr16 are somewhat larger than with irreg15, probably because the mean-square difference of irreg15 with the \ai\ QMF is slightly less than for quadr16. Therefore, we consider irred15 as the best analytic QMF.

The utility of our method to estimate the expected uncertainties of the calculations has been recently demonstrated for CO 7-0 band. In Fig. 8 of our paper \cite{Medvedev22}, the estimated errors of the predicted line intensities in the CO 7-0 band not measured at that time are about 20-30\% at $J<20$, and the measured intensities \cite[Table III]{Balashov23} are indeed within these limits. In fact, the observed-minus-calculated differences are even smaller, close to the experimental uncertainties.

\section{Comparison with the laboratory data}\label{CompLab}

\begin{figure}
    \centering
    \includegraphics[scale=0.6]{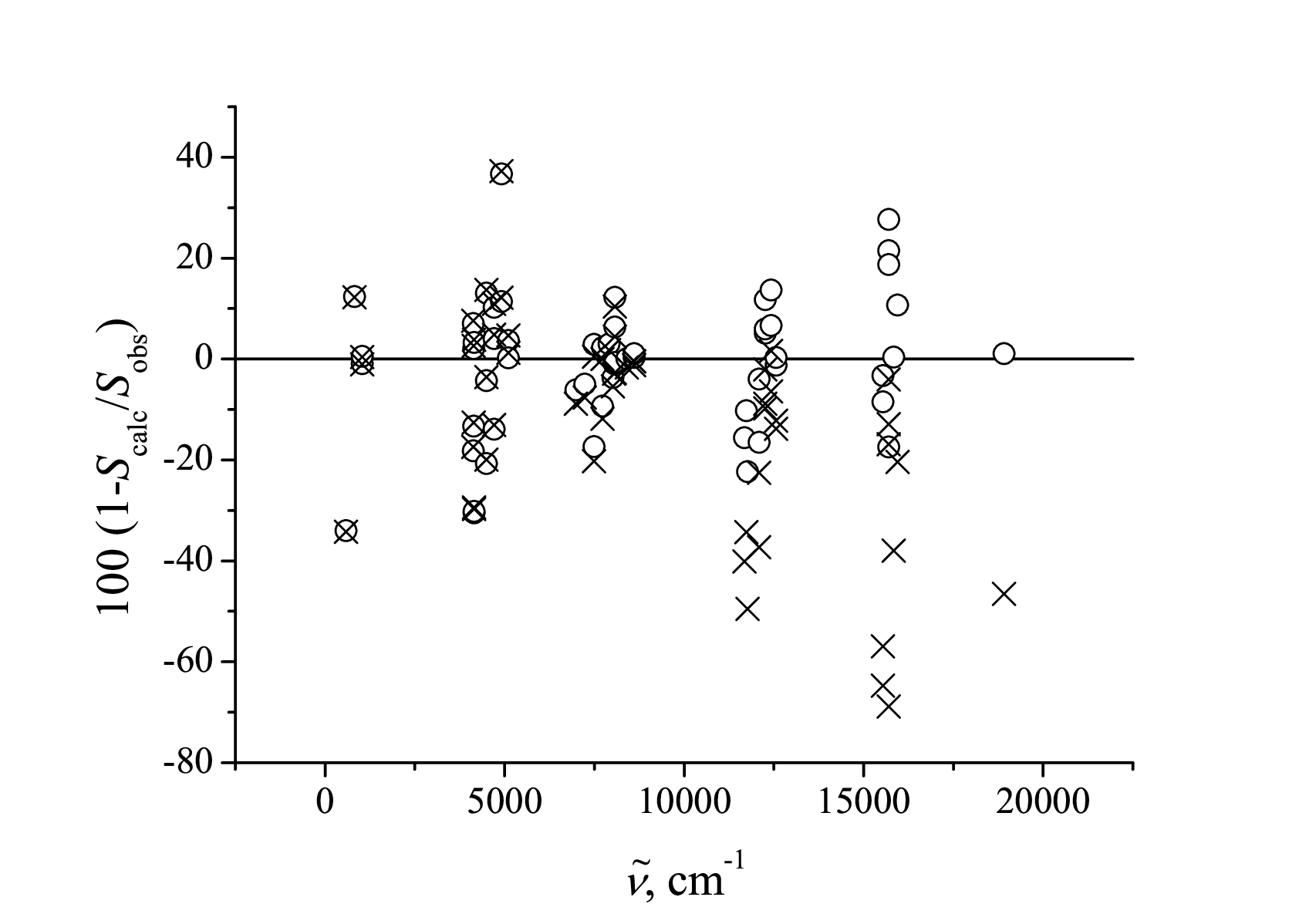}
    \caption{The relative differences of the calculated and measured \cite{Bragg82,Campargue12,Hu12,Kassi14}  \emph{v}-0 absorption band intensities. Circles, present study (the H$_2$\_spectre PEF and the irreg15exp QMF); crosses, \textbf{Roueff19} \cite{Roueff19}.}
    \label{H2obs-calc}
\end{figure}

The irreg15 and quadr16 QMFs give the same deviations of the intensities from the laboratory data \cite{Bragg82,Campargue12,Hu12,Kassi14} as those obtained by \textbf{Roueff19} \cite{Roueff19} (not shown).

In Fig. \ref{H2obs-calc} we compare the  intensities calculated with the irreg15exp QMF and those of \textbf{Roueff19} with the laboratory experiments. Since the irreg15exp QMF was fitted not only to sets I and II (the \ai\ data of Refs. \cite{Wolniewicz98} and \cite{Komasa19}, respectively), but also to set III (the  intensities measured in laboratory), the differences with experiment at $v=3$--5 are much smaller than those for the data of Roueff \emph{et al.} who used only set II (their own purely \ai\ QMF, Ref. \cite{Komasa19}). However, in view of the poor reproduction of the \ai\ data mentioned in Sec. \ref{determ}, we consider the irreg15exp function as inferior as compared to the irreg15 one. 
Hence, the disagreement with experiment at $v=3$--5 obtained with our irreg15 and by Roueff \emph{et al.} should be attributed to the experimental uncertainties as was proposed by Campargue \emph{et al.} \cite{Campargue12}. 

Comparison of the calculated intensities with the laboratory measurements is given in supplementary files 4-6.

\section{Comparison with astrophysical observations}\label{CompAstr}

The observations of the H$_2$ emission of space objects bear important spectroscopic information on the vibrational and rotational states, especially on the highly excited states: $v\le5,\Delta v\le2,J\le28$, with the upper level energies as high as 52000 K, \emph{i.e.} close to the dissociation limit \cite{Pike16,Geballe17}; $v\le10,\Delta v\le2,J\le7$ \cite{Oh16,Kaplan17}; $v\le13,\Delta v\le4,J\le11$ \cite{Le17}. Even quasi-bound states above the dissociation limit were observed \cite{Pike16} and identified \cite{Roueff23}. However, until now, to the best of our knowledge, this information has not been used to verify and improve the theoretical molecular functions of H$_2$. In this section, we summarize the spectroscopic information that can be extracted from the observational data and estimate the relevant uncertainties for their possible use in future refinements of the molecular functions.

The measured $i$-th line flux, $F_i$, is proportional to the upper-level column density, $N_i$, and the Einstein $A$ coefficient,
\begin{equation}\label{flux}
    F_i\propto N_iA_i/\lambda_i,
\end{equation}
where $\lambda_i$ is the line wavelength. For the purpose of astrophysics, this equation is used to derive unknown populations $N_i$ using the calculated values of $A_i$ and $\lambda_i$ \cite{Turner77,Wolniewicz98}. For the purpose of spectroscopy, we can use the ratio of fluxes $F_1$ and $F_2$ emitted from a common upper level with energy $E_\textrm{u}$ in order to derive the observed value of the \emph{A} ratio,
\begin{equation}\label{A1A2}
    A_1/A_2 = F_1\lambda_1/F_2\lambda_2,
\end{equation}
and then to compare the calculated and observed ratios. 

Such comparison is presented in Table \ref{Astrodata} for a number of the line pairs observed in emission from various space objects. The observed ratios of $A_1/A_2$ were calculated using the integrated line fluxes corrected for the dust extinction that are provided in appropriate references. 
The calculated ratios were obtained in the present study with use of the H2\_spectre PEF \cite{Komasa19} and three QMFs, namely, the spline-interpolated \ai\ QMF of Wolniewicz \emph{et al.} \cite{Wolniewicz98}, quadr16, and irreg15. The data of Roueff \emph{et al.} \cite{Roueff19} were also included. For all lines in Table \ref{Astrodata}, all four theoretical data sets gave the same ratios within the difference of less than 0.1\% 
(this can be seen below, in Table \ref{Table_anomalies}, where the $A$ values for some lines calculated with four QMFs are presented). Since the theoretical data were obtained with a number of different QMFs, we conclude that their precision is much higher than that of the observational data, hence the  differences between calculations and observations are fully due to the observational errors.

Table \ref{Astrodata}\footnote{Table \ref{Astrodata} in ASCII format is presented in supplementary file 9.} contains the relative difference, $\Delta$, \emph{i.e.} the observed-minus-calculated ratio divided by the experimental uncertainty. For about a half of the line pairs, it does not greatly exceed unity, hence the theory satisfactorily describes these observational data. For the other half, $\Delta>2$, which can indicate that the theory needs further refinement. However it should be noted that the observed fluxes include the dust-extinction correction, which is not independently/directly measured. Indeed, there is a number of identical line pairs observed in different sources, they are collected  in Table 
\ref{AstroAnalysis}, e.g. 6-4 S(0) and 6-4 Q(2). One can see that there is variation of $A_1/A_2$ for a given line pair between different observed sources (indicated as $\pm$err), well beyond the reported uncertainty. In principle this dispersion provides a measure of the systematic uncertainty associated with the dust-extinction correction. Comparing the differences between the calculated and observed ratios with this systematic uncertainty, we see that the theory agrees with the observations (\emph{i.e.} $\left|\textrm{obs-calc}\right|<$ err) except for three line pairs, namely those with $E_\textrm{u}=8365,23955$, and 35613.
An explanation of this discrepancy can be provided by the relatively coarse spatial resolution of the measurements under consideration. In this case the dust extinction can vary within the point spread function of the telescope. Additionally, it can also vary along each line of sight. We tried to vary the extinction correction for each observed data sets \cite{Kaplan17, Le17, Oh16}, but we could not find an appropriate unique value that could resolve discrepancies for all pair of levels within one data set. Alternatively, the dust-extinction correction can be different for different levels since in principle they can be populated in the different spatial regions of the clouds. Therefore high resolution studies are need to shed light on the observational bias in this problem, which can be done with recently launched James Webb Space Telescope (JWST).

\newpage

\begin{longtable}{|l|r|l|c|c|c|l|}
\caption{Astrophysical observations and calculated Einstein \emph{A} coefficient ratios for the H$_2$ line pairs emitted from a common upper level with energy $E_\textrm{u}$ \label{Astrodata}}\\ \hline
Line pairs	&	$E_\textrm{u}$, K	&	$\lambda$, $\mu$m	&	$A_1/A_2$, obs	&	$A_1/A_2$, calc	&	$\Delta^{a}$	&	Ref.\\
\hline 
\endfirsthead
\multicolumn{7}{c}%
{{\bfseries Table \ref{Astrodata} (continued)}} \\
\hline 
Line pairs	&	$E_\textrm{u}$, K	&	$\lambda$, $\mu$m	&	$A_1/A_2$, obs	&	$A_1/A_2$, calc	&	$\Delta^{a}$	&	Ref.\\
\hline
\endhead
\hline \multicolumn{7}{r}{{\textit{Continued on next page}}} \\ 
\endfoot
\endlastfoot
1-0 Q(2)	&	6471	&	2.413439	& $1.1658^{+0.0038}_{-0.0038}$ &	1.199	& 8.7 &	\cite{Kaplan17}$^*$\\
1-0 S(0)	&		&	2.22329	&		&		&		&	\\ \hline
1-0 Q(3)	&	6951	&	2.42373	& $0.7705^{+0.0025}_{-0.0025}$ &	0.802	& 12.5 &	\cite{Kaplan17}$^*$\\
1-0 S(1)	&		&	2.121834	&		&		&		&	\\ \hline
1-0 Q(4)	&	7584	&	2.437489	& $0.7123^{+0.0036}_{-0.0037}$ &	0.666	& 12.7 &	\cite{Kaplan17}$^*$\\
1-0 S(2)	&		&	2.033758	&		&		&		&	\\ \hline
3-1 O(5)	&	17818	&	1.522033	& $0.3764^{+0.0095}_{-0.0092}$ &	0.384	& 0.8 &	\cite{Kaplan17}$^*$\\ 
3-2 S(1)	&		&	2.386471	&		&		&		&	\\ \hline
3-1 O(6)	&	18386	&	1.581171	& $0.248^{+0.016}_{-0.016}$ &	0.244	& 0.3 &	\cite{Kaplan17}$^*$\\
3-2 S(2)	&		&	2.287045	&		&		&		&	\\ \hline
4-2 O(11)	&	27706	&	2.099586	& $0.351^{+0.081}_{-0.078}$ &	0.319	& 0.4 &	\cite{Kaplan17}$^*$\\
4-3 S(7)	&		&	2.100426	&		&		&		&	\\ \hline
5-3 O(9)	&	30063	&	2.057127	& $0.228^{+0.032}_{-0.031}$ &	0.228	& 0.0 &	\cite{Kaplan17}$^*$\\
5-3 Q(7)	&		&	1.562635	&		&		&		&	\\ \hline
6-4 Q(1)	&	31063	&	1.601534	& $1.014^{+0.035}_{-0.035}$ &	0.998	& 0.5 &	\cite{Kaplan17}$^*$\\
6-4 O(3)	&		&	1.732641	&		&		&		&	\\ \hline
6-4 S(0)	&	31304	&	1.536891	& $0.870^{+0.048}_{-0.046}$ &	0.833	& 0.8 &	\cite{Kaplan17}$^*$\\
6-4 Q(2)	&		&	1.60739	&		&		&		&	\\ \hline
6-4 S(0)	&	31304	&	1.536891	& $0.912^{+0.067}_{-0.060}$ &	0.844	& 1.1 &	\cite{Kaplan17}$^*$\\
6-4 O(4)	&		&	1.796524	&		&		&		&	\\ \hline
6-4 Q(2)	&	31304	&	1.60739	& $1.049^{+0.071}_{-0.065}$ &	1.013	& 0.6 &	\cite{Kaplan17}$^*$\\
6-4 O(4)	&		&	1.796524	&		&		&		&	\\ \hline
6-4 S(1)	&	31661	&	1.50156	& $1.466^{+0.046}_{-0.045}$ &	1.249	& 4.8 &	\cite{Kaplan17}$^*$\\
6-4 Q(3)	&		&	1.616224	&		&		&		&	\\ \hline
7-5 Q(3)	&	35613	&	1.74628	& $0.797^{+0.042}_{-0.042}$ &	0.839	& 1.0 &	\cite{Kaplan17}$^*$\\
7-5 S(1)	&		&	1.620548	&		&		&		&	\\ \hline
7-5 O(7)	&	36588	&	2.204989	& $0.376^{+0.032}_{-0.032}$ &	0.353	& 0.7 &	\cite{Kaplan17}$^*$\\
7-5 S(3)	&		&	1.56151	&		&		&		&	\\ \hline
8-6 O(5)	&	39219	&	2.210763	& $0.93^{+0.13}_{-0.11}$ &	0.831	& 0.9 &	\cite{Kaplan17}$^*$\\
8-6 S(1)	&		&	1.763952	&		&		&		&	\\ \hline
9-7 Q(1)	&	41997	&	2.073187	& $0.699^{+0.079}_{-0.072}$ &	0.863	& 2.1 &	\cite{Kaplan17}$^*$\\
9-7 O(3)	&		&	2.253724	&		&		&		&	\\ \hline 
9-7 O(4)	&	42185	&	2.345581	& $1.17^{+0.27}_{-0.24}$ &	1.596	& 1.6 &	\cite{Kaplan17}$^*$\\
9-7 S(0)	&		&	1.98735	&		&		&		&	\\ \hline
9-7 O(4)	&	42185	&	2.345581	& $0.80^{+0.21}_{-0.17}$ &	1.211	& 1.9 &	\cite{Kaplan17}$^*$\\
9-7 Q(2)	&		&	2.084098	&		&		&		&	\\ \hline
9-7 S(0)	&	42185	&	1.98735	& $0.69^{+0.16}_{-0.13}$ &	0.759	& 0.5 &	\cite{Kaplan17}$^*$\\
9-7 Q(2)	&		&	2.084098	&		&		&		&	\\ \hline
10-8 S(1)	&	45317	&	2.176855	& $1.01^{+0.17}_{-0.13}$ &	1.049	& 0.2 &	\cite{Kaplan17}$^*$\\
10-7 O(5)	&		&	1.648305	&		&		&		&	\\ \hline
6-4 S(0)	&	31304	&	1.536908	& $0.487^{+0.046}_{-0.044}$ &	0.833	& 7.5 &	\cite{Le17},$^\dagger$ A\\
6-4 Q(2)	&		&	1.607386	&		&		&		&	\\ \hline
6-4 Q(1)	&	31063	&	1.601535	& $0.873^{+0.067}_{-0.062}$ &	0.998	& 1.9 &	\cite{Le17},$^\dagger$ A\\
6-4 O(3)	&		&	1.732637	&		&		&		&	\\ \hline
7-5 S(1)	&	35613	&	1.62053	& $2.14^{+0.21}_{-0.19}$ &	1.384	& 4.0 &	\cite{Le17},$^\dagger$ A\\
7-5 O(5)	&		&	2.02204	&		&		&		&	\\ \hline
6-4 Q(5)	&	32712	&	1.64308	& $3.62^{+0.50}_{-0.41}$ &	2.106	& 3.6 &	\cite{Le17},$^\dagger$ A\\
6-4 O(7)	&		&	2.029694	&		&		&		&	\\ \hline
3-1 O(7)	&	19086	&	1.645324	& $0.174^{+0.019}_{-0.019}$ &	0.169	& 0.3 &	\cite{Le17},$^\dagger$ A\\
3-2 S(3)	&		&	2.201397	&		&		&		&	\\ \hline
11-8 Q(1)	&	47391	&	1.657105	& $0.73^{+0.12}_{-0.11}$ &	0.962	& 1.9 &	\cite{Le17},$^\dagger$ A\\
11-8 O(3)	&		&	1.760929	&		&		&		&	\\ \hline
4-2 O(7)	&	23955	&	1.756296	& $0.514^{+0.063}_{-0.060}$ &	0.408	& 1.8 &	\cite{Le17},$^\dagger$ A\\
4-3 S(3)	&		&	2.344479	&		&		&		&	\\ \hline
1-0 S(2)	&	7584	&	2.033756	& $1.202^{+0.051}_{-0.049}$ &	1.502	& 5.8 &	\cite{Le17},$^\dagger$ A\\
1-0 Q(4)	&		&	2.437491	&		&		&		&	\\ \hline
1-0 S(1)	&	6951	&	2.121831	& $0.992^{+0.029}_{-0.028}$ &	1.247	& 8.6 &	\cite{Le17},$^\dagger$ A\\
1-0 Q(3)	&		&	2.423731	&		&		&		&	\\ \hline
1-0 S(0)	&	6471	&	2.223299	& $0.723^{+0.027}_{-0.025}$ &	0.834	& 4.2 &	\cite{Le17},$^\dagger$ A\\
1-0 Q(2)	&		&	2.413436	&		&		&		&	\\ \hline
6-4 S(0)	&	31304	&	1.536908	& $0.905^{+0.096}_{-0.087}$ &	0.833	& 0.8 &	\cite{Le17},$^\dagger$ B\\
6-4 Q(2)	&		&	1.607386	&		&		&		&	\\ \hline
6-4 Q(1)	&	31063	&	1.601535	& $0.885^{+0.073}_{-0.067}$ &	0.998	& 1.5 &	\cite{Le17},$^\dagger$ B\\
6-4 O(3)	&		&	1.732637	&		&		&		&	\\ \hline
7-5 S(1)	&	35613	&	1.62053	& $1.88^{+0.20}_{-0.17}$ &	1.384	& 2.9 &	\cite{Le17},$^\dagger$ B\\
7-5 O(5)	&		&	2.02204	&		&		&		&	\\ \hline
3-1 O(7)	&	19086	&	1.645324	& $0.225^{+0.026}_{-0.025}$ &	0.169	& 2.2 &	\cite{Le17},$^\dagger$ B\\
3-2 S(3)	&		&	2.201397	&		&		&		&	\\ \hline
4-2 O(7)	&	23955	&	1.756296	& $0.659^{+0.090}_{-0.083}$ &	0.408	& 3.0 &	\cite{Le17},$^\dagger$ B\\
4-3 S(3)	&		&	2.344479	&		&		&		&	\\ \hline
1-0 S(2)	&	7584	&	2.033756	& $0.917^{+0.045}_{-0.042}$ &	1.502	& 13.1 &	\cite{Le17},$^\dagger$ B\\
1-0 Q(4)	&		&	2.437491	&		&		&		&	\\ \hline
1-0 S(1)	&	6951	&	2.121831	& $1.027^{+0.041}_{-0.039}$ &	1.247	& 5.3 &	\cite{Le17},$^\dagger$ B\\
1-0 Q(3)	&		&	2.423731	&		&		&		&	\\ \hline
1-0 S(0)	&	6471	&	2.223299	& $0.637^{+0.028}_{-0.028}$ &	0.834	& 6.9 &	\cite{Le17},$^\dagger$ B\\
1-0 Q(2)	&		&	2.413436	&		&		&		&	\\ \hline
6-4 Q(1)	&	31063	&	1.601535	& $1.16^{+0.11}_{-0.11}$ &	0.998	& 1.5 &	\cite{Le17},$^\dagger$ C\\
6-4 O(3)	&		&	1.732637	&		&		&		&	\\ \hline
7-5 S(1)	&	35613	&	1.62053	& $3.56^{+0.65}_{-0.53}$ &	1.384	& 4.1 &	\cite{Le17},$^\dagger$ C\\
7-5 O(5)	&		&	2.02204	&		&		&		&	\\ \hline
3-1 O(7)	&	19086	&	1.645324	& $0.331^{+0.055}_{-0.052}$ &	0.169	& 3.1 &	\cite{Le17},$^\dagger$ C\\
3-2 S(3)	&		&	2.201397	&		&		&		&	\\ \hline
4-2 O(7)	&	23955	&	1.756296	& $0.76^{+0.16}_{-0.13}$ &	0.408	& 2.7 &	\cite{Le17},$^\dagger$ C\\
4-3 S(3)	&		&	2.344479	&		&		&		&	\\ \hline
1-0 S(2)	&	7584	&	2.033756	& $0.618^{+0.041}_{-0.038}$ &	1.502	& 21.6 &	\cite{Le17},$^\dagger$ C\\
1-0 Q(4)	&		&	2.437491	&		&		&		&	\\ \hline
1-0 S(1)	&	6951	&	2.121831	& $0.957^{+0.058}_{-0.052}$ &	1.247	& 5.0 &	\cite{Le17},$^\dagger$ C\\
1-0 Q(3)	&		&	2.423731	&		&		&		&	\\ \hline
1-0 S(0)	&	6471	&	2.223299	& $0.699^{+0.045}_{-0.043}$ &	0.834	& 3.0 &	\cite{Le17},$^\dagger$ C\\
1-0 Q(2)	&		&	2.413436	&		&		&		&	\\ \hline
1-0 S(0)	&	6471	&	2.2233	& $0.8714^{+0.0024}_{-0.0025}$ &	0.834	& 15.0 &	\cite{Oh16}$^\ddagger$\\
1-0 Q(2)	&		&	2.41344	&		&		&		&	\\ \hline
1-0 S(1)	&	6951	&	2.12183	& $1.3909^{+0.0014}_{-0.0014}$ &	1.247	& 103.5 &	\cite{Oh16}$^\ddagger$\\
1-0 Q(3)	&		&	2.42373	&		&		&		&	\\ \hline
1-0 Q(4)	&	7584	&	2.43749	& $0.6079^{+0.0017}_{-0.0017}$ &	0.666	& 33.8 &	\cite{Oh16}$^\ddagger$\\
1-0 S(2)	&		&	2.03376	&		&		&		&	\\ \hline
1-0 Q(5)	&	8365	&	2.45475	& $139^{+26}_{-19}$ &	0.605	& 7.3 &	\cite{Oh16}$^\ddagger$\\
1-0 S(3)	&		&	1.95756	&		&		&		&	\\ \hline
2-1 S(3)	&	13890	&	2.07351	& $21.5^{+1.7}_{-1.5}$ &	18.404	& 2.1 &	\cite{Oh16}$^\ddagger$\\
2-0 O(7)	&		&	1.54641	&		&		&		&	\\ \hline
3-2 S(1)	&	17818	&	2.38645	& $2.09^{+0.14}_{-0.13}$ &	2.602	& 3.6 &	\cite{Oh16}$^\ddagger$\\
3-1 O(5)	&		&	1.52203	&		&		&		&	\\ \hline
3-1 O(7)	&	19086	&	1.64532	& $0.152^{+0.012}_{-0.012}$ &	0.169	& 1.4 &	\cite{Oh16}$^\ddagger$\\
3-2 S(3)	&		&	2.2014	&		&		&		&	\\ \hline
1-0 Q(2)	&	6471	&	2.4133	&	1.454	&	1.199	& &	\cite{Geballe17},$^\&$ PK1\\
1-0 S(0)	&		&	2.2235	&		&		&		&	\\ \hline
1-0 S(1)	&	6951	&	2.1218	&	1.006	&	1.247	& &	\cite{Geballe17},$^\&$ PK1\\
1-0 Q(3)	&		&	2.4237	&		&		&		&	\\ \hline
1-0 S(2)	&	7584	&	2.0334	&	1.02	&	1.502	& &	\cite{Geballe17},$^\&$ PK1\\
1-0 Q(4)	&		&	2.4375	&		&		&		&	\\ \hline
1-0 S(3)	&	8365	&	1.9576	&	1.875	&	1.653	& &	\cite{Geballe17},$^\&$ PK1\\
1-0 Q(5)	&		&	2.4547	&		&		&		&	\\ \hline
1-0 Q(2)	&	6471	&	2.4133	&	1.429	&	1.199	& &	\cite{Geballe17},$^\&$ PK1N\\
1-0 S(0)	&		&	2.2235	&		&		&		&	\\ \hline
1-0 S(1)	&	6951	&	2.1218	&	0.951	&	1.247	& &	\cite{Geballe17},$^\&$ PK1N\\
1-0 Q(3)	&		&	2.4237	&		&		&		&	\\ \hline
1-0 S(2)	&	7584	&	2.0334	&	0.95	&	1.502	& &	\cite{Geballe17},$^\&$ PK1N\\
1-0 Q(4)	&		&	2.4375	&		&		&		&	\\ \hline
1-0 S(3)	&	8365	&	1.9576	&	1.981	&	1.653	& &	\cite{Geballe17},$^\&$ PK1N\\
1-0 Q(5)	&		&	2.4547	&		&		&		&	\\ \hline
1-0 Q(2)	&	6471	&	2.4133	&	1.403	&	1.199	& &	\cite{Geballe17},$^\&$ PK1S\\
1-0 S(0)	&		&	2.2235	&		&		&		&	\\ \hline
1-0 S(1)	&	6951	&	2.1218	&	0.998	&	1.247	& &	\cite{Geballe17},$^\&$ PK1S\\
1-0 Q(3)	&		&	2.4237	&		&		&		&	\\ \hline
1-0 S(2)	&	7584	&	2.0334	&	0.982	&	1.502	& &	\cite{Geballe17},$^\&$ PK1S\\
1-0 Q(4)	&		&	2.4375	&		&		&		&	\\ \hline
1-0 S(3)	&	8365	&	1.9576	&	1.736	&	1.653	& &	\cite{Geballe17},$^\&$ PK1S\\
1-0 Q(5)	&		&	2.4547	&		&		&		&	\\ \hline
1-0 Q(2)	&	6471	&	2.4133	&	1.239	&	1.199	& &	\cite{Geballe17},$^\&$ 118-117\\
1-0 S(0)	&		&	2.2235	&		&		&		&	\\ \hline
1-0 S(1)	&	6951	&	2.1218	&	1.197	&	1.247	& &	\cite{Geballe17},$^\&$ 118-117\\
1-0 Q(3)	&		&	2.4237	&		&		&		&	\\ \hline
1-0 S(2)	&	7584	&	2.0334	&	1.207	&	1.502	& &	\cite{Geballe17},$^\&$ 118-117\\
1-0 Q(4)	&		&	2.4375	&		&		&		&	\\ \hline
1-0 S(3)	&	8365	&	1.9576	&	1.761	&	1.653	& &	\cite{Geballe17},$^\&$ 118-117\\
1-0 Q(5)	&		&	2.4547	&		&		&		&	\\ \hline
1-0 S(2)	&	7584	&	2.0338	&	1.767	&	1.502	& &	\cite{Pike16}$^\#$\\
1-0 Q(4)	&		&	2.4375	&		&		&		&	\\ \hline
1-0 S(1)	&	6951	&	2.1218	&	1.251	&	1.247	& &	\cite{Pike16}$^\#$\\
1-0 Q(3)	&		&	2.4237	&		&		&		&	\\ \hline
1-0 S(0)	&	6471	&	2.2233	&	1.046	&	0.834	& &	\cite{Pike16}$^\#$\\
1-0 Q(2)	&		&	2.4134	&		&		&		&	\\ \hline
\multicolumn{7}{l}{$^a$ The difference between the theoretical and observational values divided by } \\
\multicolumn{7}{l}{the observational error.} \\
\multicolumn{7}{l}{$^*$ Orion Bar, \textbf{Kaplan17} \cite{Kaplan17}.} \\
\multicolumn{7}{l}{$^{\dagger}$ NGC 7023, three regions, \textbf{Le17} \cite{Le17}.} \\
\multicolumn{7}{l}{$^\ddagger$ Orion KL, \textbf{Oh16} \cite{Oh16}.} \\ 
\multicolumn{7}{l}{$^\&$ Orion Molecular Cloud, four regions, \textbf{Geballe17} \cite{Geballe17}.} \\
\multicolumn{7}{l}{$^\#$ Herbig-Haro 7, \textbf{Pike16} \cite{Pike16}.} \\
\end{longtable}

\begin{table}[htbp]
    \caption{Analysis of the Table \ref{Astrodata} data: Error estimates for the observed intensities and comparison with theory}
    \centering
    \vspace{5pt}
    \begin{tabular}{|c|c|c|c|c|c|}
    \hline
     Line pairs & $E_\textrm{u}$   & $A_1/A_2$, obs$^a$ & $A_1/A_2$, calc & $\left|\textrm{obs-calc}\right|$ & No. of entries \\
     &  &  &   &   & in Table \ref{Astrodata}\\
     \hline
      1-0 Q(2)  & 6471  & $1.33\pm0.17^b$  &   1.20 & 0.13  & 10  \\
      1-0 S(0)  &   &   &   &  &  \\
      \hline
      1-0 S(1)  & 6951 & $1.09\pm0.21^b$   &  1.25   & 0.16  & 10  \\
      1-0 Q(3)  &  &   &   &   &   \\
       \hline
      1-0 S(2)  & 7584 & 1.17$\pm0.48^b$ &   1.50    & 0.33  &  10   \\
      1-0 Q(4)  &  &   &   &   &   \\

      \hline
      1-0 S(3)  & 8365 &  1.79$\pm0.08^b$ &   1.65    &  0.14 & 5     \\
      1-0 Q(5)  &  &   &   &   &   \\

      \hline
      3-2 S(1)  & 17818  &  $2.38\pm0.28$ & 2.60  & 0.22  & 2   \\
      3-1 O(5)  &   &   &   &   &   \\
        \hline
      3-2 S(3)  & 19086  & $4.67\pm1.90$  &  5.92 &  1.35 & 4  \\
      3-1 O(7)  &   &   &   &   &   \\
        \hline
      3-2 S(3)  & 23955  & 1.59$\pm0.35$  & 2.45  & 0.86  & 3  \\
      3-1 O(7)  &   &   &   &   &   \\
        \hline
      6-4 Q(1)  & 31063  & 0.98$\pm0.18$  &  1.00 & 0.02  & 4  \\
      6-4 O(3)  &   &   &   &   &   \\

        \hline
      6-4 Q(2)  &  31304 & $1.44\pm0.63$  & 1.20  & 0.24  & 4  \\
      6-4 S(0)  &   &   &   &   &   \\
        \hline
      7-5 S(1)  &  35613 & 2.53$\pm1.00$  &  1.38 &  1.15 &  3 \\
      7-5 O(5)  &   &   &   &   &   \\
        \hline

         \multicolumn{6}{l}{$^a$ Arithmetic average with scatter, which exceeds the individual } \\
         \multicolumn{6}{l}{uncertainties shown in Table \ref{Astrodata}.} \\
         \multicolumn{6}{l}{$^b$ The maximum and minimum of the observed $A_1/A_2$ values are omitted.} 
         
    \end{tabular}        
    \label{AstroAnalysis}
\end{table}

\newpage
\section{Anomalies in the calculated spectra}\label{anom}

Anomalies are very common phenomenon in the calculated line lists of diatomic molecules. They appear in the vibrational, vibrational-rotational, and electronic transitions, and some of them were observed in experiment \cite{Medvedev12,Chung05,Changala21}.

The rotational anomalies in the electronic spectra were first predicted by Le Roy and coworkers \cite{Brown73,LeRoy75} in the vibronic transitions of I$_2$. It was discovered that the Franck-Condon amplitude (the radial overlap integral) as function of $J$ crosses zero and changes sign at some $J$, thereby weakening the intensities of the neighboring lines. Such kind of anomalies are also present in the H$_2$ line list recently calculated  by Abgrall \emph{et al.} \cite{Abgrall93,Abgrall93L,Abgrall93W}. While the electronic transitions are beyond the scope of the present paper, there are reasons to discuss them briefly at the end of this section.

The vibrational anomalies in the $X$-$X$ transitions were described in Ref. \cite{Medvedev12}. They arise when the transition moment as function of $v^\prime$ at fixed $v^{\prime\prime},J^{\prime\prime}$ crosses zero and changes its sign, thereby making the intensities of the neighboring bands very weak as compared to the values predicted by the NIDL. The vibrational anomalies were predicted in HF \cite{Meredith73} and PN \cite{Ushakov23}, predicted and observed in CO \cite{Medvedev85JMS,Chung05}. Examples of the H$_2$ vibrational anomalies, 6--0 O(2) and 9--0 Q(1), are shown in Figs. \ref{splDVR}-\ref{errors} above.

The rotational anomalies arise in the vibrational bands of the $X$--$X$ transitions when the transition moment changes sign as function of $J$. Examples of the rotational anomalies were given in Refs. \cite[Fig. 6]{Medvedev22} (CO 7-0 R(95)) and \cite[Fig. 8 and Table 6]{Ushakov23} (PN 3-0 P(17), P(18); additionally, the 3-0 band is vibrational anomaly). Recently, a rotational anomaly was observed in the 1-0 band of CH$^+$ \cite{Changala21}.

The full list of the vibrational (type ``v") and rotational (``r") anomalies in the calculated H$_2$ $X$--$X$ vibrational bands is presented in Table \ref{Table_anomalies}, of which the full version is given in supplementary file 10. There are also mixed types (``vr") when the TQM changes sign as function of either $v$ or $J$. 
A few interesting features of the data presented in Table \ref{Table_anomalies} deserve special attention. 

The calculated intensities of the anomalies depend on the form of the QMF, as is evident from comparison of the intensities calculated in the present study and by Roueff and coauthors. This dependence is much stronger for the anomalies than for the normal lines (not shown). 
The anomaly position can also change between various QMFs, as demonstrated by two pairs of lines separated by horizontal lines at the bottom of the table.

Especially interesting is the middle part of the table showing the 1.5--2.5 $\mu$m range where a lot of   emission lines has been observed. Some observed lines located near the anomalies are shown in italic along with their calculated $A$ values taken from earlier studies in the cited references where available. There are a lot of anomalies in this region, and their $A$ coefficients are on the same order as, or 2--4 orders of magnitude lower than, for the neighboring normal lines belonging to various vibrational bands. There are two special examples where the anomalies belong to the same bands as the observed lines: 5-4 S(9) at 2.186 $\mu$m (anomaly) and 5-4 S(13) at 2.1528 $\mu$m (observed); 4-3 S(11) at 2 $\mu$m (anomaly) and 4-3 S(17) at 2.0475 $\mu$m (observed). In these pairs, the $A$ values for the anomalies are only two orders of magnitude weaker than for the observed lines. 

\begin{figure}
    \centering
    \includegraphics[scale=0.5]{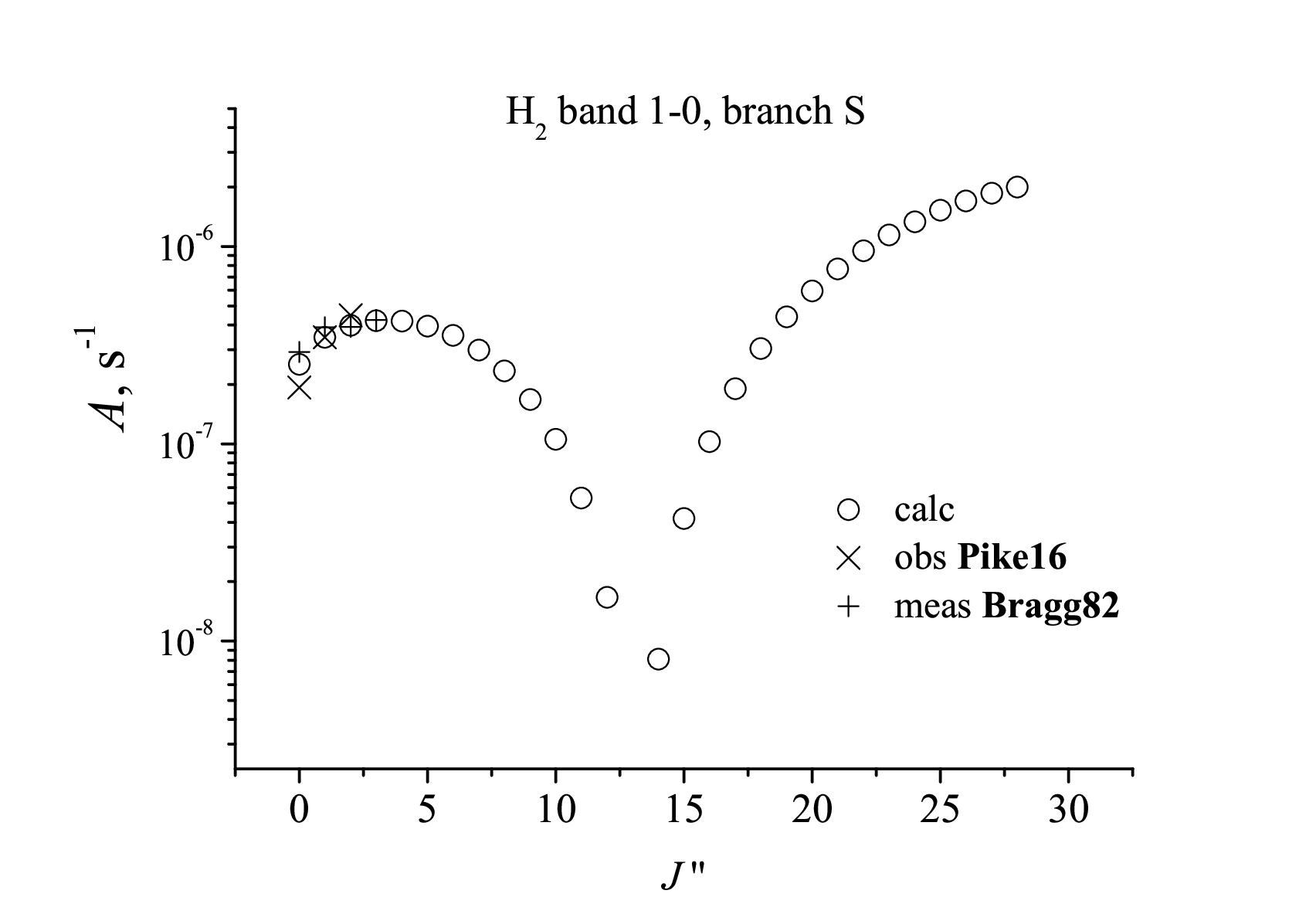}
    \caption{The 1-0 S(13) anomaly in the H$_2$ spectrum ($\lambda=1.6324$ $\mu$m, $E_\textrm{u}=22516$ K). calc, present study, \textbf{Roueff19} \cite{Roueff19}, and \textbf{Turner77} \cite{Turner77}; obs \textbf{Pike16}, recalculated from astrophysical observations \cite{Pike16}; meas \textbf{Bragg82}, recalculated from the laboratory measurements \cite{Bragg82}. The line S(1) dereddened intensity was assigned in \cite{Pike16} to be 100 in arbitrary units, therefore we first assigned Aobs = Acalc = 3.47e-7 s$^{-1}$ for this line, and then we multiplied Aobs for lines S(0) and S(2) (calculated from the relative dereddened intensities) by 3 according to the nuclear-spin statistics.}
    \label{anomalyS1-0}
\end{figure}

\begin{figure}
    \centering
    \includegraphics[scale=0.5]{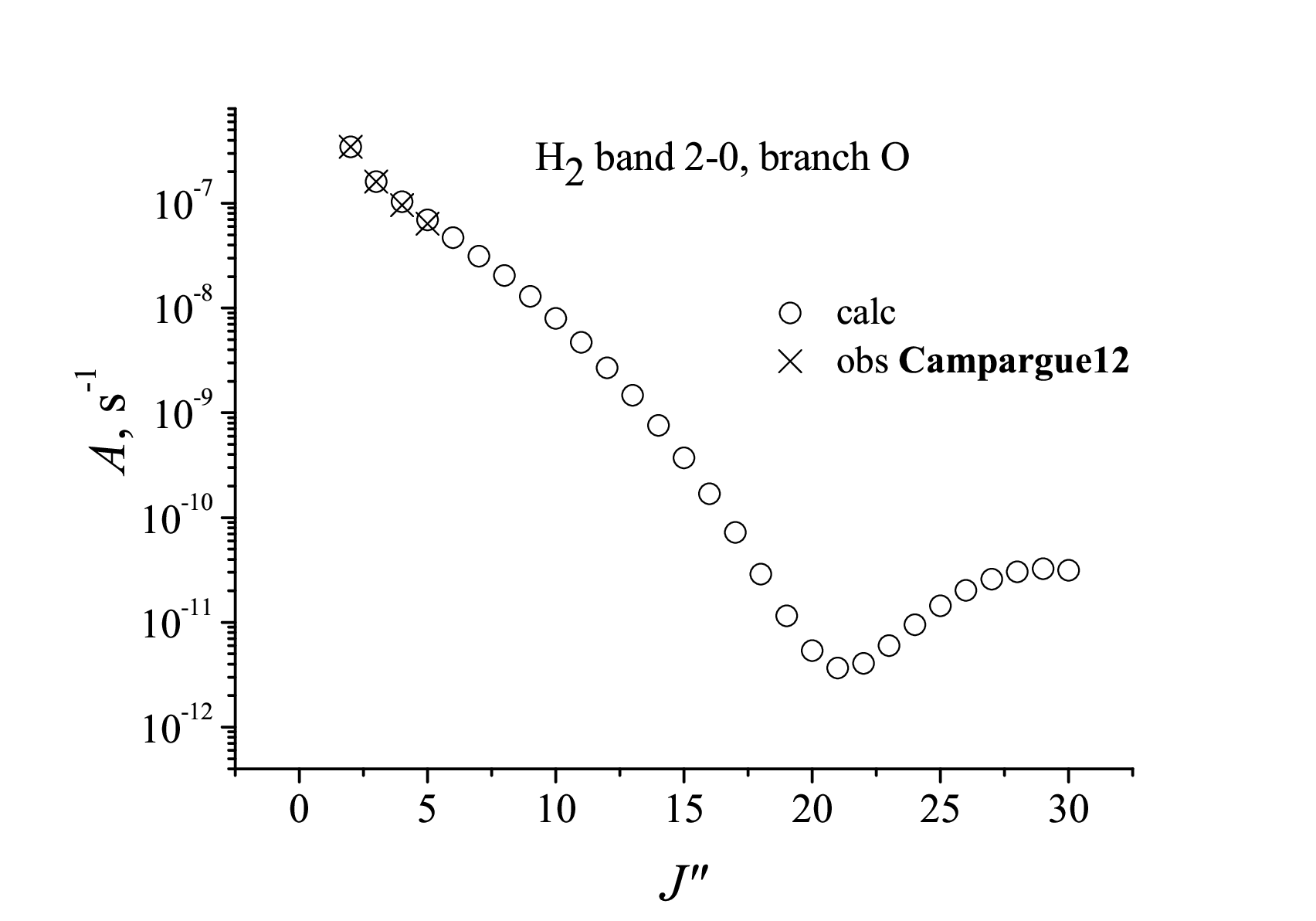}
    \caption{Example of the flat anomaly, 2-0 O(21). calc, same as in Fig. \ref{anomalyS1-0}; obs \textbf{Camparhue12}, Ref. \cite{Campargue12}.}
    \label{anomalyO20flat}
\end{figure}

Some rotational anomalies that appear in the 1-0 and 2-0 bands are shown in Figs. \ref{anomalyS1-0}, and \ref{anomalyO20flat} along with a few lines observed in space and measured in laboratory. 
As mentioned above, the $A$ values at anomalies are mostly 2-4 orders of magnitude less than at the normal lines within the same band (this is seen in Figs. \ref{anomalyS1-0} and \ref{anomalyO20flat}),
which makes them difficult to observe. Yet, there are other factors affecting the observed intensity that vary in wide limits. For instance, the number density in astrophysical objects emitting the H$_2$ lines vary by three orders of magnitude, $n_H=6.3\cdot10^2$--$10^5$ cm$^{-3}$ \cite{Kaplan17}. The large number of such objects with widely variable conditions give hope that some anomalies could be observed in future, especially with aforementioned JWST.

\begin{figure}
    \centering
    \includegraphics[scale=0.5]{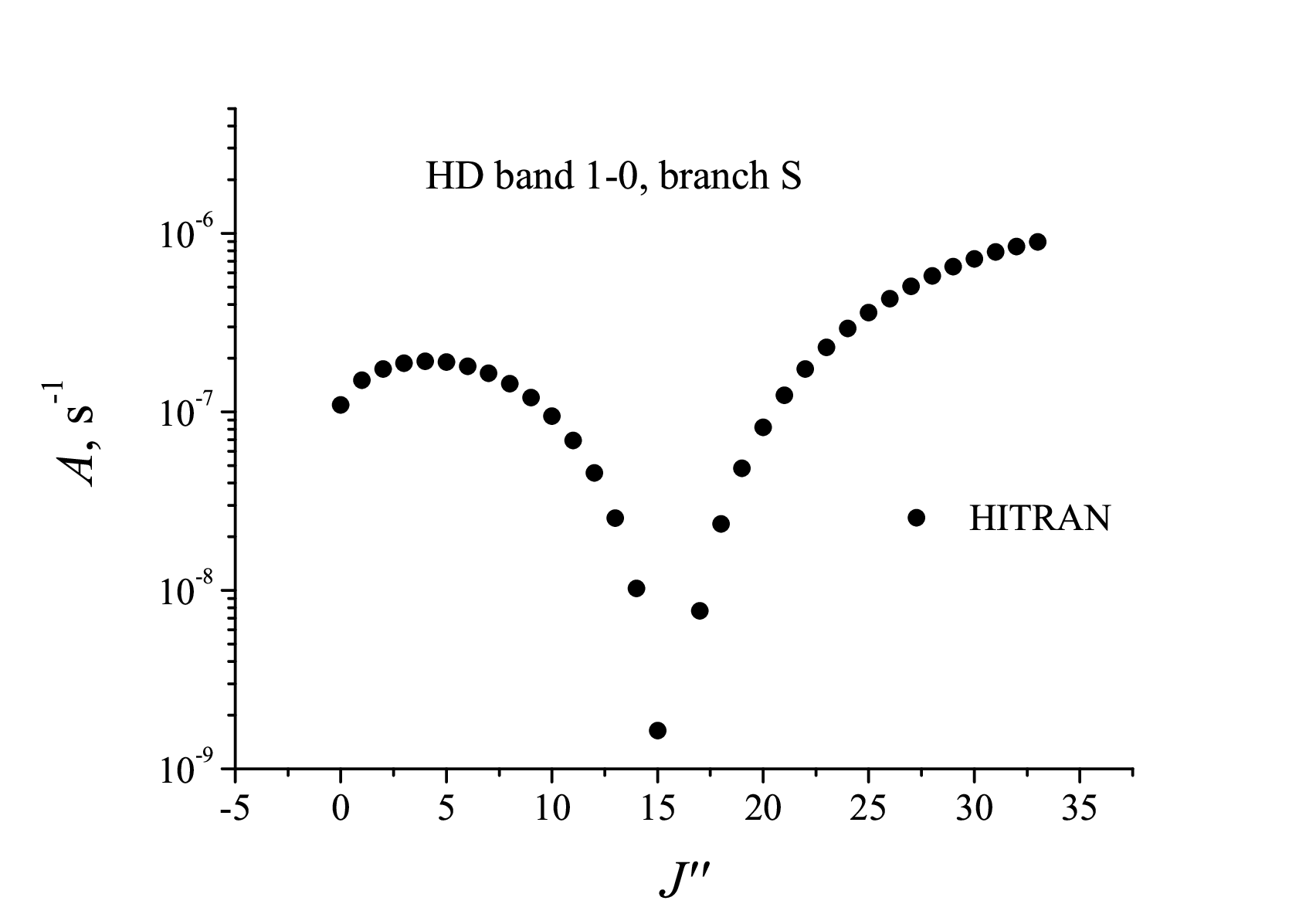}
    \caption{Isotope effect on the 1-0 S(13) anomaly in H$_2$ shown in Fig. \ref{anomalyS1-0}: the anomaly jumped to $J"=15$. Data of Ref. \cite{HITRAN2020}.}
    \label{HDanomaly}
\end{figure}

Figure \ref{HDanomaly} shows the isotope effect on the 1--0 S(13) anomaly. In HD, the anomaly jumped from $J=13$ to 15, and the intensity dropped down by a few times. Note, however, that HD observations in astrophysical sources is a limiting task, with only a few detections of the strongest $J=1\to0$ line being observed to date. 

\begin{figure}[htbp]
    \centering
    \includegraphics[scale=0.5]{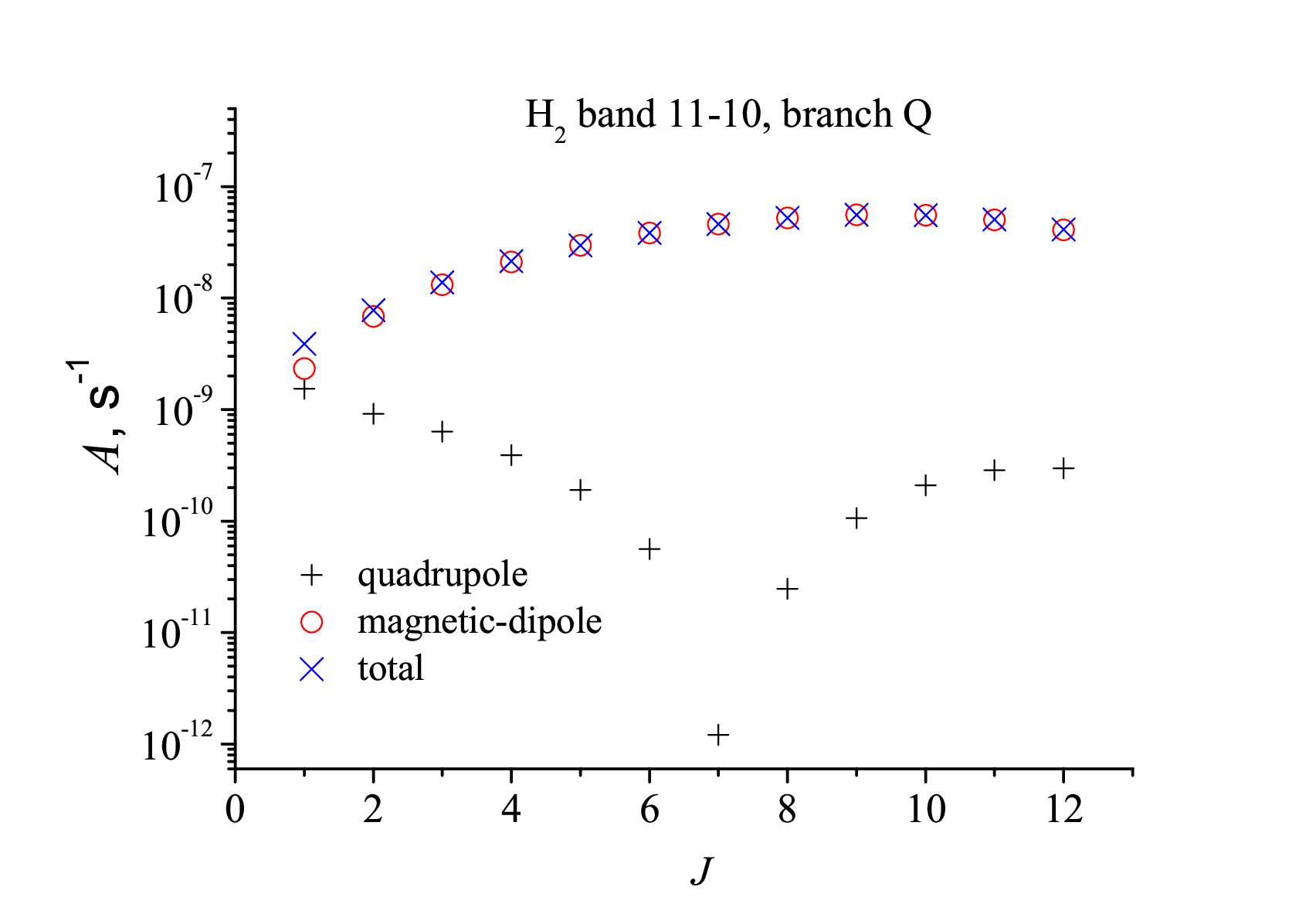}
    \caption{Example of an anomaly, 11-10 Q(7), hidden under a magnetic-dipole transition in the calculated spectrum of H$_2$, data of \textbf{Roueff19} \cite{Roueff19}.}
    \label{anomalyQ11-10}
\end{figure}

Figure \ref{anomalyQ11-10} shows that the magnetic-dipole transitions (with selection rule $\Delta J=0$) can overlap the anomaly in the Q branch of the electric quadrupole spectrum, thereby making the anomaly not observable.

Sometimes, a special kind of rotational anomaly arises when the transition moment decreases with changing $J$ but turns over to increase without changing sign. As shown in Fig. \ref{anomalyO20flat}, the variation of $A$ on the logarithmic scale becomes smooth, therefore we call such an anomaly ``flat" as distinct from the ``keen", cusp-shaped ones (Figs. \ref{anomalyS1-0} and \ref{HDanomaly}) that are due to the sign change of the transition moment. The effect of the TQM sign change at the vibrational anomalies is demonstrated in  supplementary file 7 (not shown for the rotational anomalies).

The rest of this section is devoted to vibronic transitions, which seem to be beyond the scope of the present study, so much the more that no relevant calculations were performed by us. The reason to discuss them here is that they contain the keen anomalies (having the beak shape on the semi-logarithmic scale) of exactly the same nature as the keen anomalies in the \emph{X}--\emph{X} transitions, namely they all are manifestations of the destructive interference that makes them sensitive to the parameters of the Schr\"odinger equation.

\begin{figure}[htbp]
    \centering
    \includegraphics[scale=0.5]{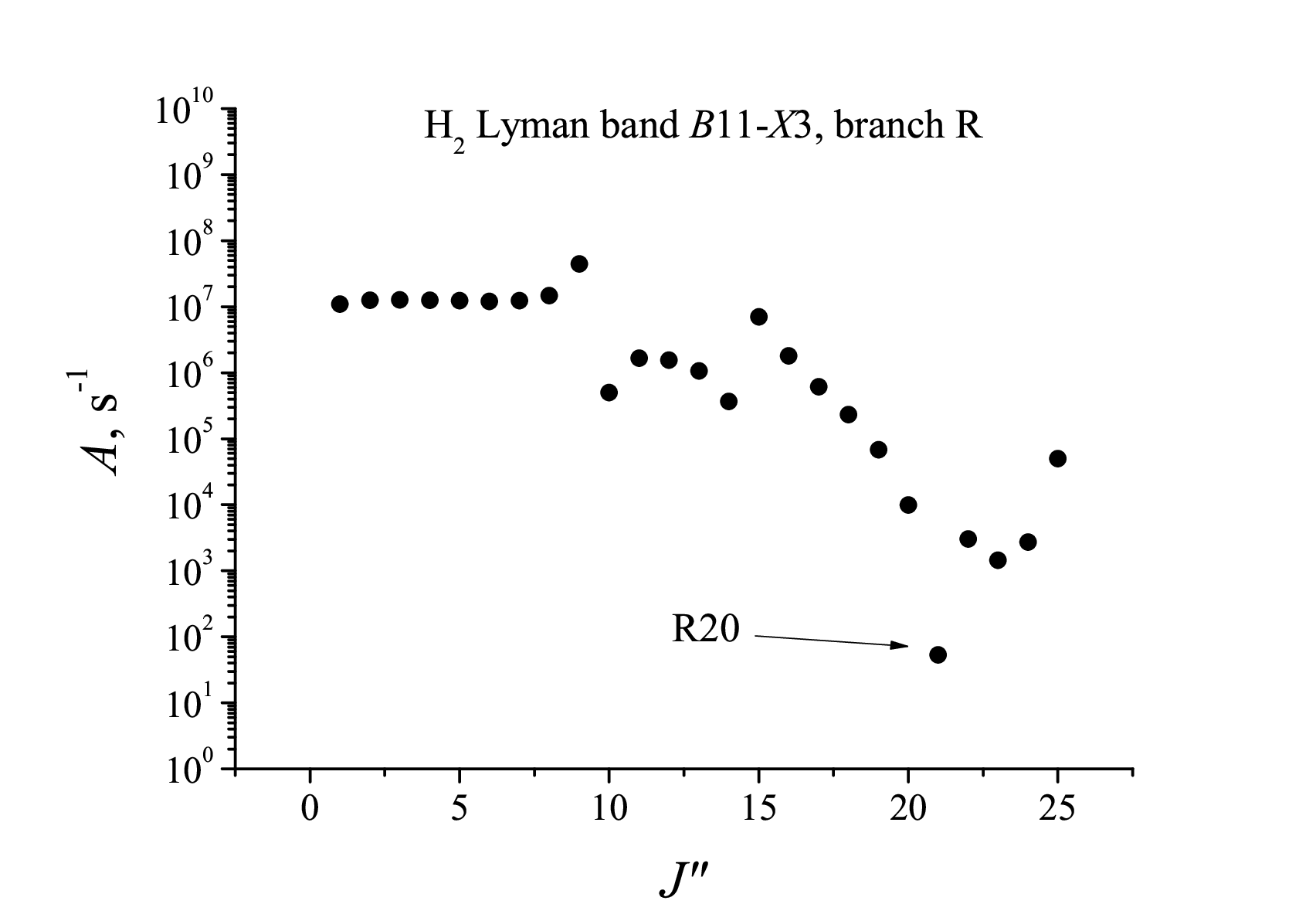}
    \caption{Example of the keen anomaly, R(20), in the calculated H$_2$ vibronic spectrum \cite{Abgrall93}. Jumps at $J=9$ and 14 are anomalies of different kinds.}
    \label{anomLyman}
\end{figure}

Figure \ref{anomLyman} shows an example of the keen anomaly in the Lyman spectrum calculated by Abgrall
\emph{et al.} \cite{Abgrall93,Abgrall93L}. In the footnote to Table II of Ref. \cite{Abgrall93}, the authors write: ``Some R and P lines are missing due to weak Franck-Condon factors." These  lines missing in the observed spectrum are exactly what Le Roy with coauthors predicted \cite{Brown73,LeRoy75}, therefore we call such kind of vibronic anomaly \textbf{ the Le Roy anomaly}. More such anomalies are presented in supplementary file 11.

Another important features in Fig. \ref{anomLyman} are jumps at $J^{\prime\prime}=9$ and 14, which are obviously not the keen anomalies. There are many anomalies of different kinds in the calculated Lyman and Werner bands of H$_2$, \emph{e.g.} jumps in the $A$ values at particular $J$ caused by avoided crossings or intensity borrowing \cite{Abgrall93}, yet the keen anomalies deserve special attention. The first key point here is that \textbf{the keen anomalies are manifestation of a special interference effect}. Indeed, when $J$ changes, the rotational distortion changes the balance between the positive and negative contributions into the transition integral so that the transition moment can approach and then cross zero at some $J$, -- a typical interference phenomenon. And, as a consequence of this destructive interference, the intensity of the weak line, the anomaly, becomes very sensitive to various perturbations. In particular, what can be especially interesting with H$_2$, it can be sensitive to variations of the proton-to-electron mass ratio, $\Delta\mu$.

As we discovered earlier \cite{Medvedev18}, the anomalies are very sensitive to the form of the potential and moment functions. When observed, they could be used, in principle, to improve the theoretical molecular functions. Moreover, they could be also sensitive to the proton-to-electron mass ratio, $\mu$. These are motivations to discuss here the keen anomalies in more detail.

The search for possible variations of $\mu$ in time or in strong gravitational fields is based on the calculations of line shifts in the observed spectra \cite{Ubachs16}, the respective sensitivity coefficients were calculated in Refs. \cite{Meshkov06} and \cite{Salumbides15}. \textbf{An alternative method could be search of the keen anomalies sensitive to such variations}. We did not find any keen anomalies, sensitive enough to $\Delta\mu$, in the H$_2$ $X$--$X$ spectrum calculated by Roueff \emph{et al.} \cite{Roueff19}, yet this does not mean they do not exist in reality, as discussed below. 

By our request, E. Roueff and H. Abgrall made calculations to find a sensitive anomaly in the Lyman band system. Here is citation from Roueff's letter (bold face are ours). ``As you mention in your mail, we indeed find some strong anomalies in the intensities of some electronic transitions of the $B$--$X$ system arising from the computations of the  transition matrix element and possible cancellation due to \textbf{destructive interference} between the wave functions. These anomalies manifest themselves as huge variations in the transition emission probabilities for which \textbf{the dependence of the proton-to-electron mass ratio is indeed larger than the mean value}. However, the possible experimental or observational measurement seems quite improbable to me. Indeed, for example, we get for $B$--$X$, $\lambda$ = 1254.7, $v^\prime$ = 11, $J^\prime = 21 \rightarrow v^{\prime\prime} = 3, J^{\prime\prime} = 20, A = 0.478$ s$^{–1}$,\footnote{Misprint: $A=47$ s$^{-1}$.} a variation in $A$ of 2\% for $\Delta\mu/\mu$ = 5e–5." In other words, a relatively huge (in comparison with other studies) variation of $\Delta\mu/\mu$ resulted in a negligible variation in $A$. The above-mentioned anomaly is shown in Fig. \ref{anomLyman}.

Our objection to the above reasoning is based on the second key point: \textbf{The weak sensitivity of the keen anomalies found by us and by the colleagues relates to the artificial theoretical functions} rather than the true potential and quadrupole moment, which are unknown. Using the model functions, we can determine only locations and types of the anomalies, but not their precise intensities; the true intensities can differ from the calculated ones by times and orders of magnitude. The sensitivity of the destructive-interference effect to $\Delta\mu/\mu$ depends on the accuracy with which the interference is complete. When the transition matrix element is extremely small, the sensitivity will be extremely high, and \emph{vice versa}. Of course, this creates a difficulty to observe such weak lines, but it is a matter of experimental technique, which develops very rapidly towards increasing the measurement precision. When Einstein discovered gravitational waves, he could not imagine that one can ever detect a vibration of a 4th-km-long ruler with the amplitude of a few thousandth of the proton diameter.

Finally, it should be stressed that our classification of the keen anomalies in the vibronic spectra is purely speculative because no calculations on the electronic transitions were made by us in order to determine whether the anomalies shown in Fig. \ref{anomLyman} and in supplementary file 11 are indeed associated with the sign change of the TQM. Therefore, before looking for sensitivity of a particular anomaly, one has to verify that the TQM sign change indeed takes place and that, additionally, no stronger magnetic-dipole transition overlaps the given anomaly. 

\vspace{20pt}

\begin{longtable}{|c|c|c|c|r|c|c|l|}
\caption{Anomalies in the calculated H$_2$ spectrum caused by sign change of the transition matrix element as function of $v$ and $J$ (extract; the full table is given in supplementary file 10})\\ \hline

\multicolumn{4}{|c|}{} &  \multicolumn{4}{c|}{$A$, s$^{-1}$} \\ \hline
line	&	type$^a$	&	$E{_\textrm{u}}^b$	&	$\lambda,\mu$m	&
	irreg15$^c$	&	\ai\ \cite{Wolniewicz98}$^d$	&	\ai\ \cite{Komasa19}$^e$	&	\textbf{Roueff19} \cite{Roueff19}	 \\
\hline 
\endfirsthead
\multicolumn{8}{c}%
{{\bfseries Table \ref{Table_anomalies} (continued)}} \\
\hline 

\multicolumn{4}{|c|}{} &  \multicolumn{4}{c|}{$A$, s$^{-1}$} \\ \hline
line	&	type$^a$	&	$E{_\textrm{u}}^b$	&	$\lambda,\mu$m	&
	irreg15$^c$	&	\ai\ \cite{Wolniewicz98}$^d$	&	\ai\ \cite{Komasa19}$^e$	&	\textbf{Roueff19} \cite{Roueff19}	 \\

\hline
\endhead
\hline \multicolumn{8}{r}{{\textit{Continued on next page}}} \\ 
\endfoot
\endlastfoot


13-12 O(4)	&	r	&	50995	&	13.536	&	4.8553e-013	&	4.8546e-013	&	4.9091e-013	&	4.382e-013	 \\
13-12 O(3)	&	v	&	50908	&	11.976	&	1.3324e-011	&	1.3324e-011	&	1.3277e-011	&	1.366e-011	 \\ \hline
\multicolumn{8}{|c|}{$\cdots\cdots\cdots\cdots$} \\ \hline

4-1 O(26)	&	v	&	49878	&	2.485	&	2.0469e-009	&	2.0457e-009	&	2.0506e-009	&	2.05e-009	 \\
11-9 S(7)	&	vr	&	50183	&	2.485	&	3.4126e-009	&	3.4147e-009	&	3.429e-009	&	3.483e-009	 \\
7-4 O(20)	&	v	&	49418	&	2.478	&	1.0714e-009	&	1.0726e-009	&	1.0759e-009	&	1.069e-009	 \\
\emph{1–0 Q(6)}$^f$ &    &  9286  &   2.4755    &      \multicolumn{4}{c|}{\textbf{Geballe17} \cite{Geballe17}} \\

5-2 O(24)	&	v	&	49573	&	2.471	&	3.1231e-009	&	3.1228e-009	&	3.1363e-009	&	3.135e-009	 \\
6-3 O(22)	&	v	&	49420	&	2.467	&	2.868e-009	&	2.8691e-009	&	2.8862e-009	&	2.881e-009	 \\
2-0 O(17)	&	v	&	27265	&	2.458	&	7.0875e-011	&	7.1418e-011	&	7.1424e-011	&	7.203e-011	 \\
\emph{1-0 Q(5)}$^f$ &    &    &    2.45475    &      \multicolumn{4}{c|}{\textbf{Oh16} \cite{Oh16}} \\

11-9 S(6)	&	v	&	49676	&	2.448	&	4.9573e-009	&	4.9553e-009	&	4.9205e-009	&	4.87e-009	 \\
\emph{1-0 Q(4)}$^f$ &    &    &   2.437491    &      \multicolumn{4}{c|}{\textbf{Le17} \cite{Le17}} \\

11-9 S(5)	&	v	&	49198	&	2.429	&	3.763e-008	&	3.7626e-008	&	3.7488e-008	&	3.739e-008	 \\
6-5 S(7)	&	r	&	35989	&	2.428	&	1.1468e-009	&	1.1475e-009	&	1.1463e-009	&	1.15e-009	 \\
\emph{1–0 Q(3)}$^f$    &       &   6952    &   2.4237  &    \multicolumn{4}{c|}{\textbf{Pike16} \cite{Pike16}; \textit{A} = 2.78e-007 \textbf{Turner77}  \cite{Turner77}} \\

11-9 S(4)	&	v	&	48758	&	2.426	&	9.3066e-008	&	9.3062e-008	&	9.28e-008	&	9.27e-008	 \\
6-5 S(8)	&	v	&	37012	&	2.391	&	1.0786e-008	&	1.0783e-008	&	1.0789e-008	&	1.078e-008	 \\
\emph{3–2 S(1)}$^f$ &    &  17819  &   2.3863    &      \multicolumn{4}{c|}{\textbf{Geballe17} \cite{Geballe17}} \\

2-0 O(16)	&	v	&	25569	&	2.333	&	1.6751e-010	&	1.6831e-010	&	1.6832e-010	&	1.694e-010	 \\
\emph{10-8 Q(2)}$^f$ &    &    &    2.337306     &      \multicolumn{4}{c|}{\textbf{Le17} \cite{Le17}} \\

8-5 O(17)	&	v	&	48556	&	2.329	&	4.289e-011	&	4.2562e-011	&	4.3533e-011	&	4.577e-011	 \\
4-1 O(25)	&	v	&	48344	&	2.311	&	2.8962e-009	&	2.8932e-009	&	2.8972e-009	&	2.897e-009	 \\
7-4 O(19)	&	v	&	48267	&	2.303	&	1.9168e-009	&	1.9186e-009	&	1.934e-009	&	1.926e-009	 \\
5-2 O(23)	&	v	&	48158	&	2.298	&	4.5584e-009	&	4.5565e-009	&	4.5702e-009	&	4.569e-009	 \\
6-3 O(21)	&	v	&	48133	&	2.294	&	4.4127e-009	&	4.4134e-009	&	4.4379e-009	&	4.432e-009	 \\
\emph{3–2 S(25)}$^f$    &       &    51939   &  2.2663   &    \multicolumn{4}{c|}{\textbf{Pike16} \cite{Pike16}; \textit{A} = 3.07e-006 \textbf{Turner77}  \cite{Turner77}} \\

5-4 S(9)	&	vr	&	34288	&	2.186	&	5.5675e-009	&	5.568e-009	&	5.5693e-009	&	5.551e-009	 \\
\emph{5–4 S(13)}$^f$    &       &   39539    &   2.1528  &    \multicolumn{4}{c|}{\textbf{Pike16} \cite{Pike16}; \textit{A} = 5.08e-007 \textbf{Turner77}  \cite{Turner77}} \\

8-5 O(16)	&	v	&	47551	&	2.178	&	2.4859e-011	&	2.4549e-011	&	2.3562e-011	&	2.52e-011	 \\
5-4 S(10)	&	v	&	35526	&	2.166	&	5.2832e-008	&	5.283e-008	&	5.2835e-008	&	5.28e-008	 \\
4-1 O(24)	&	v	&	46782	&	2.163	&	3.8428e-009	&	3.8375e-009	&	3.8405e-009	&	3.841e-009	 \\
10-8 S(9)	&	v	&	49541	&	2.161	&	8.0449e-009	&	8.0459e-009	&	8.1621e-009	&	8.225e-009	 \\
7-4 O(18)	&	v	&	47109	&	2.155	&	2.7777e-009	&	2.7797e-009	&	2.8055e-009	&	2.796e-009	 \\
\emph{2-1 S(2)}$^f$ &    &    &    2.154225    &      \multicolumn{4}{c|}{\textbf{Le17} \cite{Le17}} \\

5-2 O(22)	&	v	&	46721	&	2.151	&	6.1442e-009	&	6.1399e-009	&	6.1515e-009	&	6.15e-009	 \\
6-3 O(20)	&	v	&	46831	&	2.147	&	6.0818e-009	&	6.0812e-009	&	6.1077e-009	&	6.102e-009	 \\
\emph{4–3 S(6)}$^f$ &    &  26615  &   2.1460    &      \multicolumn{4}{c|}{\textbf{Geballe17} \cite{Geballe17}} \\

10-8 S(8)	&	vr	&	48865	&	2.123	&	4.6464e-009	&	4.6467e-009	&	4.5481e-009	&	4.517e-009	 \\
10-8 S(7)	&	v	&	48213	&	2.098	&	4.936e-008	&	4.9364e-008	&	4.9051e-008	&	4.9e-008	 \\
12-9 S(8)	&	v	&	51947	&	2.086	&	2.983e-009	&	2.9818e-009	&	2.9553e-009	&	2.883e-009	 \\
10-8 S(6)	&	v	&	47595	&	2.085	&	1.3364e-007	&	1.3365e-007	&	1.3321e-007	&	1.332e-007	 \\
10-8 S(5)	&	v	&	47022	&	2.083	&	2.4264e-007	&	2.4265e-007	&	2.422e-007	&	2.423e-007	 \\
\emph{2-1 S(3)}$^f$ &    &  13890  &   2.073482    &      \multicolumn{4}{c|}{\textbf{Kaplan17} \cite{Kaplan17}; \textit{A} = 5.75e-007 \textbf{Wolniewicz98}  \cite{Wolniewicz98}} \\
\emph{12-9 O(3)}$^f$ &    &    &    2.069969    &      \multicolumn{4}{c|}{\textbf{Le17} \cite{Le17}} \\

8-5 O(15)	&	v	&	46557	&	2.05	&	5.9279e-011	&	5.8746e-011	&	5.5134e-011	&	5.771e-011	 \\
4-1 O(23)	&	v	&	45202	&	2.035	&	4.8545e-009	&	4.8467e-009	&	4.8487e-009	&	4.849e-009	 \\
7-4 O(17)	&	v	&	45954	&	2.028	&	3.4565e-009	&	3.4582e-009	&	3.4886e-009	&	3.478e-009	 \\
5-2 O(21)	&	v	&	45273	&	2.024	&	7.7927e-009	&	7.7854e-009	&	7.7942e-009	&	7.793e-009	 \\
\emph{7-5 O(5)}$^f$ &    &    &    2.022040    &      \multicolumn{4}{c|}{\textbf{Le17} \cite{Le17}} \\

6-3 O(19)	&	v	&	45524	&	2.02	&	7.71e-009	&	7.7075e-009	&	7.7312e-009	&	7.725e-009	 \\
4-3 S(10)	&	vr	&	31465	&	2.014	&	1.9822e-009	&	1.9848e-009	&	1.9852e-009	&	1.964e-009	 \\
4-3 S(11)	&	v	&	32854	&	2	&	4.1269e-008	&	4.1278e-008	&	4.128e-008	&	4.121e-008	 \\
\emph{4–3 S(17)}$^f$    &    &   42022    &   2.0475    &       \multicolumn{4}{c|}{\textbf{Pike16} \cite{Pike16}; \textit{A} = 1.34e-006 \textbf{Turner77}  \cite{Turner77}} \\
\emph{1–0 S(3)}$^f$ &    &  8365  &   1.9576    &      \multicolumn{4}{c|}{\textbf{Geballe17} \cite{Geballe17}} \\

10-6 O(16)	&	v	&	51596	&	1.951	&	1.7576e-009	&	1.7603e-009	&	1.74e-009	&	1.722e-009	 \\
\emph{2-1 S(5)}$^f$ &    &    &    1.94487    &      \multicolumn{4}{c|}{\textbf{Oh16} \cite{Oh16}} \\

8-5 O(14)	&	v	&	45584	&	1.938	&	2.4683e-010	&	2.4572e-010	&	2.3574e-010	&	2.412e-010	 \\
9-7 S(11)	&	v	&	48987	&	1.937	&	2.1414e-008	&	2.1409e-008	&	2.1622e-008	&	2.168e-008	 \\
4-1 O(22)	&	v	&	43612	&	1.923	&	5.8951e-009	&	5.885e-009	&	5.8862e-009	&	5.887e-009	 \\
7-4 O(16)	&	v	&	44812	&	1.917	&	3.7642e-009	&	3.7651e-009	&	3.7937e-009	&	3.783e-009	 \\
5-2 O(20)	&	v	&	43822	&	1.912	&	9.4048e-009	&	9.3943e-009	&	9.4005e-009	&	9.399e-009	 \\
6-3 O(18)	&	v	&	44223	&	1.909	&	9.1162e-009	&	9.1112e-009	&	9.1298e-009	&	9.123e-009	 \\
9-7 S(10)	&	vr	&	48144	&	1.897	&	9.0535e-010	&	9.0694e-010	&	8.7707e-010	&	8.722e-010	 \\
9-7 S(9)	&	v	&	47318	&	1.869	&	4.1735e-008	&	4.1747e-008	&	4.1656e-008	&	4.167e-008	 \\
3-2 S(11)	&	vr	&	28555	&	1.868	&	2.6272e-010	&	2.6385e-010	&	2.6398e-010	&	2.542e-010	 \\
3-2 S(12)	&	v	&	30097	&	1.858	&	2.9354e-008	&	2.9366e-008	&	2.9367e-008	&	2.927e-008	 \\
9-7 S(8)	&	v	&	46521	&	1.85	&	1.3711e-007	&	1.3713e-007	&	1.3716e-007	&	1.373e-007	 \\
9-7 S(7)	&	v	&	45762	&	1.84	&	2.7337e-007	&	2.734e-007	&	2.7365e-007	&	2.739e-007	 \\
8-5 O(13)	&	v	&	44641	&	1.839	&	9.0176e-010	&	8.9977e-010	&	8.7895e-010	&	8.895e-010	 \\
4-1 O(21)	&	v	&	42020	&	1.823	&	6.9242e-009	&	6.9123e-009	&	6.913e-009	&	6.915e-009	 \\
7-4 O(15)	&	v	&	43693	&	1.819	&	3.5617e-009	&	3.5615e-009	&	3.584e-009	&	3.573e-009	 \\
5-2 O(19)	&	v	&	42377	&	1.813	&	1.0871e-008	&	1.0858e-008	&	1.0862e-008	&	1.086e-008	 \\
6-3 O(17)	&	v	&	42936	&	1.811	&	1.0114e-008	&	1.0107e-008	&	1.012e-008	&	1.012e-008	 \\
10-6 O(15)	&	v	&	50921	&	1.802	&	3.1503e-009	&	3.155e-009	&	3.1701e-009	&	3.149e-009	 \\
11-8 S(10)	&	v	&	51758	&	1.794	&	9.4044e-009	&	9.4071e-009	&	9.263e-009	&	9.173e-009	 \\
\emph{11-8 O(3)}$^f$ &    &    &     1.760929   &      \multicolumn{4}{c|}{\textbf{Le17} \cite{Le17}} \\

8-5 O(12)	&	v	&	43739	&	1.752	&	2.669e-009	&	2.6661e-009	&	2.6315e-009	&	2.651e-009	 \\
2-1 S(12)	&	r	&	25569	&	1.742	&	5.3761e-011	&	5.3882e-011	&	5.3859e-011	&	5.93e-011	 \\
2-1 S(13)	&	v	&	27265	&	1.736	&	1.809e-008	&	1.809e-008	&	1.809e-008	&	1.8e-008	 \\
4-1 O(20)	&	v	&	40433	&	1.734	&	7.896e-009	&	7.8833e-009	&	7.8837e-009	&	7.885e-009	 \\
\emph{6-4 O(3)}$^f$ &    &    &    1.732637    &      \multicolumn{4}{c|}{\textbf{Le17} \cite{Le17}} \\

8-6 S(12)	&	vr	&	47551	&	1.732	&	1.4013e-009	&	1.3982e-009	&	1.3851e-009	&	1.38e-009	 \\
7-4 O(14)	&	v	&	42605	&	1.731	&	2.8123e-009	&	2.811e-009	&	2.8261e-009	&	2.816e-009	 \\
5-2 O(18)	&	v	&	40946	&	1.725	&	1.2074e-008	&	1.2059e-008	&	1.2062e-008	&	1.206e-008	 \\
6-3 O(16)	&	v	&	41671	&	1.724	&	1.053e-008	&	1.052e-008	&	1.0529e-008	&	1.052e-008	 \\
\emph{1-0 S(8)}$^f$ &    &    &    1.71466    &      \multicolumn{4}{c|}{\textbf{Oh16} \cite{Oh16}} \\

8-6 S(11)	&	v	&	46557	&	1.701	&	2.1541e-008	&	2.1551e-008	&	2.1668e-008	&	2.172e-008	 \\
10-6 O(14)	&	v	&	50231	&	1.681	&	4.1372e-009	&	4.143e-009	&	4.2133e-009	&	4.191e-009	 \\
8-6 S(10)	&	v	&	45584	&	1.678	&	1.0572e-007	&	1.0573e-007	&	1.0607e-007	&	1.062e-007	 \\
\emph{2-0 O(9)}$^f$ &    &  15763  &   1.679641    &      \multicolumn{4}{c|}{\textbf{Kaplan17} \cite{Kaplan17}; \textit{A} = 1.29e-008 \textbf{Wolniewicz98}  \cite{Wolniewicz98}} \\

8-6 S(9)	&	v	&	44641	&	1.663	&	2.4328e-007	&	2.4329e-007	&	2.4382e-007	&	2.442e-007	 \\
\emph{11-8 Q(1)}$^f$ &    &    &    1.657105    &      \multicolumn{4}{c|}{\textbf{Le17} \cite{Le17}} \\

8-6 S(8)	&	v	&	43739	&	1.654	&	4.1862e-007	&	4.1862e-007	&	4.1926e-007	&	4.198e-007	 \\
7-4 O(13)	&	v	&	41558	&	1.653	&	1.6498e-009	&	1.648e-009	&	1.6563e-009	&	1.648e-009	 \\
4-1 O(19)	&	v	&	38859	&	1.653	&	8.7593e-009	&	8.747e-009	&	8.7472e-009	&	8.749e-009	 \\
5-2 O(17)	&	v	&	39537	&	1.645	&	1.2892e-008	&	1.2876e-008	&	1.2878e-008	&	1.288e-008	 \\
6-3 O(15)	&	v	&	40438	&	1.644	&	1.0219e-008	&	1.0207e-008	&	1.0213e-008	&	1.02e-008	 \\
\emph{1-0 S(11)}$^f$    &    &   18979    &   1.650413    &       \multicolumn{4}{c|}{\textbf{Kaplan17} \cite{Kaplan17}; \textit{A} = 5.37e-008 \textbf{Wolniewicz98}  \cite{Wolniewicz98}} \\

1-0 S(13)	&	r	&	22516	&	1.632	&	4.5463e-010	&	4.5889e-010	&	4.5886e-010	&	4.735e-010	 \\
1-0 S(14)	&	v	&	24367	&	1.63	&	8.16e-009	&	8.1445e-009	&	8.1446e-009	&	8.087e-009	 \\
7-5 S(14)	&	v	&	47109	&	1.607	&	1.5097e-008	&	1.5088e-008	&	1.4966e-008	&	1.492e-008	 \\
10-7 S(12)	&	v	&	51596	&	1.6	&	1.0993e-008	&	1.1001e-008	&	1.0923e-008	&	1.087e-008	 \\
\emph{7-5 S(2)}$^f$ &    &    &    1.588290    &      \multicolumn{4}{c|}{\textbf{Le17} \cite{Le17}} \\

7-4 O(12)	&	v	&	40559	&	1.582	&	4.6704e-010	&	4.6559e-010	&	4.6875e-010	&	4.635e-010	 \\
10-6 O(13)	&	v	&	49541	&	1.58	&	4.3071e-009	&	4.3126e-009	&	4.4248e-009	&	4.403e-009	 \\
4-1 O(18)	&	v	&	37306	&	1.579	&	9.4571e-009	&	9.4464e-009	&	9.4466e-009	&	9.445e-009	 \\
7-5 S(13)	&	vr	&	45954	&	1.575	&	3.8377e-009	&	3.8403e-009	&	3.9022e-009	&	3.939e-009	 \\
6-3 O(14)	&	v	&	39245	&	1.573	&	9.1045e-009	&	9.0927e-009	&	9.0965e-009	&	9.086e-009	 \\
5-2 O(16)	&	v	&	38158	&	1.572	&	1.3204e-008	&	1.3189e-008	&	1.319e-008	&	1.319e-008	 \\
\emph{5-3 O(2)}$^f$ &    &  26606  &   1.560736    &      \multicolumn{4}{c|}{\textbf{Kaplan17} \cite{Kaplan17}; \textit{A} = 2.24E-006 \textbf{Wolniewicz98}  \cite{Wolniewicz98}} \\

7-5 S(12)	&	v	&	44812	&	1.549	&	5.9038e-008	&	5.9038e-008	&	5.9247e-008	&	5.943e-008	 \\
12-7 O(12)	&	v	&	51947	&	1.54	&	8.3512e-010	&	8.3761e-010	&	8.4954e-010	&	8.323e-010	 \\
7-5 S(11)	&	v	&	43693	&	1.53	&	1.7356e-007	&	1.7354e-007	&	1.7383e-007	&	1.742e-007	 \\
\emph{3-1 O(5)}$^f$ &    &    &    1.52203    &      \multicolumn{4}{c|}{\textbf{Oh16} \cite{Oh16}} \\

7-4 O(11)	&	vr	&	39618	&	1.518	&	2.3598e-011	&	2.4024e-011	&	2.3508e-011	&	2.47e-011	 \\
7-5 S(10)	&	v	&	42605	&	1.516	&	3.3583e-007	&	3.3578e-007	&	3.3607e-007	&	3.366e-007	 \\
\emph{5-3 Q(4)}$^f$ &    &    &    1.515792    &      \multicolumn{4}{c|}{\textbf{Le17} \cite{Le17}} \\

4-1 O(17)	&	v	&	35783	&	1.511	&	9.9274e-009	&	9.9198e-009	&	9.92e-009	&	9.922e-009	 \\
\emph{4-2 O(3)}$^f$ &    &  22079  &   1.509865    &      \multicolumn{4}{c|}{\textbf{Kaplan17} \cite{Kaplan17}; \textit{A} = 7.76e-007 \textbf{Wolniewicz98}  \cite{Wolniewicz98}} \\

7-5 S(9)	&	v	&	41558	&	1.508	&	5.3073e-007	&	5.3065e-007	&	5.309e-007	&	5.317e-007	 \\
6-3 O(13)	&	v	&	38100	&	1.507	&	7.2187e-009	&	7.2077e-009	&	7.2101e-009	&	7.202e-009	 \\
5-2 O(15)	&	v	&	36818	&	1.506	&	1.2904e-008	&	1.289e-008	&	1.2891e-008	&	1.289e-008	 \\
\emph{6-4 S(1)}$^f$ &    &  31661  &    1.50156   &      \multicolumn{4}{c|}{\textbf{Kaplan17} \cite{Kaplan17}; \textit{A} = 1.15e-006 \textbf{Wolniewicz98}  \cite{Wolniewicz98}} \\

10-6 O(12)	&	v	&	48865	&	1.496	&	3.5165e-009	&	3.5203e-009	&	3.6386e-009	&	3.619e-009	 \\
6-4 S(15)	&	vr	&	45524	&	1.477	&	1.2438e-009	&	1.2444e-009	&	1.2239e-009	&	1.198e-009	 \\
9-6 S(14)	&	v	&	51507	&	1.463	&	6.435e-009	&	6.4423e-009	&	6.5505e-009	&	6.547e-009	 \\ \hline
\multicolumn{8}{|c|}{$\cdots\cdots\cdots\cdots$} \\ \hline
12-3 O(2)	&	v	&	49354	&	0.45	&	1.6556e-011	&	1.1558e-011	&	1.1441e-011	&	1.111e-011	 \\
8-1 O(2)	&	v	&	38604	&	0.448	&	1.2222e-011	&	7.4728e-012	&	7.7483e-012	&	8.672e-012	 \\

10-2 O(2)	&	v	&	44819	&	0.44	&		&	2.3926e-011	&		&		 \\
11-1 Q(2)	&	v	&	47535	&	0.35	&	1.6571e-014	&		&		&		 \\
9-0 Q(4)	&	v	&	42827	&	0.35	&	1.8971e-014	&		&		&		 \\
11-1 Q(1)	&	v	&	47391	&	0.349	&	1.1775e-016	&		&		&		 \\

9-0 Q(3)	&	vr	&	42463	&	0.347	&	9.1824e-016	&		&		&		 \\
9-0 Q(2)	&	v	&	42185	&	0.345	&	3.1132e-015	&	1.9326e-014	&	1.5311e-014	&	1.521e-014	 \\
9-0 Q(1)	&	v	&	41998	&	0.344	&	1.8542e-014	&	8.6009e-015	&	5.3482e-015	&	5.269e-015	 \\

12-1 Q(4)	&	vr	&	49932	&	0.34	&	5.3309e-016	&		&		&		 \\
12-1 Q(3)	&	v	&	49706	&	0.337	&	3.75e-014	&		&		&		 \\
12-1 Q(3)	&	vr	&	49706	&	0.337	&	            &	3.7163e-015 &	2.0719e-015	&	3.778e-015	  \\

10-0 Q(6)	&	vr	&	46500	&	0.334	&	3.263e-016	&		&		&		 \\
12-1 Q(2)	&	v	&	49532	&	0.334	&		&	5.5923e-015	&	8.9173e-015	&	6.355e-015	 \\
13-1 Q(5)	&	vr	&	51473	&	0.334	&	2.0924e-015	&		&		&		 \\
12-1 Q(1)	&	v	&	49414	&	0.333	&		&	3.9967e-014	&	5.1183e-014	&	5.132e-014	 \\

10-0 Q(5)	&	v	&	46038	&	0.33	&	8.2368e-015	&	2.2102e-015	&		&		 \\
10-0 Q(5)	&	vr	&	46038	&	0.33	&	         	&	         &	1.2674e-15	&	1.2410e-15	 \\

13-1 Q(4)	&	vr	&	51284	&	0.329	&		&	3.6783e-016	&	8.77e-016	&	8.85e-016	 \\

10-0 Q(4)	&	v	&	45641	&	0.327	&		&	2.6532e-015	&	4.3113e-015	&	4.381e-015	 \\
10-0 Q(3)   &   v    &   45317   &    0.325  &      &   1.8397E-14  &    2.3179E-14    &  2.3380E-14  \\

\hline
11-0 Q(7) $^g$	&	vr	&	49198	&	0.323	&	1.8539e-015	&		&		&		 \\
11-0 Q(6)   &   vr  &   48758   &   0.318   &       &  1.1498e-15   &  2.0069E-15     &   2.0410E-15   \\
\hline
12-0 Q(8) $^g$	&	vr	&	51218	&	0.317	&	7.4371e-016	&		&		&		 \\
12-0 Q(7)   &   vr  &   50856   &   0.311   &       &  8.7092e-16     &  1.4616E-15     &   1.4800E-15  \\

\hline 
\multicolumn{8}{l}{$^a$ Anomaly types: v, vibrational; r, rotational; vr, vibrational-rotational (see text).} \\
\multicolumn{8}{l}{$^b$ The energy of the upper state in Kelvin.} \\
\multicolumn{8}{l}{$^c$ Present study, the irreg15 TQM.} \\
\multicolumn{8}{l}{$^d$ Present study, the spline-interpolated \ai\ TQM of \textbf{Wolniewicz} \cite{Wolniewicz98}.} \\
\multicolumn{8}{l}{$^e$ Present study, the spline-interpolated TQM of \textbf{Komasa19} \cite{Komasa19}.} \\
\multicolumn{8}{l}{$^f$ Observed lines near the anomalies are shown in italic. The reference to the source} \\
\multicolumn{8}{l}{ is given. The calculated $A$ values for these lines from the source are cited.} \\
\multicolumn{8}{l}{$^g$ When QMF is changed, the anomaly can disappear (empty spaces) or move to a neighbouring $J$} \\
\multicolumn{8}{l}{(two examples are shown by horizontal lines.)} \\

\label{Table_anomalies}    
\end{longtable}

\newpage

\newpage
\section{Conclusions}\label{concl}

This study was motivated by the fact that the \emph{X-X} ro-vibrational transition intensities were calculated with use of the spline-interpolated quadrupole-moment function (QMF) \cite{Roueff19}, which had been compromised in CO \cite{Medvedev15,Medvedev16} and other molecules \cite{Medvedev22c}. The second motivation was that the numerous available experimental and observational data were not used in construction of the QMF. In order to resolve these issues, we constructed a few analytic functions fitted to either the \ai\ data only or to both the \ai\ and experimental data. We used these analytic QMFs to calculate the transition intensities and found the following.
 
1. In the case of H$_2$, as distinct from other molecules, the use of the spline-interpolated QMF does not deteriorate the overtone intensities and does not lead to nonphysical saturation owing to the high precision of the \ai\ data (at least six decimal places) and the high density of the \ai\ grid.

2. The use of the laboratory data does not help increase the quality of the analytic QMF. While the observed-minus-calculated differences can be essentially decreased by fitting the QMF to the experimental data, this is reached by a large deviation of the QMF from the \ai\ data that seems improbable.

3. The same is true of the astrophysical observations in view of low precision of the data, since there is likely large systematic uncertainties are associated with the dust extinction. 

4. The theoretical \emph{A} values are used in astrophysics to determine the column densities for individual lines and the temperature of the emitting objects \cite{Pike16,Kaplan17,Roueff23,Le17,Geballe17,Neufeld21}. Our new data do not affect the results of this analysis.

5. The NIDL theory applies to H$_2$, \emph{i.e.} the transition moment is factorised as $T_0B_0$, where $T_0$ is the exponential factor depending on the potential alone and pre-factor $B_0$ depending on both the potential and the QMF. As always, the pre-factor is responsible for the anomalies, \emph{i.e.} the lines with intensities lower than those predicted by the NIDL for the ``normal" lines. The anomalies arise because $B_0$ can cross zero and change its sign at some transitions. The difference with other molecules is that $B_0$ is no longer a slow function of the overtone number as compared to $T_0$, therefore it affects the NIDL slope.

6. Numerous anomalies resulting from destructive interference (the ``keen" anomalies) are identified in the calculated line lists. Some of them are situated within the recently observed spectral regions, \emph{e.g.} 1.5-2.5 $\mu$m, and some are close to such regions. The intensities of the anomalies are very sensitive to the QMF representation, as illustrated by comparison of our results with those of Roueff \emph{et al.} \cite{Roueff19}.

7. As a side result, we mention the Le Roy anomalies in the ro-vibronic (electronic) spectra of H$_2$. Le Roy with co-authors \cite{Brown73,LeRoy75} predicted that the transition matrix element of I$_2$ can cross zero as function of $J$ and change its sign, exactly as in our case of the \emph{X-X} transitions. Here, we demonstrated that numerous Le Roy anomalies are seen in the Lyman and Werner series. 

8. Since the keen anomalies in both the vibration-rotation and vibronic spectra are sensitive to the potential and the moment, we assumed that they might be also sensitive to the proton-to-electron mass ratio. Sample calculations by us as well as by E. Roueff and H. Abgrall performed by our request (unpublished) did reveal enhanced sensitivity of the anomalies as compared to the normal lines, yet insufficient to be observed.

9. The insensitivity of the keen anomalies relates to the model molecular functions whereas the actual functions can make some anomalies sensitive. The only thing that theory can do is to predict the position of the anomaly; one also has to make sure that the particular anomaly is due to the sign change of the transition matrix element in order that the  anomaly were of the interference nature. 

\section*{Supplementary material}

File 1 \textbf{Comparing sets I and II QMFs.pdf}.

File 2 \textbf{H2\_laboratory\_data.txt}.

File 3 \textbf{QMFs\_em.f}. FORTRAN codes for three analytic quadrupole-moment functions: irreg15, irreg16, and irreg15exp.

File 4 \textbf{Fit\_OUT\_Irreg15\_Paper\_em1.txt}. The output file for the FORTRAN code using irreg15.

File 5 \textbf{Fit\_OUT\_Quadr16\_Paper\_em.txt}. The output file for the  FORTRAN code using quadr16.

File 6 \textbf{Fit\_TDM\_DMF-0\_001\_em1.txt}. The output file for the  FORTRAN code using irreg15exp.

File 7 \textbf{TQMs\_em.txt}. The transition quadrupole moments computed with three QMFs for the S(0), Q(1), and O(2) lines at $0\le v^{\prime\prime}\le v^\prime\le14$.

File 8 \textbf{parameters\_em.txt}. The fitted parameters of the irreg15, irreg15exp, and quadr16 QMFs.

File 9 \textbf{H2\_astrophysical\_data\_Table1.txt}.

File 10 \textbf{Paper\_on\_H2\_2023\_Table3\_full.pdf}.

File 11 \textbf{Anomalies in Lyman and Werner series.pdf}. List of keen anomalies found by us in Refs. \cite{Abgrall93,Abgrall93L,Abgrall93W}.

\section*{Acknowledgements}

We are grateful to J. Komasa for \ai\ data, E. Roueff and H. Abgrall for the calculations discussed in Sec. \ref{anom}
. The work by VGU and ESM was carried out in accordance with the state
task, state registration number AAAA-A19-119071190017-7.

\bibliography{Paper_on_H2_2023}

\begin{thebibliography}{10}
\expandafter\ifx\csname url\endcsname\relax
  \def\url#1{\texttt{#1}}\fi
\expandafter\ifx\csname urlprefix\endcsname\relax\def\urlprefix{URL }\fi
\expandafter\ifx\csname href\endcsname\relax
  \def\href#1#2{#2} \def\path#1{#1}\fi

\bibitem{HITRAN2020}
I.~E. {Gordon}, L.~S. {Rothman}, R.~J. {Hargreaves}, R.~{Hashemi}, E.~V.
  {Karlovets}, F.~M. {Skinner}, E.~K. {Conway}, C.~{Hill}, R.~V. {Kochanov},
  Y.~{Tan}, P.~{Wcis{\l}o}, A.~A. {Finenko}, K.~{Nelson}, P.~F. {Bernath},
  M.~{Birk}, V.~{Boudon}, A.~{Campargue}, K.~V. {Chance}, A.~{Coustenis}, B.~J.
  {Drouin}, J.-M. {Flaud}, R.~R. {Gamache}, J.~T. {Hodges}, D.~{Jacquemart},
  E.~J. {Mlawer}, A.~V. {Nikitin}, V.~I. {Perevalov}, M.~{Rotger},
  J.~{Tennyson}, G.~C. {Toon}, H.~{Tran}, V.~G. {Tyuterev}, E.~M. {Adkins},
  A.~{Baker}, A.~{Barbe}, E.~{Can{\`{e}}}, A.~G. {Cs{'{a}}sz{'{a}}r},
  A.~{Dudaryonok}, O.~{Egorov}, A.~J. {Fleisher}, H.~{Fleurbaey},
  A.~{Foltynowicz}, T.~{Furtenbacher}, J.~J. {Harrison}, J.-M. {Hartmann},
  V.-M. {Horneman}, X.~{Huang}, T.~{Karman}, J.~{Karns}, S.~{Kassi},
  I.~{Kleiner}, V.~{Kofman}, F.~{Kwabia-Tchana}, N.~N. {Lavrentieva}, T.~J.
  {Lee}, D.~A. {Long}, A.~A. {Lukashevskaya}, O.~M. {Lyulin}, V.~Y. {Makhnev},
  W.~{Matt}, S.~T. {Massie}, M.~{Melosso}, S.~N. {Mikhailenko}, D.~{Mondelain},
  H.~S.~P. {M{"{u}}ller}, O.~V. {Naumenko}, A.~{Perrin}, O.~L. {Polyansky},
  E.~{Raddaoui}, P.~L. {Raston}, Z.~D. {Reed}, M.~{Rey}, C.~{Richard},
  R.~{T{'{o}}bi{'{a}}s}, I.~{Sadiek}, D.~W. {Schwenke}, E.~{Starikova},
  K.~{Sung}, F.~{Tamassia}, S.~A. {Tashkun}, J.~{Vander Auwera}, I.~A.
  {Vasilenko}, A.~A. {Vigasin}, G.~L. {Villanueva}, B.~{Vispoel}, G.~{Wagner},
  A.~{Yachmenev}, S.~N. {Yurchenko}, The {HITRAN2020} molecular spectroscopic
  database, Journal of Quantitative Spectroscopy and Radiative Transfer 277
  (2022) 107949.
\newblock \href {https://doi.org/10.1016/j.jqsrt.2021.107949}
  {\path{doi:10.1016/j.jqsrt.2021.107949}}.

\bibitem{Ubachs16}
W.~Ubachs, J.~Bagdonaite, E.~J. Salumbides, M.~T. Murphy, L.~Kaper,
  {Colloquium: Search for a drifting proton-electron mass ratio from
  ${\mathsf{H}}_{\mathsf{2}}$}, Rev. Mod. Phys. 88 (2016) 021003.
\newblock \href {https://doi.org/10.1103/RevModPhys.88.021003}
  {\path{doi:10.1103/RevModPhys.88.021003}}.

\bibitem{Pike16}
R.~E. Pike, T.~R. Geballe, M.~G. Burton, A.~Chrysostomou, {Highly excited H$_2$
  in Herzberg-Haro 7: Formation pumping in shocked molecular gas?}, Astrophys.
  J. 822 (2016) 82 (13pp).
\newblock \href {https://doi.org/10.3847/0004-637X/822/2/82}
  {\path{doi:10.3847/0004-637X/822/2/82}}.

\bibitem{Piszczatowski09}
K.~Piszczatowski, G.~{\L}ach, M.~Przybytek, J.~Komasa, K.~Pachucki,
  B.~Jeziorski, {Theoretical Determination of the Dissociation Energy of
  Molecular Hydrogen}, J. Chem. Theory Comp. 5 (2009) 3039--3048.
\newblock \href {https://doi.org/10.1021/ct900391p}
  {\path{doi:10.1021/ct900391p}}.

\bibitem{Pachucki10}
K.~Pachucki, {Born-Oppenheimer potential for H$_2$}, Phys. Rev. A 82 (2010)
  032509.
\newblock \href {https://doi.org/10.1103/PhysRevA.82.032509}
  {\path{doi:10.1103/PhysRevA.82.032509}}.

\bibitem{Komasa19}
J.~Komasa, M.~Puchalski, P.~Czachorowski, G.~{\L}ach, K.~Pachucki,
  Rovibrational energy levels of the hydrogen molecule through nonadiabatic
  perturbation theory, Phys. Rev. A 100 (2019) 032519.
\newblock \href {https://doi.org/10.1103/PhysRevA.100.032519}
  {\path{doi:10.1103/PhysRevA.100.032519}}.

\bibitem{Turner77}
J.~Turner, K.~Kirby-Docken, A.~Dalgarno, The quadrupole vibration-rotation
  transition probabilities of molecular hydrogen, Astrophys. J. Suppl. Ser. 35
  (1977) 281--292.

\bibitem{Wolniewicz98}
L.~Wolniewicz, I.~Simbotin, A.~Dalgarno, {Quadrupole transition probabilities
  for the excited rovibrational states of H$_2$}, Astrophys. J. Suppl. Ser.
  115~(2) (1998) 293--313.
\newblock \href {https://doi.org/10.1086/313091} {\path{doi:10.1086/313091}}.

\bibitem{Campargue12}
A.~Campargue, S.~Kassi, K.~Pachucki, J.~Komasa, {The absorption spectrum of
  H$_2$: CRDS measurements of the (2-0) band, review of the literature data and
  accurate \emph{ab initio} line list up to 35000 cm$^{-1}$}, Phys. Chem. Chem.
  Phys. 14 (2012) 802--815.
\newblock \href {https://doi.org/10.1039/c1cp22912e}
  {\path{doi:10.1039/c1cp22912e}}.

\bibitem{Roueff19}
E.~Roueff, H.~Abgrall, P.~Czachorowski, K.~Pachucki, M.~Puchalski, J.~Komasa,
  The full infrared spectrum of molecular hydrogen, A\&A 630 (2019) A58.
\newblock \href {https://doi.org/10.1051/0004-6361/201936249}
  {\path{doi:10.1051/0004-6361/201936249}}.

\bibitem{Medvedev15}
E.~S. Medvedev, V.~V. Meshkov, A.~V. Stolyarov, I.~E. Gordon, Peculiarities of
  high-overtone transition probabilities in carbon monoxide revealed by
  high-precision calculation, J. Chem. Phys. 143~(16) (2015) 154301.
\newblock \href {https://doi.org/10.1063/1.4933136}
  {\path{doi:10.1063/1.4933136}}.

\bibitem{Medvedev22}
E.~S. Medvedev, V.~G. Ushakov, Irregular semi-empirical dipole-moment function
  for carbon monoxide and line lists for all its isotopologues verified for
  extremely high overtone transitions, J. Quant. Spectrosc. Radiat. Transfer
  288 (2022) 108255.
\newblock \href {https://doi.org/10.1016/j.jqsrt.2022.108255}
  {\path{doi:10.1016/j.jqsrt.2022.108255}}.

\bibitem{Ushakov23}
V.~G. Ushakov, M.~Semenov, S.~N. Yurchenko, A.~Y. Ermilov, E.~S. Medvedev,
  {Improved potential-energy and dipole-moment functions of the ground
  electronic state of phosphorus nitride}, J. Mol. Spectrosc. 395 (2023)
  111804.
\newblock \href {https://doi.org/10.1016/j.jms.2023.111804}
  {\path{doi:10.1016/j.jms.2023.111804}}.

\bibitem{Medvedev22c}
E.~S. Medvedev, V.~G. Ushakov, Selection of the model functions for
  calculations of high-overtone intensities in the vibrational-rotational
  spectra of diatomic molecules, Opt. spectrosc. 130~(9) (2022) 1334--1342.
\newblock \href {https://doi.org/10.21883/OS.2022.09.53292.3428-22 [in Russian.
  English translation is available at doi: 10.21883/EOS.2022.09.54822.3428-22]}
  {\path{doi:10.21883/OS.2022.09.53292.3428-22 [in Russian. English translation
  is available at doi: 10.21883/EOS.2022.09.54822.3428-22]}}.

\bibitem{Medvedev12}
E.~S. Medvedev, {Towards understanding the nature of the intensities of
  overtone vibrational transitions}, J. Chem. Phys. 137 (2012) 174307.
\newblock \href {https://doi.org/10.1063/1.4761930}
  {\path{doi:10.1063/1.4761930}}.

\bibitem{Fink65}
U.~Fink, T.~A. Wiggins, D.~H. Rank, Frequency and intensity measurements on the
  quadrupole spectrum of molecular hydrogen, J. Mol. Spectrosc. 18~(4) (1965)
  384--395.
\newblock \href {https://doi.org/10.1016/0022-2852(65)90044-5}
  {\path{doi:10.1016/0022-2852(65)90044-5}}.

\bibitem{Margolis73}
J.~S. Margolis,
  \href{https://www.sciencedirect.com/science/article/pii/0022285273902038}{{Measurement
  of some 1-0 H$_2$ quadrupole transition strengths}}, J. Mol. Spectrosc.
  48~(2) (1973) 409--410.
\newblock \href {https://doi.org/https://doi.org/10.1016/0022-2852(73)90203-8}
  {\path{doi:https://doi.org/10.1016/0022-2852(73)90203-8}}.
\newline\urlprefix\url{https://www.sciencedirect.com/science/article/pii/0022285273902038}

\bibitem{Chackerian75a}
C.~Chackerian, L.~P. Giver, {Density-dependent frequency shift of the hydrogen
  S$_2$(1) quadrupole line}, J. Mol. Spectrosc. 58~(3) (1975) 339--343.
\newblock \href {https://doi.org/10.1016/0022-2852(75)90215-5}
  {\path{doi:10.1016/0022-2852(75)90215-5}}.

\bibitem{Chackerian75b}
C.~Chackerian, L.~P. Giver, {Measurement of weak ir absorption with a tunable
  laser: the hydrogen S$_2$(1) line strength}, Applied Optics 14~(8) (1975)
  1993--1996.
\newblock \href {https://doi.org/10.1364/AO.14.001993}
  {\path{doi:10.1364/AO.14.001993}}.

\bibitem{Bergstralh78}
J.~T. Bergstralh, J.~W. Brault, {Intensity and pressure shift of H$_2$ (4,0)
  S(1) quadrupole line}, Astrophys. J. 224 (1978) L39--L41.

\bibitem{Reid78}
J.~Reid, A.~R.~W. McKellar, {Observation of the ${S}_{0}(3)$ pure rotational
  quadrupole transition of ${\mathrm{H}}_{2}$ with a tunable diode laser},
  Phys. Rev. A 18 (1978) 224--228.
\newblock \href {https://doi.org/10.1103/PhysRevA.18.224}
  {\path{doi:10.1103/PhysRevA.18.224}}.

\bibitem{Trauger78}
J.~T. Trauger, M.~E. Mickelson, L.~E. Larson, {Laboratory absorption strengths
  for the H$_2$ (4,0) and (3,0) S(1) lines}, Astrophys. J. 225 (1978)
  L157--L160.

\bibitem{Brault80}
J.~W. Brault, W.~H. Smith, {Determination of the H$_2$ 4-0 S(1) quadrupole line
  strength and pressure shift}, Astrophys. J. 235 (1980) L177--L178.

\bibitem{Bragg82}
S.~L. Bragg, J.~W. Brault, W.~H. Smith, Line positions and strengths in the
  h$_2$ quadrupole spectrum, Astrophys. J. 263 (1982) 999--1004.

\bibitem{Jennings82}
D.~E. Jennings, J.~W. Brault, {The Ground State of Molecular Hydrogen},
  Astrophys. J. 256 (1982) L29--L31.

\bibitem{Ferguson93}
D.~W. Ferguson, K.~N. Rao, M.~E. Mickelson, L.~E. Larson, {An Experimental
  Study of the 4-0 and 5-0 Quadrupole Vibration Rotation Bands of H$_2$ in the
  Visible}, J. Mol. Spectrosc. 160~(2) (1993) 315--325.
\newblock \href {https://doi.org/10.1006/jmsp.1993.1178}
  {\path{doi:10.1006/jmsp.1993.1178}}.

\bibitem{Reuter94}
D.~C. Reuter, J.~M. Sirota, Astrophys. J. 428 (1994) L77–L79.

\bibitem{Gupta06}
M.~Gupta, T.~Owano, S.~Baer, D, A.~O'Keefe, {Quantitative determination of the
  Q(1) quadrupole hydrogen absorption in the near infrared via off-axis ICOS},
  Chem. Phys. Lett. 418~(1) (2006) 11--14.
\newblock \href {https://doi.org/10.1016/j.cplett.2005.10.081}
  {\path{doi:10.1016/j.cplett.2005.10.081}}.

\bibitem{Robie06}
D.~C. Robie, J.~T. Hodges, {Line positions and line strengths for the
  $3\leftarrow$ electric quadrupole band of H$_2\Sigma_g^+$}, J. Chem. Phys.
  124 (2006).
\newblock \href {https://doi.org/10.1063/1.2145925}
  {\path{doi:10.1063/1.2145925}}.

\bibitem{Hu12}
S.-M. Hu, H.~Pan, C.-F. Cheng, Y.~R. Sun, X.-F. Li, J.~Wang, A.~Campargue,
  A.-W. Liu, {The $v = 3\leftarrow 0$ S(0)-S(3) electric quadrupole transitions
  of H$_2$ near 0.8 $\mu$m}, Astrophys. J. 749 (2012) 76.
\newblock \href {https://doi.org/10.1088/0004-637X/749/1/76}
  {\path{doi:10.1088/0004-637X/749/1/76}}.

\bibitem{Kassi14}
S.~Kassi, A.~Campargue, {Electric quadrupole transitions and collision-induced
  absorption in the region of the first overtone band of H$_2$ near 1.25
  $\mu$m}, J. Mol. Spectrosc. 300 (2014) 55--59.
\newblock \href {https://doi.org/10.1016/j.jms.2014.03.022}
  {\path{doi:10.1016/j.jms.2014.03.022}}.

\bibitem{Oh16}
H.~Oh, T.-S. Pyo, K.~Kaplan, I.-S. Yuk, B.-G. Park, G.~Mace, C.~Park, M.-Y.
  Chun, S.~Pak, K.-M. Kim, J.~S. Oh, U.~Jeong, Y.~S. Yu, J.-J. Lee, H.~Kim,
  N.~Hwang, H.-I. Lee, H.~A.~N. Le, S.~Lee, D.~T. Jaffe, Three-dimensional
  shock structure of the orion kl outflow with igrins, Astrophys. J. 833 (2016)
  275.
\newblock \href {https://doi.org/10.3847/1538-4357/833/2/275}
  {\path{doi:10.3847/1538-4357/833/2/275}}.

\bibitem{Geballe17}
T.~R. Geballe, M.~G. Burton, R.~E. Pike, {Very High Excitation Lines of H$_2$
  in the Orion Molecular Cloud Outflow}, Astrophys. J. 837 (2017) 83 (7pp).
\newblock \href {https://doi.org/10.3847/1538-4357/aa619e}
  {\path{doi:10.3847/1538-4357/aa619e}}.

\bibitem{Kaplan17}
K.~F. Kaplan, H.~L. Dinerstein, H.~Oh, G.~N. Mace, H.~Kim, K.~R. Sokal, M.~D.
  Pavel, S.~Lee, S.~Pak, C.~Park, J.~Sok~Oh, D.~T. Jaffe, {Excitation of
  Molecular Hydrogen in the Orion Bar Photodissociation Region from a Deep
  Near-infrared IGRINS Spectrum}, Astrophys. J. 838 (2017) 152 (13pp).
\newblock \href {https://doi.org/10.3847/1538-4357/aa5b9f}
  {\path{doi:10.3847/1538-4357/aa5b9f}}.

\bibitem{Le17}
H.~A.~N. Le, S.~Pak, K.~Kaplan, G.~Mace, S.~Lee, M.~Pavel, U.~Jeong, H.~Oh,
  H.-I. Lee, M.-Y. Chun, I.-S. Yuk, T.-S. Pyo, N.~Hwang, K.-M. Kim, C.~Park,
  J.~Sok~Oh, Y.~S. Yu, B.-G. Park, Y.~C. Minh, D.~T. Jaffe, {Fluorescent H$_2$
  Emission Lines from the Reflection Nebula NGC 7023 Observed with IGRINS},
  Astrophys. J. 841 (2017) 13 (18pp).
\newblock \href {https://doi.org/10.3847/1538-4357/aa6bf7}
  {\path{doi:10.3847/1538-4357/aa6bf7}}.

\bibitem{Pachucki11}
K.~Pachucki, J.~Komasa, Magnetic dipole transitions in the hydrogen molecule,
  Phys. Rev. A 83 (2011) 032501.
\newblock \href {https://doi.org/10.1103/PhysRevA.83.032501}
  {\path{doi:10.1103/PhysRevA.83.032501}}.

\bibitem{Meshkov06}
V.~V. Meshkov, A.~V. Stolyarov, A.~V. Ivanchik, D.~A. Varshalovich, {\emph{Ab
  initio} nonadiabatic calculation of the sensitivity coefficients for the
  $X^1\Sigma^+_g \rightarrow B^1\Sigma^+_u$; $C^1\Pi_u$ lines of H$_2$ to the
  proton-to-electron mass ratio}, JETP Lett. 83 (2006) 303--307.
\newblock \href {https://doi.org/10.1134/S0021364006080017}
  {\path{doi:10.1134/S0021364006080017}}.

\bibitem{Medvedev18}
E.~S. Medvedev, V.~G. Ushakov, High sensitivity of the anomalies in the
  rotational and ro-vibrational bands of carbon monoxide to small changes in
  the molecular potential and dipole moment, J. Mol. Spectrosc. 349 (2018)
  60--64.
\newblock \href {https://doi.org/10.1016/j.jms.2018.04.008}
  {\path{doi:10.1016/j.jms.2018.04.008}}.

\bibitem{Meshkov08}
V.~V. Meshkov, A.~V. Stolyarov, R.~J. Le~Roy, {Adaptive analytical mapping
  procedure for efficiently solving the radial Schr\"odinger equation}, Phys.
  Rev. A 78 (2008) 052510.
\newblock \href {https://doi.org/10.1103/PhysRevA.78.052510}
  {\path{doi:10.1103/PhysRevA.78.052510}}.

\bibitem{Balashov23}
A.~A. Balashov, K.~Bielska, G.~Li, A.~A. Kyuberis, S.~W{\'o}jtewicz,
  J.~Domys{\l}awska, R.~Ciury{\l}o, N.~F. Zobov, D.~Lisak, J.~Tennyson, O.~L.
  Polyansky, {Measurement and calculation of CO (7-0) overtone line
  intensities}, J. Chem. Phys. 158 (2023) 234306.
\newblock \href {https://doi.org/10.1063/5.0152996}
  {\path{doi:10.1063/5.0152996}}.

\bibitem{Roueff23}
E.~Roueff, M.~G. Burton, T.~R. Geballe, H.~Abgrall, {Analysis of the first
  infrared spectrum of quasi-bound H$_2$ line emission in Herbig-Haro 7},
  Astronomy \& Astrophysics (2023, manuscript no. 45358arxaa).

\bibitem{Chung05}
C.-Y. Chung, J.~F. Ogilvie, Y.-P. Lee, {Detection of Vibrational-Rotational
  Band 0-3 of $^{12}$C$^{16}$O $X^1\Sigma^+$ with Cavity Ringdown Absorption
  near 0.96 $\mu$m}, J. Phys. Chem. A 109 (2005) 7854--7858.
\newblock \href {https://doi.org/10.1021/jp052035x}
  {\path{doi:10.1021/jp052035x}}.

\bibitem{Changala21}
P.~B. Changala, D.~A. Neufeld, B.~Godard, {Anomalous Intensities in the
  Infrared Emission of CH$^+$ Explained by Quantum Nuclear Motion and Electric
  Dipole Calculations}, Astrophys. J. 917 (2021) 16.
\newblock \href {https://doi.org/10.3847/1538-4357/ac05c8}
  {\path{doi:10.3847/1538-4357/ac05c8}}.

\bibitem{Brown73}
J.~D. Brown, G.~Burns, R.~J. Le~Roy, {Improved spectroscopic data synthesis for
  I$_2(B ^3\Pi{0_u}^+)$ and predictions of $J$ dependence for
  $B(^3\Pi_{0u}^+)-X(^1\Sigma{_g}^+)$ transition intensities}, Can. J. Phys. 51
  (1973) 1664--1677.

\bibitem{LeRoy75}
R.~J. Le~Roy, E.~R. Vrscay, {Periodicity of the oscillatory $J$ dependence of
  diatomic molecule Franck–Condon factors}, Canad. J. Phys. 53~(16) (1975).
\newblock \href {https://doi.org/10.1139/p75-198} {\path{doi:10.1139/p75-198}}.

\bibitem{Abgrall93}
H.~Abgrall, E.~Roueff, F.~Launay, J.-Y. Roncin, J.-L. Subtil, {The Lyman and
  Werner Band Systems of Molecular Hydrogen}, J. Mol. Spectrosc. 157 (1993)
  512--523.
\newblock \href {https://doi.org/10.1006/jmsp.1993.1040}
  {\path{doi:10.1006/jmsp.1993.1040}}.

\bibitem{Abgrall93L}
H.~Abgrall, E.~Roueff, F.~Launay, J.-Y. Roncin, J.-L. Subtil, {The Lyman Band
  Systems of Molecular Hydrogen}, Astron, Astrophys. Suppl. Ser. 101 (1993)
  273--321.

\bibitem{Abgrall93W}
H.~Abgrall, E.~Roueff, F.~Launay, J.-Y. Roncin, J.-L. Subtil, {The Werner Band
  Systems of Molecular Hydrogen}, Astron, Astrophys. Suppl. Ser. 017 (1993)
  323--362.

\bibitem{Meredith73}
R.~E. Meredith, F.~G. Smith, Computation of electric dipole matrix elements for
  hydrogen fluoride, J. Quant. Spectrosc. Radiat. Transfer 13 (1973) 89--114.
\newblock \href {https://doi.org/10.1016/0022-4073(73)90105-2}
  {\path{doi:10.1016/0022-4073(73)90105-2}}.

\bibitem{Medvedev85JMS}
E.~Medvedev, Determination of a new molecular constant from overtone
  vibrational spectra, J. Mol. Spectrosc. 114 (1985) 1--12.
\newblock \href {https://doi.org/10.1016/0022-2852(85)90330-3}
  {\path{doi:10.1016/0022-2852(85)90330-3}}.

\bibitem{Salumbides15}
E.~J. Salumbides, J.~Bagdonaite, H.~Abgrall, E.~Roueff, W.~Ubachs, {H$_2$ Lyman
  and Werner band lines and their sensitivity for a variation of the
  proton-electron mass ratio in the gravitational potential of white dwarfs},
  Mon. Not. Royal Astr. Soc. 450 (2015) 1237--1245.
\newblock \href {https://doi.org/10.1093/mnras/stv656}
  {\path{doi:10.1093/mnras/stv656}}.

\bibitem{Medvedev16}
E.~S. Medvedev, V.~V. Meshkov, A.~V. Stolyarov, V.~G. Ushakov, I.~E. Gordon,
  Impact of the dipole-moment representation on the intensity of high
  overtones, J. Mol. Spectrosc. 330 (2016) 36--42.
\newblock \href {https://doi.org/10.1016/j.jms.2016.06.013}
  {\path{doi:10.1016/j.jms.2016.06.013}}.

\bibitem{Neufeld21}
D.~A. Neufeld, B.~Godard, P.~B. Changala, A.~Faure, T.~R. Geballe, R.~Gusten,
  K.~M. Menten, H.~Wiesemeyer, {Observations and analysis of CH$^+$ vibrational
  emissions from the young, carbon-rich planetary nebula NGC 7027: A textbook
  example of chemical pumping}, Astrophys. J. 917 (2021) 15.
\newblock \href {https://doi.org/10.3847/1538-4357/ac05c9}
  {\path{doi:10.3847/1538-4357/ac05c9}}.

\end{thebibliography}
\bibliographystyle{elsarticle-num}

\end{document}